\documentclass[preprint]{elsarticle}

\pdfoutput=1
\usepackage{graphicx}
\usepackage{cancel}

\newcommand{\met}{$\cancel{\it{E}}_{T}$}

\journal{Journal of \LaTeX\ Templates}

\begin{document}

\begin{frontmatter}

\title{Impact of Detector Simulation in Particle Physics Collider Experiments}

\author[mymainaddress]{V.~Daniel Elvira\corref{mycorrespondingauthor}}
\cortext[mycorrespondingauthor]{Corresponding author}
\ead{daniel@fnal.gov}

\address[mymainaddress]{Fermi National Accelerator Laboratory, P.~O.~Box 500, Batavia, IL 60510-5011, USA}

\begin{abstract}
Through the last three decades, accurate simulation of the interactions of 
particles with matter and modeling of detector geometries has proven to be of 
critical importance to the success of the international high-energy physics 
(HEP) experimental programs. For example, the detailed detector modeling and 
accurate physics of the Geant4-based simulation software of the CMS and 
ATLAS particle physics experiments at the European Center of Nuclear
Research (CERN) Large Hadron Collider (LHC) 
was a determinant factor for these collaborations to deliver physics results 
of outstanding quality faster than any hadron collider experiment ever before.
 
This review article highlights the impact of detector simulation on particle 
physics collider experiments. It presents numerous examples of the use of 
simulation, from 
detector design and optimization, through software and computing development 
and testing, to cases where the use of simulation samples made a difference 
in the precision of the physics results and publication turnaround, from
data-taking to submission. It also 
presents estimates of the cost and economic impact of simulation in the CMS 
experiment. 

Future experiments will
collect orders of magnitude more data with increasingly complex detectors,
taxing heavily the performance of 
simulation and reconstruction software. 
Consequently, exploring solutions to speed up simulation and reconstruction
software to satisfy the growing demand of computing resources 
in a time of flat budgets is
a matter that deserves immediate attention. 
The article ends with a short 
discussion on the potential solutions that are being considered, based on 
leveraging core count growth in multicore machines, using new generation 
coprocessors, and re-engineering HEP code for concurrency and parallel 
computing. 
\end{abstract}

\begin{keyword}
\texttt simulation\sep high energy physics \sep particle physics \sep Geant4 \sep software \sep computing
\end{keyword}

\end{frontmatter}


\section{Introduction} \label{Introduction}

Accurate software modeling is essential to design, build 
and commission the highly 
sophisticated detectors utilized in experimental particle physics and 
cosmology. It is also a fundamental tool to analyze and interpret 
the resulting experimental data.

In particle physics, an ``event'' is composed of all the data collected in 
a single occurrence of an experiment. For example, in astroparticle
physics an event may be defined as all the data produced by 
a very energetic cosmic ray particle as it interacts with the atmosphere.
In particle colliders, an event includes all the data produced in a
beam crossing, when particle interactions occur, with the caveat that
some of the collected information may belong to previous crossings due to
the finite speed of detector electronics.  
In neutrino experiments, an event occurs when a neutrino interacts with a
nucleus of an atom in the detector.  
Simulation software in high-energy physics (HEP) experiments is designed to 
produce, in an ideal scenario, events which are identical to those resulting 
from the actual experiment. The output data format is typically the same
for simulated and real events, so that event processing and physics analysis is
performed in the same way and with the same tools.
 
\begin{figure}[htbp]
\centering
\includegraphics[width=0.8\linewidth]{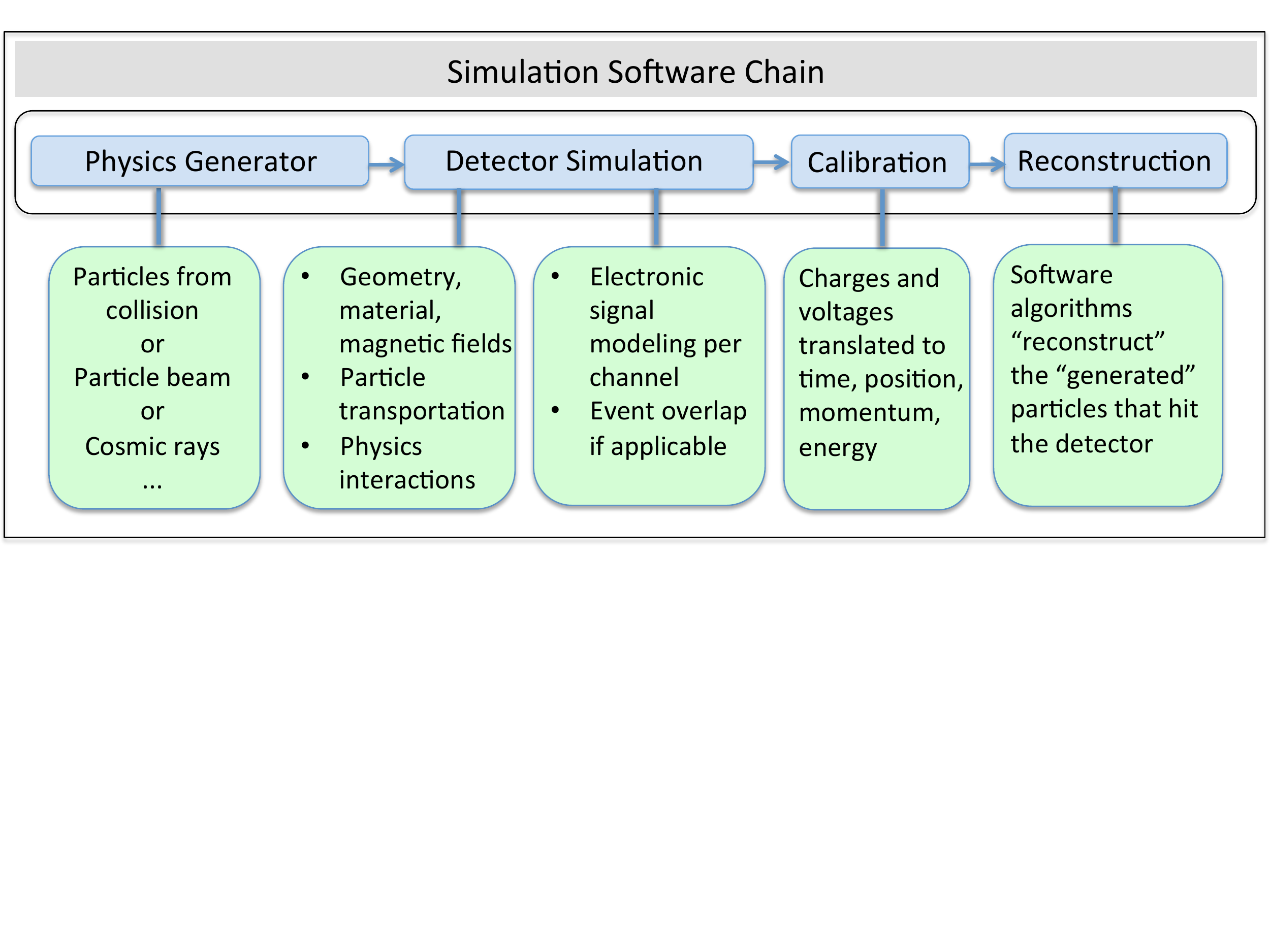}
\vspace{-3cm}
\caption{Generic simulation software chain in a typical HEP experiment.}
\label{simuchain}
\end{figure}

The typical simulation software package in HEP experiments consists 
of a chain of modules that starts with a 
generator of the physics processes of interest. Generators provide the final 
state particles in a hard collision, a cosmic ray particle shower, a neutrino 
beam with the desired energy and angular distribution, or any other
set of particles to be observed in the detector. 
A second module simulates 
the passage of the generated particles through the detector material and 
its magnetic field. In most contemporary experiments, this detector simulation 
module is based on the Geant4 simulation toolkit~\cite{geant4,geant4R},
a software package that provides the tools to describe the detector
geometry and materials, and incorporates a large number of models to
simulate electromagnetic and hadronic interactions of particles with matter.
The next module simulates the detector electronics and the calibration of
individual channels. At the end of the chain, the same algorithms used to 
identify and reconstruct individual particles and physics 
observables in real data are applied to 
simulated events. Fig.~\ref{simuchain}
describes a generic simulation software chain and the functionality of each 
module for a typical HEP experiment.

There was a time, however, when experiments modeled their detectors 
using simple analytic or 
back-of-the-envelope calculations, toy simulations, or parametrizations based 
on theoretical predictions, or experimental data of the passage of particles 
through matter and 
electromagnetic fields. The era of detailed detector 
simulation started in the late 1970's and early 1980's when the 
Electron Gamma Shower software (EGS)~\cite{egs} was developed, and the GEANT 
team released GEANT3~\cite{geant3}, a software toolkit that allowed the 
experiments to describe 
complex geometry, propagate particles through these geometries, and trace the 
incident and secondary 
particles as they interact with the different materials according to physics 
models implemented as part of the toolkit. It did not take too long before 
GEANT3 was widely used at the European Center for Nuclear Research (CERN), 
the German Deutsches Elektronen-Synchrotron (DESY) and the US Fermi National
Accelerator Laboratory (FNAL) experiments. 
Although initially of limited use because of the insufficient speed of
computers in those days, these 
GEANT3-based full detector simulation software applications became the norm 
through the nineteen nineties 
and revolutionized the way the particle physics community plans experiments, 
design detectors, and perform 
physics measurements.

It should not be necessary to clarify that particles cannot be 
discovered through 
simulation. For example, the Higgs boson can only be “observed” in 
simulated data if its 
production mechanisms and decay modes are coded in the event generator
package that feeds the detector simulation module. What a simulation 
does is to teach physicists what mark the Higgs boson would leave in the real 
detector if it were present in the data sample. For instance, the Standard 
Model of Elementary Particles and their Interactions (SM) predicts that the 
Higgs boson decays into two photons with certain kinematic properties. Using 
simulation, physicists designed detectors and data analysis procedures that 
targeted Higgs searches with the goal to identify events with the 
characteristics predicted by theories. In 2012, the observation of events of
this kind 
in real data signaled a discovery by the ATLAS~\cite{atlashiggs} and 
CMS~\cite{cmshiggs} experiments at the CERN Large Hadron Collider (LHC). 
Real-life physics 
measurement procedures are more complex than the example presented here 
but the general idea is always the 
same in what pertains to the role of simulation.

During the last three decades, simulation has proven to be of critical 
importance to the success of HEP experimental programs. For example, 
the detailed detector modeling and accurate physics of the 
CMS~\cite{cmssimulationL, cmssimulationB, cmssimulationE} and 
ATLAS~\cite{atlassimulationC, atlassimulationM, atlassimulationR}  
Geant4-based simulation software was a determinant factor for these 
experiments to deliver physics results of outstanding quality faster than 
any hadron collider experiment ever before. Simulation software at the LHC 
experiments is more accurate and yet runs much faster than their predecessors 
at the Tevatron, resulting in a faster analysis turnaround and smaller 
systematic uncertainties in the measurements. As an example, the CMS experiment 
simulated, 
reconstructed and stored more than ten billion events during the Run 1, 
2010-2012, data-taking period. This effort required more than half of the 
total computing resources allocated to the experiment. Simulation samples of
better quality and in larger quantities, evolving detector and computing 
technology, and a wealth of experience from pre-LHC experiments on calibration 
and data analysis techniques, improved significantly the precision of the 
measurements in the current generation of experiments and shortened the 
time between data-taking and public results or journal submission. 

This review article focuses on collider experiments and places the emphasis on 
the detector simulation part of the simulation chain,
particularly on the Geant4-based module. It explains the concepts 
of Full and Fast Simulation and the tuning of the many physics
models involved. It presents numerous examples where the use of detector 
simulation made a difference in the precision of the physics 
measurements and publication turnaround. For this, it borrows heavily 
but not exclusively from the CMS (LHC) and D0 (Tevatron) experiments, 
drawing from the 
author's
personal experience. The last two sections include estimates of the cost 
and economic impact of simulation in HEP and introduce concepts 
that will shape the detector simulation efforts of the future. 

\section{Toy, Parametrized, and Full Simulation} \label{ToyParaFull}

The common classification of simulation code in ``full'' and ``fast'' 
is misleading since it refers to the speed of the software 
application, a relative concept, rather than to its nature. Instead, it is more 
useful to 
introduce the concepts of Toy Simulation (ToySim), Parametrized Simulation 
(ParSim), and Full Simulation (FullSim). Often, physicists refer to simulation 
software and simulated data samples as Monte Carlo (MC) Simulation and MC 
samples. The expression makes an analogy between
randomness in 
casino games and randomness built in the methods to integrate the
equations of motion for particles traversing matter and magnetic fields within
the detector. Physics interactions occurring with 
different probabilities in the detector material, due to
quantum mechanics, introduces a second source of randomness.

A Toy Simulation is a basic tool that may consist of a few simple 
analytical equations. It may be used to demonstrate large physics effects or 
biases in a measurement as a proof of principle, but it is often not 
accurate enough to make predictions of the size of these effects or biases. 
ToySim does 
not involve a detector geometry description or the detail of particle shower 
development. Typically, ToySim events take a small fraction of a second to 
generate.

Most modern Full Simulation applications in HEP are based on Geant4. 
They include 
detailed geometry and magnetic fields descriptions and accurate modeling of 
electromagnetic and hadronic particle showers provided by the many collections 
of physics models either adopted or developed, as well as optimized and 
validated by the Geant4 
team. FullSim is the slowest but the option of choice for most studies, 
as fast and large scale computing became available and made it possible for
big experiments to 
generate tens of billions of 
complex detector events per year. FullSim of complex HEP detectors typically 
takes between a few seconds and a few minutes per event to generate. 

A Parametrized Simulation involves a geometry description, parametrizations 
of the energy response of single particles measured in data or 
extracted from Full 
Simulation or theoretical calculations, a mechanism to randomize the results of 
the parametrizations, and magnetic field maps. The goal is to make the ParSim 
much faster than the FullSim, typically a couple of orders of magnitude, and 
almost as accurate. The accuracy limitations of ParSim tools are typically 
more severe in describing particle shower shapes and related detector 
effects such as energy leakage 
beyond the detector boundaries, and regions of phase space where data 
are not available. ParSim tools may also be based on GEANT3 or Geant4 
up-to-the-point of the first interaction of primary incident 
particles with matter, 
after which tools such as GFLASH\cite{gflash} may be used to describe 
shower 
shapes and response with better or worse accuracy, depending on the number of 
parameters used to describe the showers and the time performance cost the
experiment is willing to pay for accuracy. GFLASH-based 
simulations, or a simulation that uses particle shower libraries constructed 
from FullSim events, are sometimes used in combination with FullSim to model 
sub-detectors with high particle occupancy, for example those located near the 
beam pipe in collider experiments. ParSim is commonly used for detector design 
studies that require to test many geometry scenarios, and to generate signal 
samples for new physics that involve scans over a large sector of theory 
parameter space. In most cases, the output of ParSim 
applications has the same format as the output from FullSim and the 
data coming from the real detector. Instead, an event in a 
ToySim sample is often a collection of particles with their four-vector 
position and momentum. ParSim events typically take on the order of a second 
to a few seconds to generate.

An example of a ParSim is the CDF Fast Simulation software package, also 
known as QFL, developed in 1989. (In the absence of public documentation, the
information about the CDF QFL and GFLASH parametrized simulations was obtained 
from private communication with Soon Young Jun, Kenichi Hatakeyama, and 
Marjorie Shapiro.)
QFL was based on a detailed geometry description and 
accurate parametrization of single particle showers. Pion information was 
extracted from minimum bias and track triggers in the 0.75-20~GeV range and 
test beam pion data were utilized in the higher 57-145~GeV energy range. This 
approach provided a suitable solution in the 1989-1990 (Run 0) and 
1991-1996 (Run 1) data taking periods to deal 
with the fact that GEANT3-based simulations were still too slow to be 
practical. In 2002, QFL was replaced with a hybrid application that utilized 
GEANT3 to model the detector geometry and track particles, but only up to 
the point when the incident particles interact with the detector material
for the first time, with subsequent showers 
modeled with GFLASH parametrizations tuned to test beam and collider data. 
In contrast, the D0 experiment followed a 
FullSim approach from the 
start of Run 1 in 1991, since the absence of a solenoidal magnetic field within 
the tracker made it challenging to tune a ParSim tool with high precision. 
However, in the interest of speed, D0 introduced approximations 
such as a “mixed plate” geometry option, based on average material, 
rather than a “full plate” option,
based  on the actual material detail. In addition, showers 
were truncated once 95\% of the energy of the shower was released in the 
detector material, approximation that had a significant impact on the 
description of hadron energy response linearity, and on shower shapes. Twenty 
years later, the availability of more advanced computing and software systems 
has 
allowed the ATLAS and CMS experiments to develop FullSim applications and 
generate tens of billions of Geant4-based events per year with unprecedented 
geometry, material and magnetic field detail, as well as significantly improved
physics models. In CMS, the amount of CPU (Central Processing Unit in a
computer) time spent per event during the
2009-2013 (Run 1) period ranged 
between ~15 seconds for the simplest events to three minutes for more 
complex events such as those with top quarks or many high momentum jets. 
In contrast, it took up to one hour per event to generate the Monte Carlo 
sample utilized in the $W$ boson mass measurement at the Tevatron D0 
experiment a decade earlier. In the early 1990's, most of the FullSim MC 
samples used by the Tevatron experiments consisted of a few hundreds of 
thousands events and included approximations which introduced severe 
limitations in their utilization.

\section{Simulation Software Tools} \label{SimuToolsPhysics}

At the core of the impressive agreement between simulation and data at the 
LHC is Geant4, the detector simulation toolkit developed, maintained, and 
supported by the Geant4 Collaboration, and currently used by most
HEP experiments. The first production version of 
Geant4, the Object Oriented C++ incarnation of the GEANT family, was
released in 1998. Since then, its areas of application have extended to 
include high-energy, nuclear and accelerator physics, as well as medical 
science and treatment, and space exploration. 
The GEANT saga started in 1975 with the release of GEANT1,
a very basic framework to drive a simulation 
program providing a user-defined output with histograms. 
GEANT2 was
released in 1976 as an extension of GEANT1. It had a
more complete set of physics models, including electromagnetic (EM) showers 
based
on a subset of the Electron Gamma Shower (EGS)~\cite{egs} package, 
multiple scattering, particle decay, and 
energy loss.
GEANT2 was used by several Super Proton Synchrotron (SPS) experiments at CERN.
The breakthrough came in 1980 with GEANT3, an evolution of the
GEANT software that contained a data 
structure to describe complex geometries at the level required by the
experiments planned for the 1980's. 
GEANT3 was first used in the OPAL experiment at the CERN Large Electron-Positron
Collider (LEP) and then adopted by other LEP experiments, such as L3
and ALEPH. Experiments at DESY and FNAL soon followed suit. 

Other simulation tools worth mentioning are FLUKA~\cite{fluka} 
and MARS~\cite{mars}. FLUKA is a fully integrated particle physics simulation
package with many 
applications in HEP and engineering, 
shielding, detector and telescope design, cosmic ray studies, dosimetry, 
medical physics and radio-biology. MARS is a software package for
the simulation of particle transport and interactions with matter 
in accelerator, detector, spacecraft and shielding components. It is widely
used to model radiation shielding enclosures. 

The success of Geant4-based simulation at the LHC was not due to magic,
but the result of many years of hard work and partnership between 
the experiments and the Geant4 Collaboration. It involved a lengthy process to
develop, optimize, and validate the many physics models available for use
in Geant4 to describe the interaction of particles with the detector material. 
Different fora, such us meetings of the Geant4 physics 
groups and dedicated workshops centered on the topic of Geant4 physics 
validation, served as vehicles of communication, discussion, and information 
exchange. 

\subsection{Geant4 in a Nutshell} \label{G4nutshell}

Geant4 is a toolkit because experimenters assemble their simulation
package by selecting, implementing and integrating different elements
such as geometry (from available Geant4 shapes), materials, magnetic fields,
a method of integration of the equation of motion, and a physics list composed
of a subset of the many available physics models. These models
describe interactions with matter for different types of incident particles 
with energies as low as 250~eV and as high as 100~TeV. 
Geant4 provides interfaces to 
communicate with the experiment's software framework, which connects to various
services and the other modules in the simulation chain. A detailed description
of the Geant4 toolkit may be found elsewhere~\cite{geant4}. In this section,
the focus is on the Geant4 physics and its validation.

\subsection{The Physics of Geant4} \label{PhyG4}

The Geant4 simulation tool kit is not only a software engine to propagate 
particles through a geometric representation of a detector. It comes with a 
remarkably complete library of physics models to simulate the interactions of 
particles with matter. Electrons, muons, and charged hadrons interact 
electromagnetically with matter through processes such as ionization, 
bremsstrahlung, pair production, and multiple scattering. 
Examples of photon interactions are the photoelectric, Compton, conversion, 
and Rayleigh scattering processes. 
Hadrons, such as pions, kaons, protons, and neutrons are 
abundantly produced in HEP collisions and interact strongly with nuclei in the 
detector material. Although QCD is the theory that describes all hadronic 
interactions, perturbative calculations may be applied only to a small region 
of phase-space, while hadronization and nucleus interactions are 
non-perturbative and may only be described by approximate models. A hadronic 
shower is the result of the interaction of a single hadron with the detector 
material. It consists of a cascade of strong interactions producing large 
numbers of secondary particles of diminishing energies. The development of a 
hadronic shower covers a large range of energy scales, from the hundreds of 
GeV down to, in the case of neutrons, thermal energies. 
Hadronic showers are difficult to 
model and are of critical importance to simulate events with quark-initiated 
or gluon-initiated jets in the experiments.

\begin{figure}[htbp]
\centering
\includegraphics[width=0.8\linewidth]{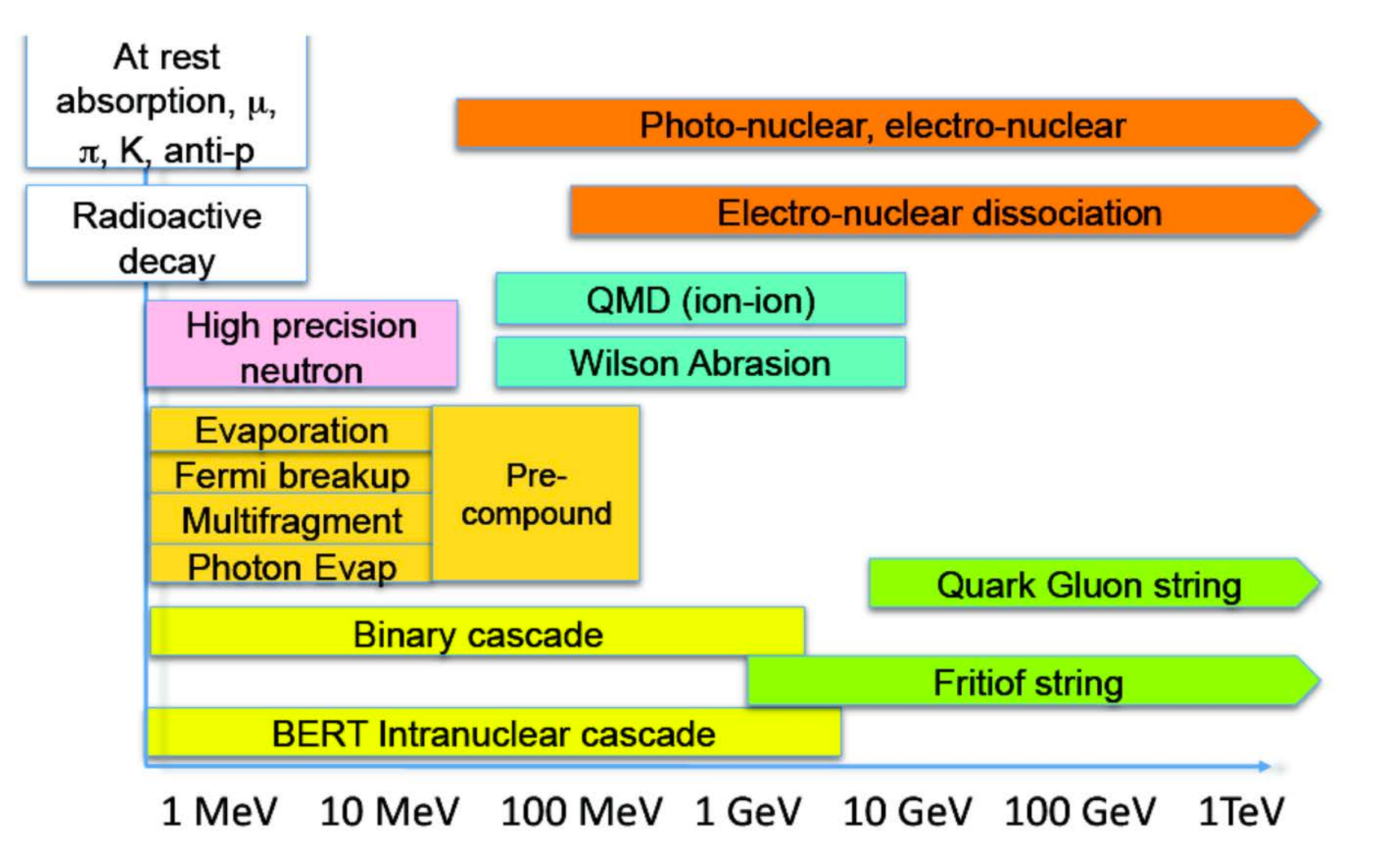}
\caption{Partial inventory of Geant4 hadronic physics models. 
``Physics lists'' are assembled from a selection of models which are
valid in different energy ranges for different particle types.}
\label{G4PhysMod}
\end{figure}

As illustrated in Fig.~\ref{G4PhysMod}, Geant4 provides a rich inventory of 
hadronic physics models, typically assembled in “physics lists” where energy 
ranges and model-to-model transition regions are defined and optimized for 
different incident particles.

\subsection{Physics Validation of Geant4} \label{ValidG4}

The task of improving the Geant4 physics models from comparisons 
between MC predictions and dedicated or thin-target experiments is part of the
Geant4 development process. Thin-target experiments consist of directing 
beams of particles of different types onto thin targets made of the materials
typically used in HEP experiments. The measured cross-sections of different
nuclear interactions, angular distributions, and particle multiplicities are
then  used to validate individual models at the microscopic, single-interaction 
level; examples are the CALICE\cite{calice}, HARP\cite{harp1,harp2,harp3}, 
NA49\cite{na49a,na49b,na49c}, and NA61\cite{na61a,na61b} experiments. 
Selecting the physics models to be used in a Geant4 application is not a 
one-size-fits-all operation, in the 
sense that some models may represent the data better than others for a given
particle type, detector material, and energy range. The reason is that these
models typically depend on parameters which are adjusted to the available
experimental data, and not all particles, energy ranges, and target materials
are present in the currently available thin-target experimental data-sets.
Therefore, it is essential for particle physics experiments to validate 
their Geant4-based simulation software by comparing MC predictions with
test beam or collider data. Test beam experiments are designed to
study the performance of realistic detector prototypes or solid
angle slices of the actual detectors. The data collected
are not only essential to understand, optimize and calibrate the detectors, but
are also of critical importance to validate the experiment's simulation 
software.
Experiments also contribute to the MC validation process by comparing Geant4 
predictions with in situ measurements performed using data collected during 
their physics runs. Examples of the latter are studies
to understand the modeling of single charged tracks, jet 
response and resolution, and shower shapes. 

\subsubsection{Thin-target Experiments} \label{thintarget}

\begin{figure}[htbp]
\centering
\includegraphics[width=0.9\linewidth]{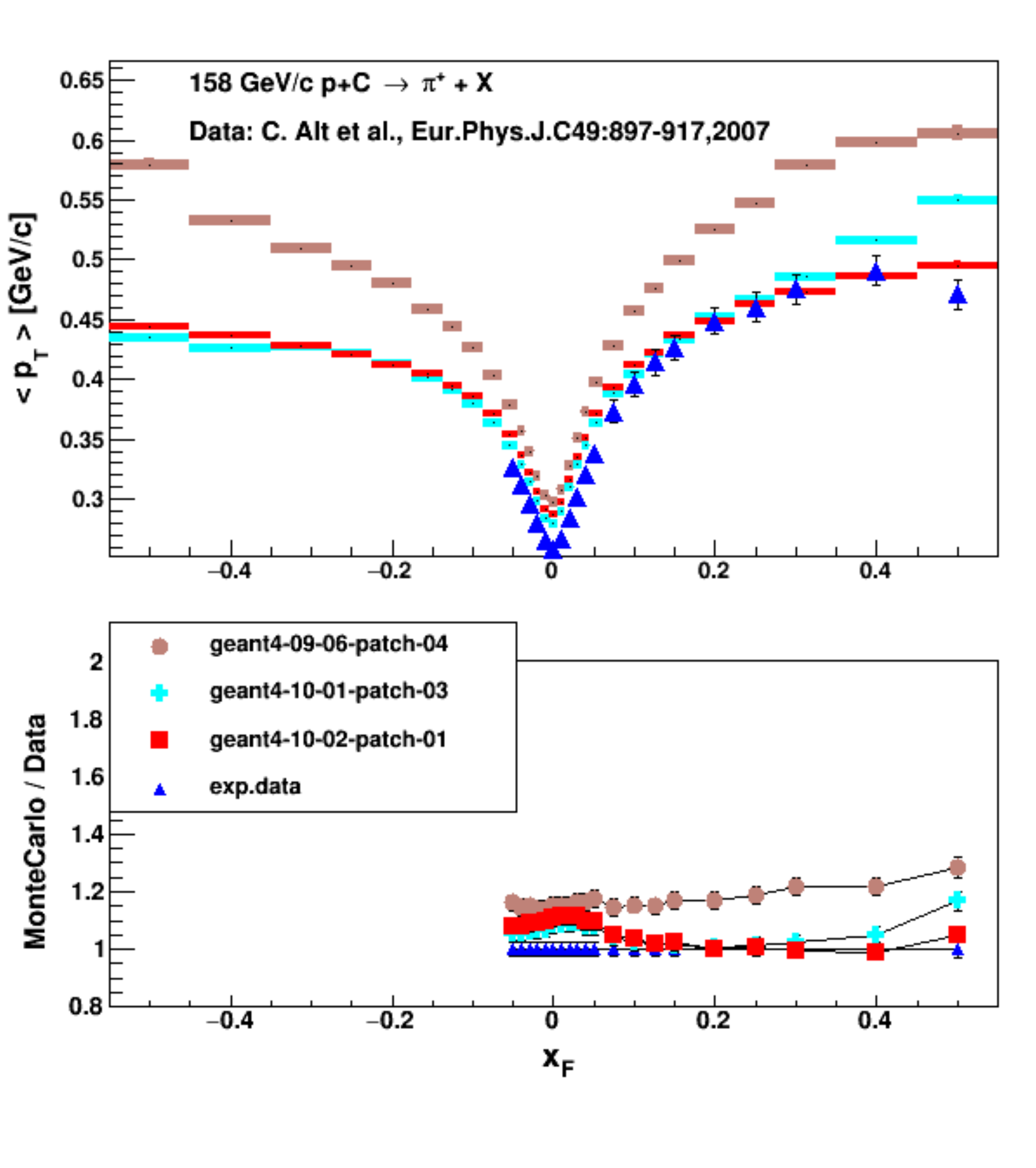}
\caption{Comparison between NA49~\cite{na49a} results and Geant4 predictions
for successive Geant4 versions for which the FTF and Bertini models have been 
improved. The $\pi^{+}$ average momentum in the plane transverse to 
the particle beam, $p_T$, is presented as a function of Feynman $x$ ($x_F$), 
for events with a final state $\pi^{+}$ originating from a 158~GeV/c proton 
beam that hits a carbon target (${\rm p+C} \rightarrow \pi^{+}{\rm +X}$). 
$x_F$ is 
defined as the ratio between the measured longitudinal momentum of the 
pion and the maximum value allowed by the kinematics of the collision, 
$x_F=p^{\pi}_{z}/p^{\pi}_{z \thinspace max}$.}
\label{g4thin1a}
\end{figure}

\begin{figure}[htbp]
\centering
\includegraphics[width=0.9\linewidth]{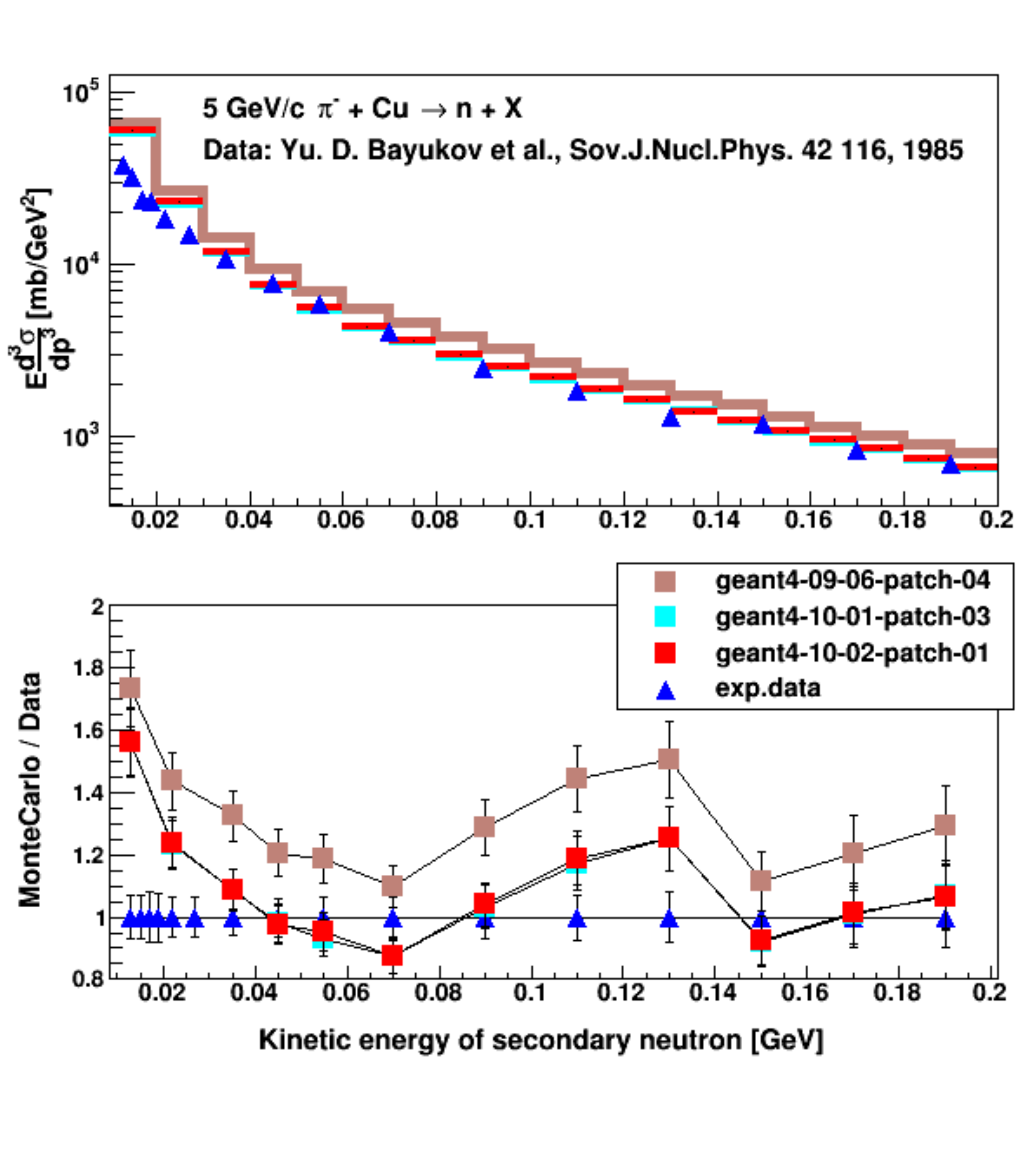}
\caption{Comparisons between ITEP-771~\cite{itep} 
experiment results and Geant4 predictions, for successive Geant4
versions for which the FTF and Bertini models have been improved.
The $\pi^{-} \thinspace {\rm (5 GeV)+Cu} \rightarrow {\rm n + X}$ 
cross section is
shown as a function of the neutron kinetic energy.}
\label{g4thin1b}
\end{figure}

\begin{figure}[htbp]
\centering
\includegraphics[width=1.0\linewidth]{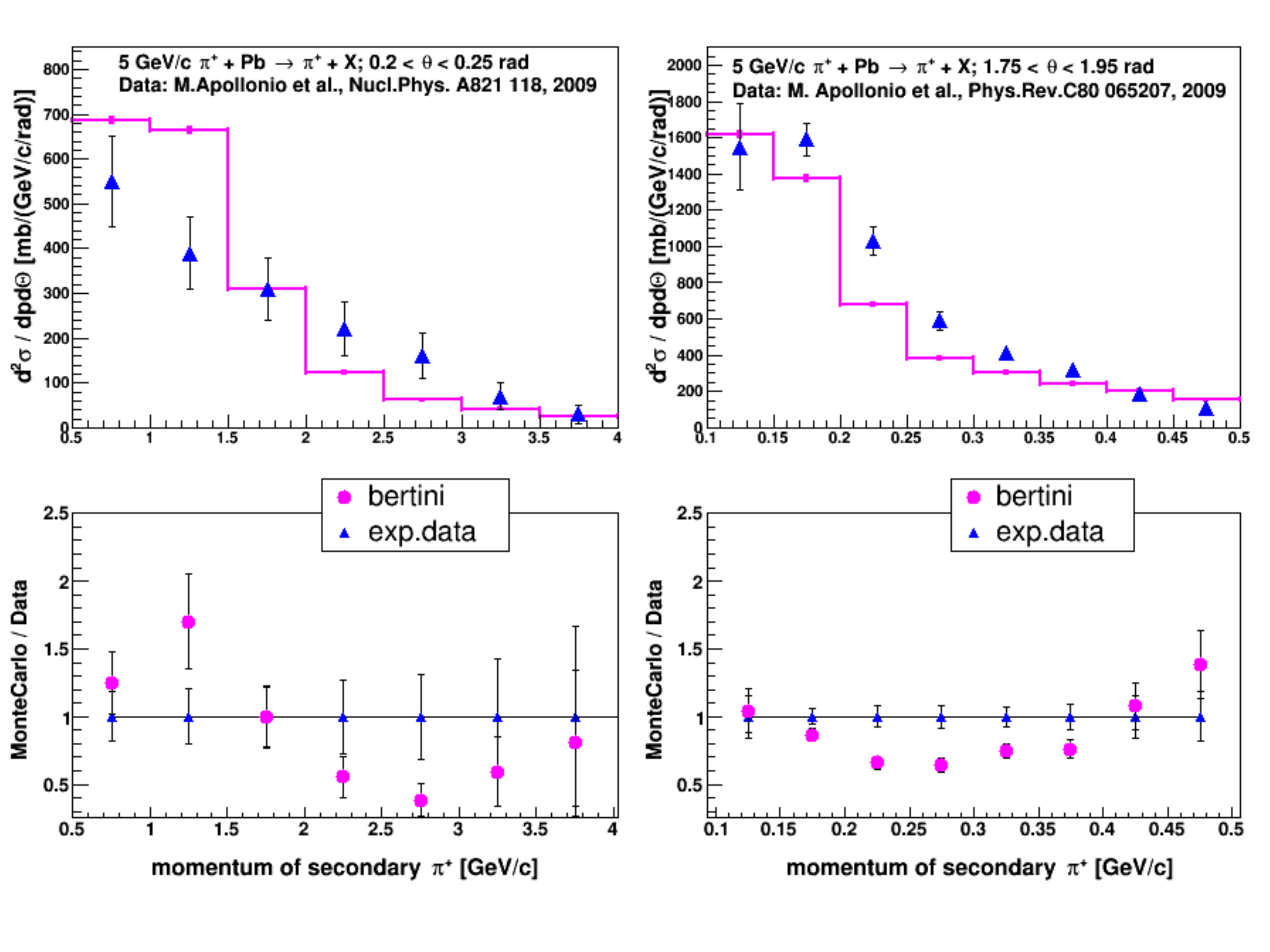}
\caption{Polar angle distributions of the outgoing pion with respect to the 
direction of the incident pion beam versus the momentum of the secondary pion 
in $\pi^{+} \thinspace {\rm (5 GeV)+Pb} \rightarrow \pi^{+} {\rm + X}$ 
collisions 
recorded by the HARP experiment~\cite{harp1,harp2}. The comparison is made 
using Geant4 version 10.2.p01 with the Bertini model for two polar 
angle ranges.}
\label{g4thin2}
\end{figure}

Figs.~\ref{g4thin1a},~\ref{g4thin1b},~\ref{g4thin2} illustrate the validation 
procedure to 
evaluate the accuracy of the FRITIOF Precompound (FTFP)~\cite{fritiof} and 
Bertini Cascade~\cite{bertini} models in Geant4. The Geant4 FTFP 
model handles the formation of strings in the hadron-nucleon 
collision and the subsequent de-excitation of the remnant nucleus.
The Geant4 Bertini Model generates the final state for hadron inelastic 
scattering by simulating the intra-nuclear cascade. This cascade results
from the collision of incident hadrons with protons and neutrons in the 
nucleus of the target material, which produce secondary 
particles that interact with other nucleons.
Fig.~\ref{g4thin1a} shows results for a thin-target experiment with
a final state $\pi^{+}$ originating from a 158~GeV/c proton beam that hits a 
carbon target (${\rm p+C} \rightarrow \pi^{+}{\rm +X}$). 
The observable is the $\pi^{+}$
average momentum in the plane transverse to the particle beam, $p_T$, 
as a function of Feynman $x$ ($x_F$), defined as the ratio between the measured 
longitudinal momentum of the pion and the maximum value allowed by the 
kinematics of the collision, $x_F=p^{\pi}_{z}/p^{\pi}_{z \thinspace max}$. The 
improvement in the agreement between the Geant4 prediction 
and the NA49~\cite{na49a} experimental data is clearly visible, 
as updates to the FTFP physics model are incorporated to successive Geant4 
releases. Fig.~\ref{g4thin1b} shows a Geant4-to-data comparison of the
ITEP-771~\cite{itep} experiment measurement of the 
$\pi^{-} \thinspace {\rm (5 GeV)+Cu} \rightarrow {\rm n + X}$ 
cross section as a 
function of the neutron kinetic energy. 
A trend of improvement in the agreement 
between data and MC is observed for different versions of Geant4, as updates to
the Bertini model are incorporated. 
Fig.~\ref{g4thin2} shows the polar angle distribution of the outgoing pion with 
respect to the direction of the incident pion beam versus the momentum of the 
secondary pion in 
$\pi^{+} \thinspace {\rm (5 GeV)+Pb} \rightarrow \pi^{+} {\rm + X}$ 
collisions recorded by the HARP experiment~\cite{harp1,harp2}. 
The data is compared to a Geant4 prediction,
version 10.2.p01, based on the the Bertini model. 
Figs.~\ref{g4thin1a},~\ref{g4thin1b},~\ref{g4thin2} 
are just examples of the many comparison plots available in the Geant4 software 
validation suite that is used to validate and improve physics models during 
the development process of a new Geant4 release. 

\subsubsection{HEP Experiments} \label{hepexperiments}

\begin{figure}[htbp]
\centering
\includegraphics[width=0.9\linewidth]{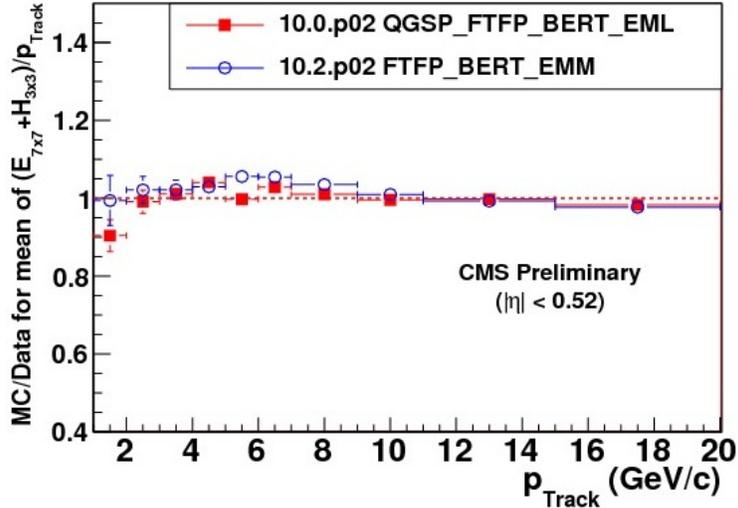}
\vspace{-3.5cm}
\caption{MC-to-data comparison results for 
isolated charged tracks in CMS min-bias events~\cite{cmsg4valid}. 
The vertical axis displays 
the MC-over-data ratio of the ratio of the energy measured in the 
calorimeters over the momentum measured in the tracker, for a single isolated 
track. This ratio is measured as a function of the track momentum, 
$p_{\rm Track}$. The simulation is performed for two different choices of Geant4
physics lists. Error bars are statistical uncertainties only.}
\label{cmstrack}
\end{figure}

\begin{figure}[htbp]
  \centering
\vspace{-1.5cm}
  \begin{minipage}{.7\textwidth}
    \centering
    \includegraphics[width=\textwidth]{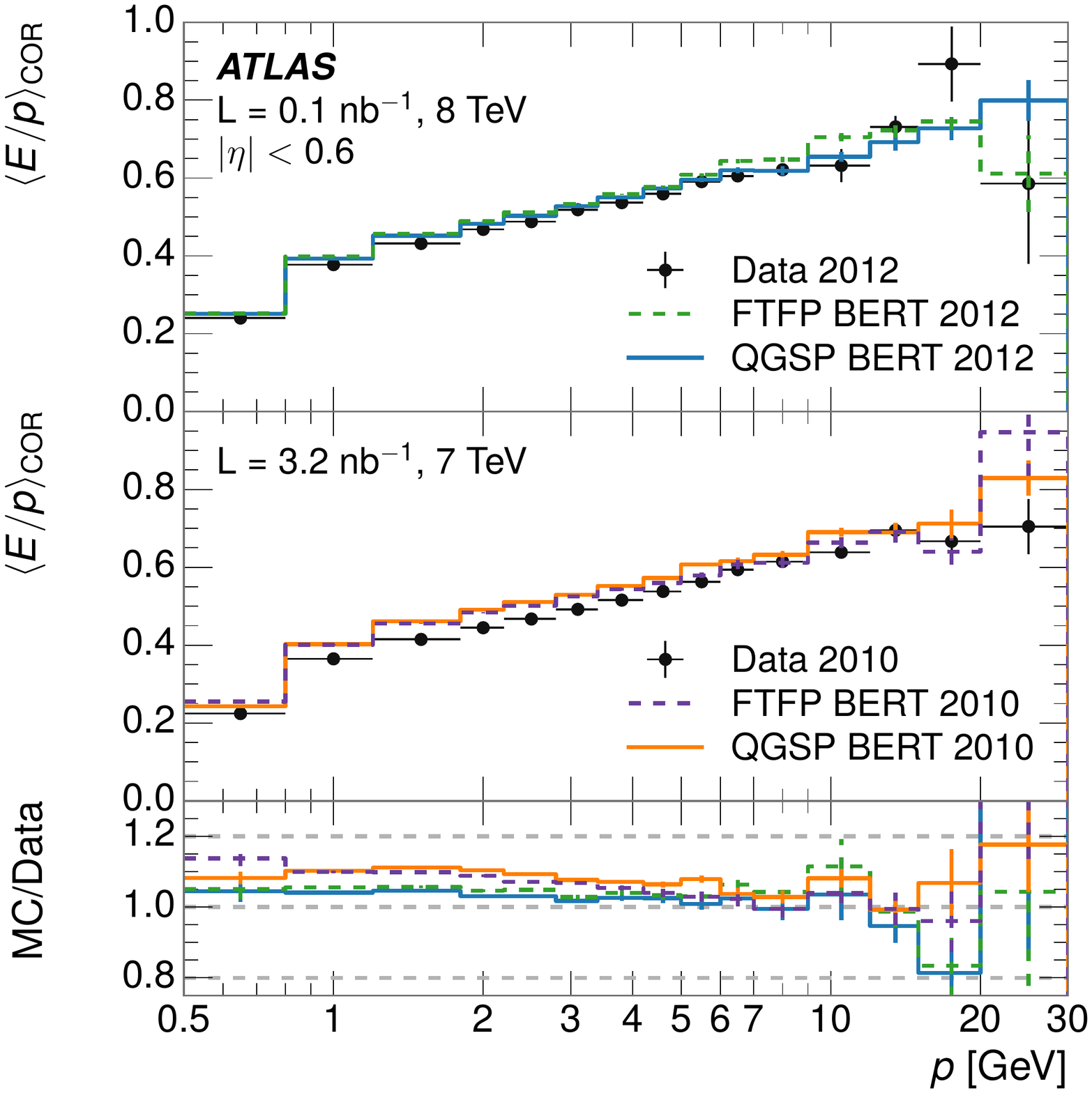}
    \vspace{-3.0cm}
  \end{minipage}
  \begin{minipage}{.7\textwidth}
    \centering
    \includegraphics[width=\textwidth]{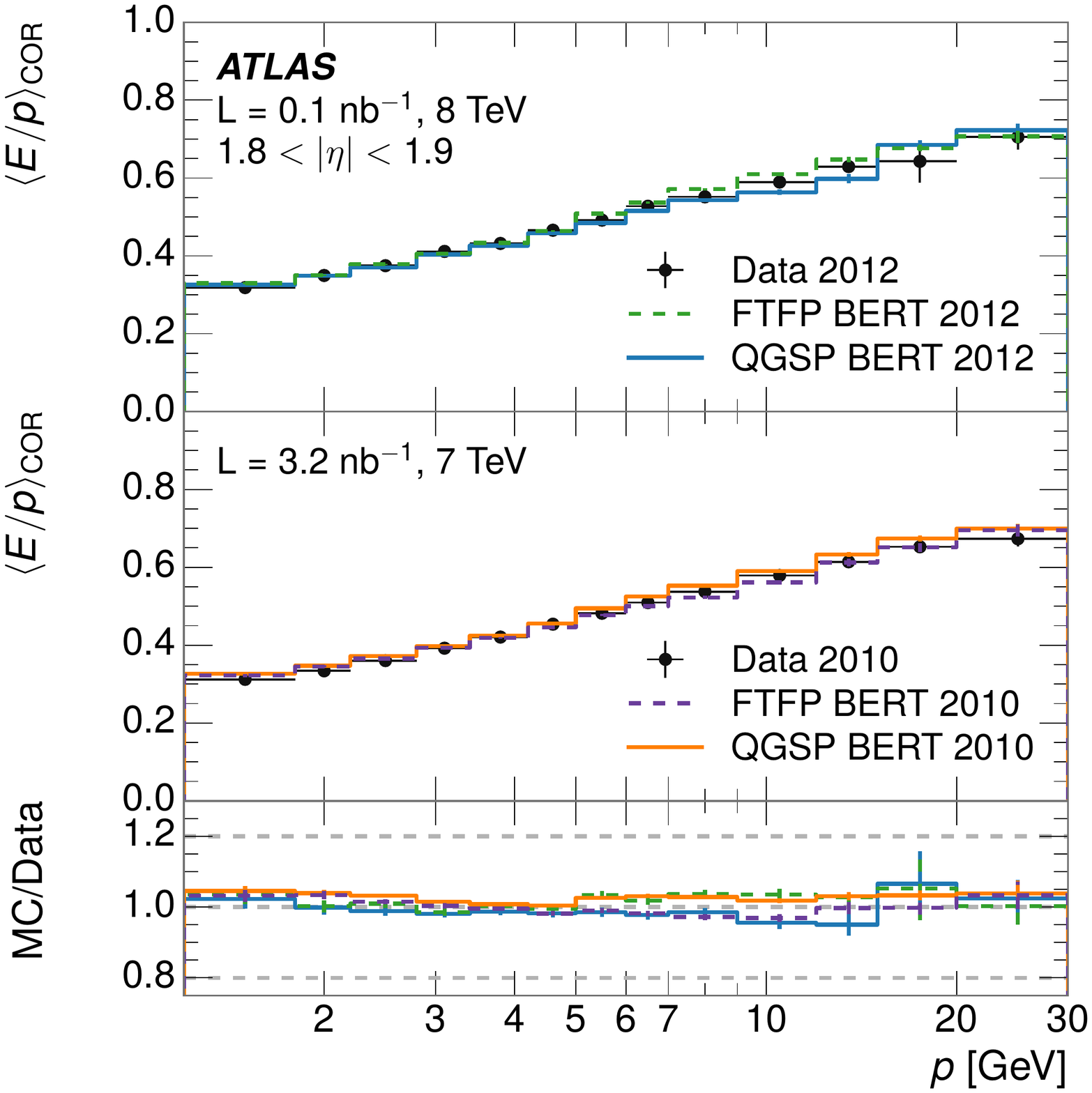}
  \end{minipage}
 \vspace{-1.5cm}
\caption{ATLAS measurement of $E/p$ from min-bias data, with $E$ the energy
deposited by an isolated charged track in the calorimeter, and $p$ the
momentum measured in the tracker. The  background
subtracted mean ratio, $<E/p>_{\rm COR}$, is plotted as a function of the
track momentum $p$ in two pseudorapidity regions, $|\eta|<0.6$ and 
$1.8<|\eta|<1.9$, using the 2010 and 2012 data samples. The measurements are
compared to Geant4-based simulation predictions in the $p=0.5-30$~GeV range, 
using the FTFP\_BERT and QGSP\_BERT physics lists~\cite{atlasg4valid}.}
\label{atlastrack}
\end{figure}

HEP experiments validate their Geant4-based simulation software 
using data collected during their physics runs or in dedicated test beam 
experiments. Fig.~\ref{cmstrack} shows CMS MC-to-data comparison results for 
isolated charged tracks in min-bias events, defined as a beam crossing with
the requirement of a hard collision~\cite{cmsg4valid}. 
The vertical axis displays the MC-to-data ratio of the ratio of the 
energy measured in the 
calorimeters over the momentum measured in the tracker, for a single isolated 
track. This ratio is measured as a function of the track momentum, 
$p_{\rm Track}$. The energy was measured in a $7\times7$ cell cluster in
the electromagnetic calorimeter (ECAL) and a $3\times3$ cell cluster in the 
hadronic calorimeter (HCAL) in the region covered by a polar angle such that 
the pseudorapidity of the particles, 
$\eta \equiv - {\rm tan} (\theta/2) < 0.52$. The squares and circles 
correspond to different versions
of Geant4, 10.0.p02 and 10.2.p02, and collections of physics
models or physics lists. The FTFP\_BERT\_EMM list is the CMS experiment default
Geant4 physics list (as of May 2017), based on the Bertini and FTFP models.
As illustrated, the simulation models the track data within less than 5$\%$
in the 1-20~GeV/c $p_{\rm Track}$ range.
Fig.~\ref{atlastrack} shows a similar measurement performed by the ATLAS
experiment. ATLAS measured the $E/p$ from min-bias data, with $E$ the energy
deposited by an isolated charged track in the calorimeter and $p$ the
momentum measured in the tracker. The  background
subtracted mean ratio, $<E/p>_{\rm COR}$, is plotted as a function of the
track momentum $p$ in two pseudorapidity regions, $|\eta|<0.6$ and 
$1.8<|\eta|<1.9$, using the 2010 and 2012 data samples. The measurements are
compared to Geant4-based simulation predictions in the $p=0.5-30$~GeV range, 
using the FTFP\_BERT and QGSP\_BERT physics lists. In the kinematic region with
small enough statistical uncertainties, the study proved that the simulation
models the data to within 5\%~\cite{atlasg4valid}.
The previous examples are an illustration of how the increasing speed of 
computers during the last couple of decades allowed 
the LHC experiments to generate large-enough samples of 
simulated events to test different sets of Geant4 competing physics models
and select those that describe the data best. 

\begin{figure}[htbp]
  \centering
  \begin{minipage}{.9\textwidth}
    \centering
    \includegraphics[width=\textwidth]{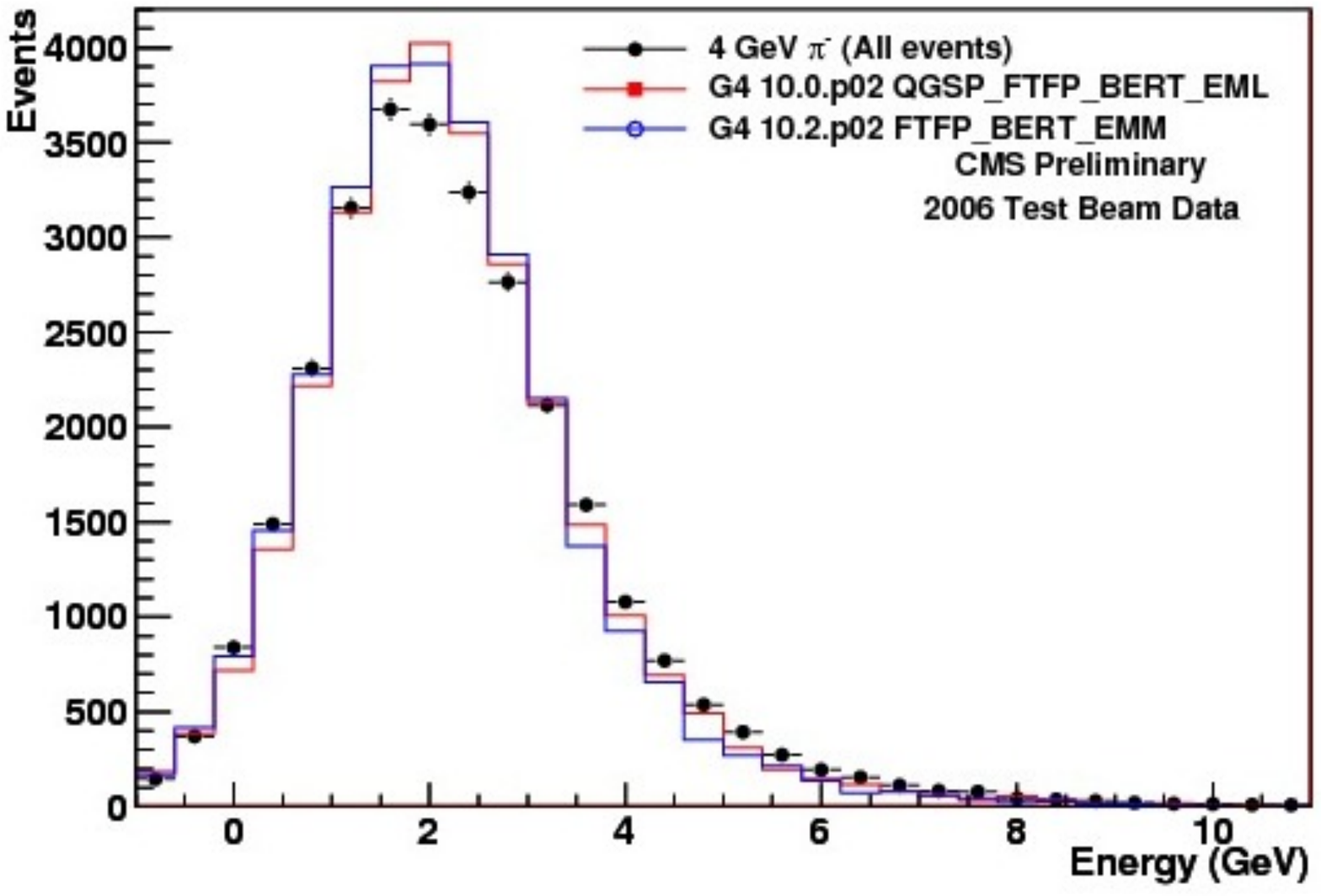}
    \vspace{-7cm}
  \end{minipage}
  \begin{minipage}{.8\textwidth}
    \centering
    \includegraphics[width=\textwidth]{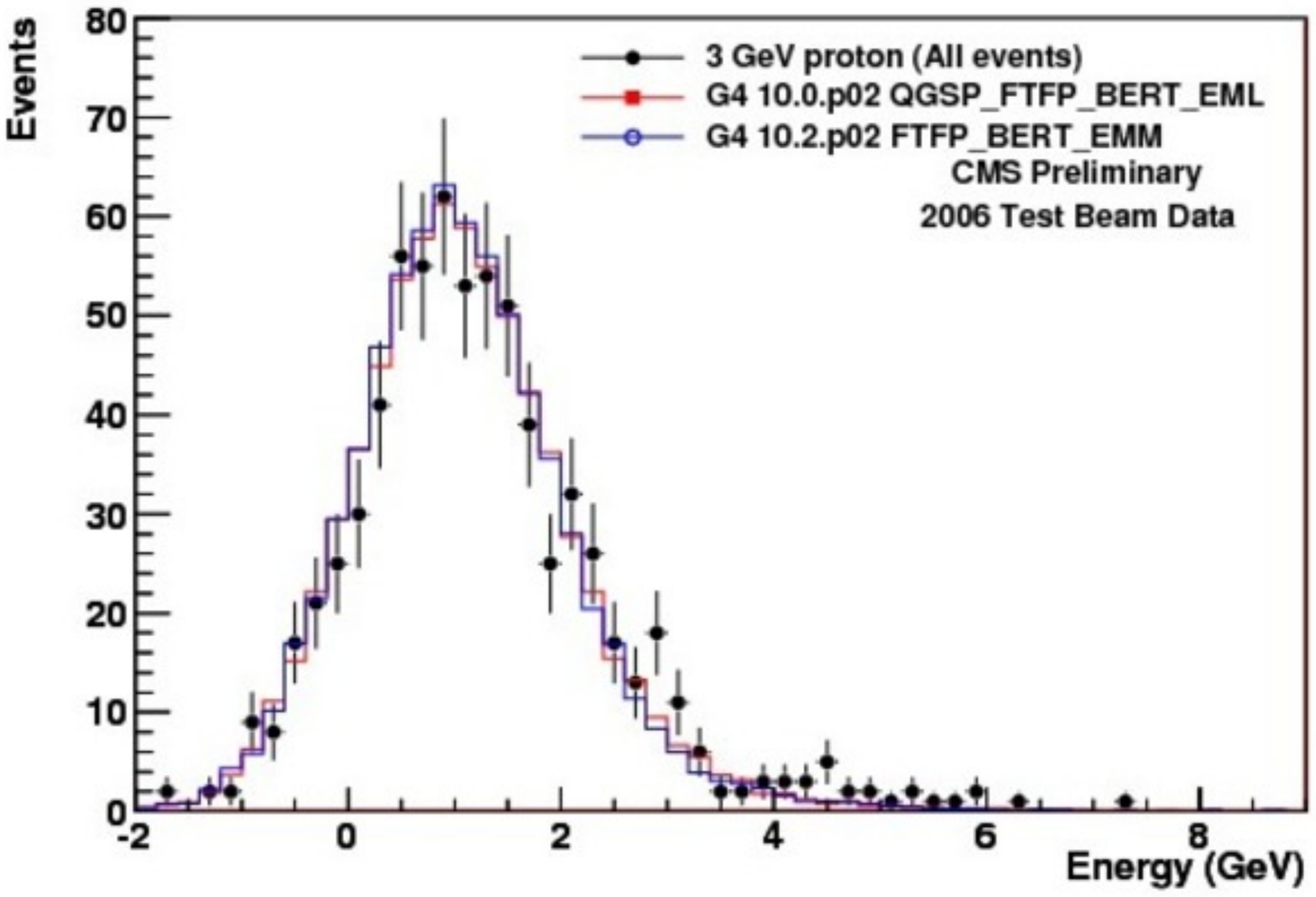}
  \end{minipage}
 \vspace{-3cm}
\caption{Comparisons of Geant4 and 2006 CMS test beam data measurements
of the response distribution, or response function, for 4~GeV pions and 
3~GeV protons incident onto a solid angle slice of the
CMS ECAL and ECAL calorimeters~\cite{cmsg4valid}. 
The simulation is performed for two different choices of Geant4 physics lists.}
\label{cmshadronresponse1}
\end{figure}

\begin{figure}[htbp]
  \centering
\vspace{-1cm}
  \begin{minipage}{.85\textwidth}
    \centering
    \includegraphics[width=\textwidth]{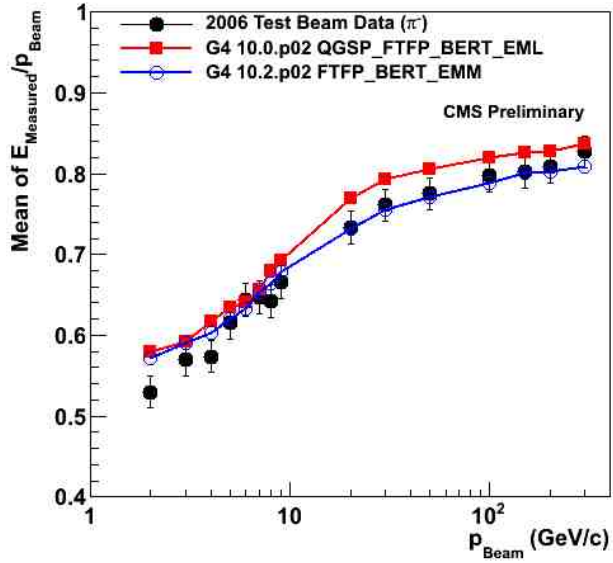}
    \vspace{-5cm} 
 \end{minipage}
  \begin{minipage}{.75\textwidth}
    \centering
    \includegraphics[width=\textwidth]{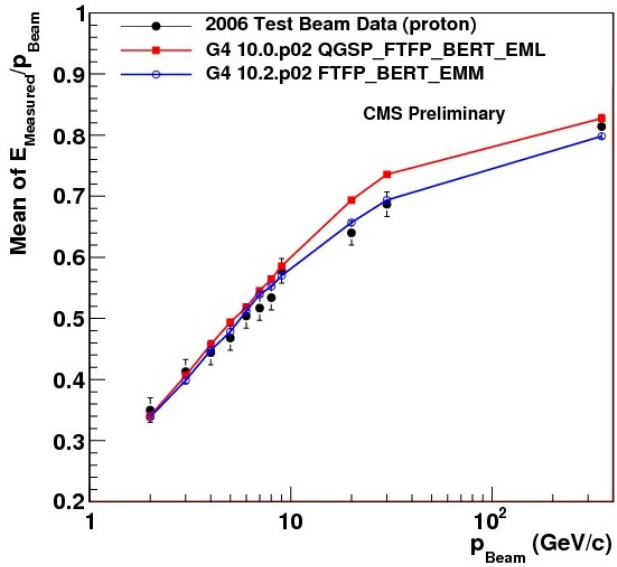}
  \end{minipage}
  \vspace{-2cm}
\caption{Mean pion and proton response as a function of the beam 
momentum, $p_{\rm beam}$, as measured in the 2006 CMS test beam 
experiment~\cite{cmsg4valid}. The simulation is performed for two
different choices of physics lists. The error bars are statistical
uncertainties only.}
\label{cmshadronresponse2}
\end{figure}

\begin{figure}[htbp]
  \centering
  \begin{minipage}{.75\textwidth}
    \centering
    \includegraphics[width=\textwidth]{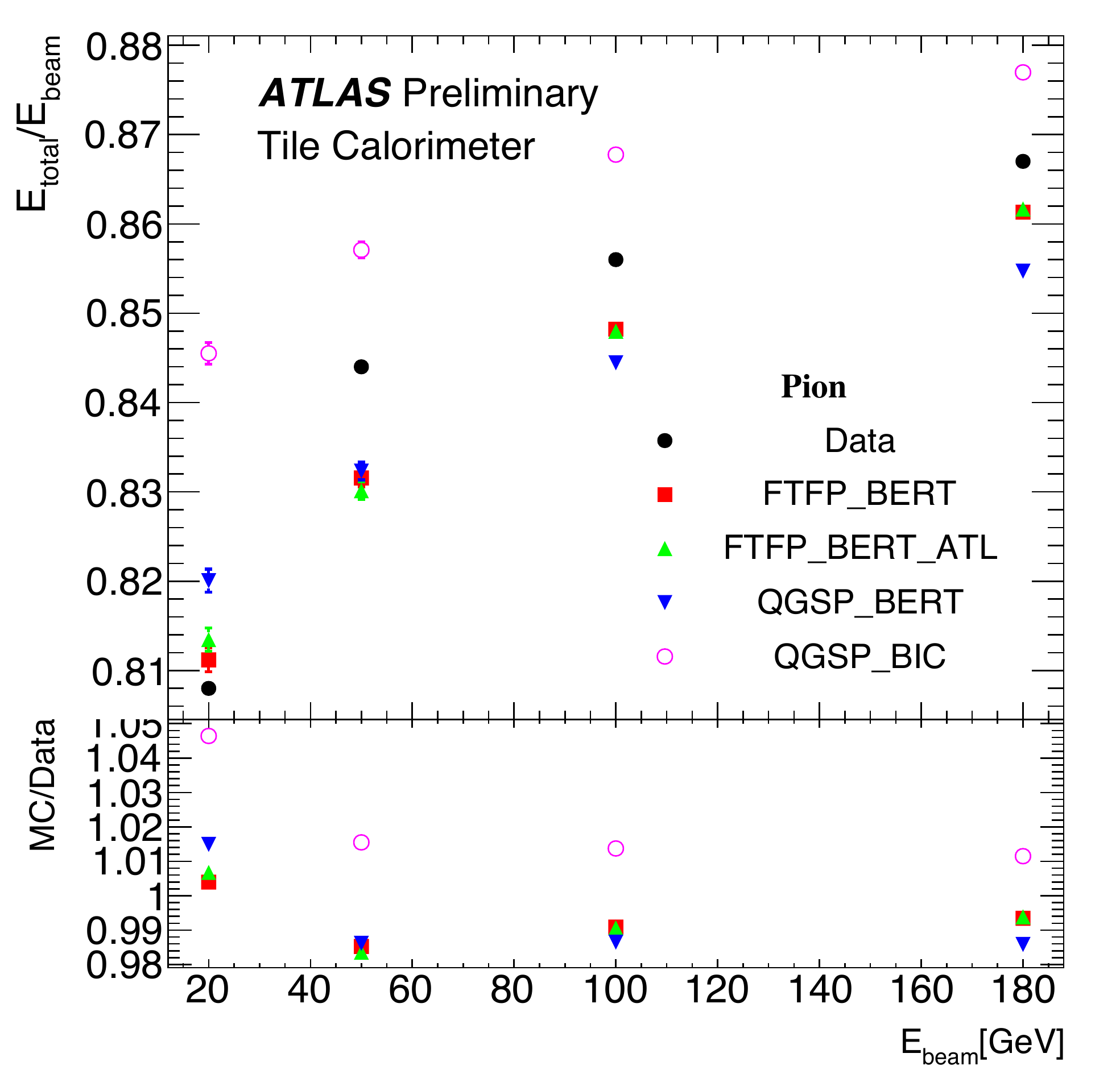}
 \end{minipage}
  \begin{minipage}{.75\textwidth}
    \centering
    \includegraphics[width=\textwidth]{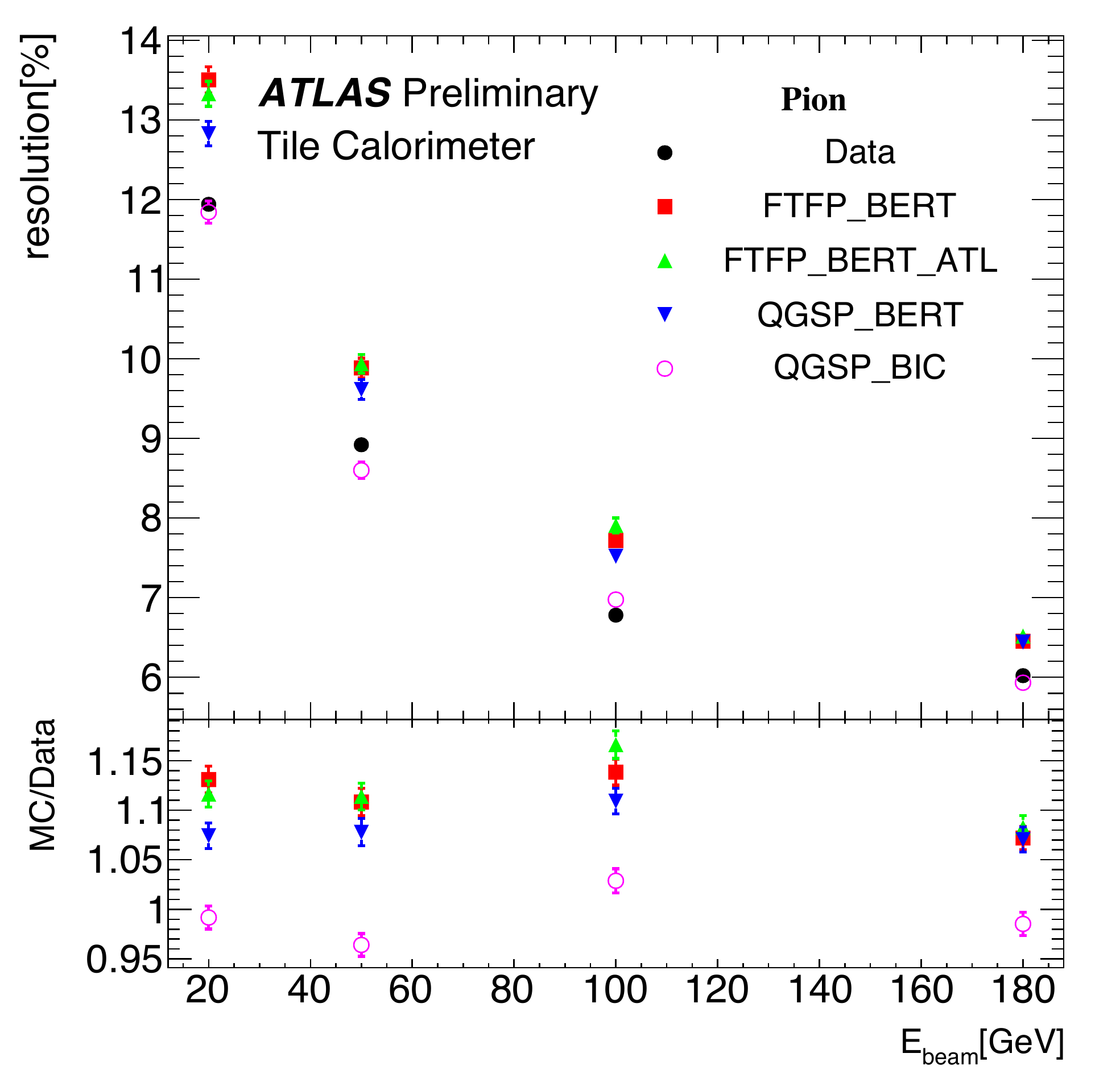}
  \end{minipage}
\caption{ATLAS calorimeter energy response, 
$E_{\rm total}/E_{\rm beam}$, and percentage resolution as a function of the
beam energy $E_{\rm beam}$. These 2000-2003 test beam results are compared
to predictions from
a Geant4 version 10.1 simulation using different physics list 
options~\cite{atlastb}. The error bars are statistical only.}
\label{atlaspiontb}
\end{figure}

Another set of experimental results utilized by modern collider experiments to
discriminate between different Geant4 physics lists, and offer the
Geant4 collaboration guidance on how to assemble them from individual
physics models, is the set of test beam single particle energy 
response and resolution measurements performed for different particles 
such as electrons, protons, and pions.
Fig.~\ref{cmshadronresponse1} depicts comparisons of MC and data measurements
of the response distribution, or response function,
for 4~GeV pions and 3~GeV protons incident onto a solid angle slice of the
CMS ECAL and HCAL calorimeters~\cite{cmsg4valid}. 
Fig.~\ref{cmshadronresponse2} shows
the mean pion and proton response as a function of the beam momentum, 
$p_{\rm beam}$, for 2006 test beam data and the same two Geant4 software
versions and physics lists as in Fig.~\ref{cmshadronresponse1}. 
The agreement is excellent, within uncertainties, 
over the whole range of particle momenta. For pions below $\sim$5~GeV, there
seems to be a trend with the prediction overestimating the data by a few
percent, although data and MC agree within uncertainties above 
$p_{\rm beam}>$3~GeV. 
For positive pions with beam energies of 20, 50, 100, and 180~GeV,
Fig.~\ref{atlaspiontb} depicts the ATLAS calorimeter energy response, 
$E_{\rm total}/E_{\rm beam}$, and percentage resolution as a function of the
beam energy $E_{\rm beam}$. These 2000-2003 test beam results are compared
to predictions from
a Geant4 version 10.1 simulation using different physics list 
options~\cite{atlastb}. The error bars are statistical
only. The MC-to-data ratios show agreements within less than 2\% for 
energy response and 10-15\% for energy resolution.

\begin{figure}[htbp]
\centering
\includegraphics[width=0.9\linewidth]{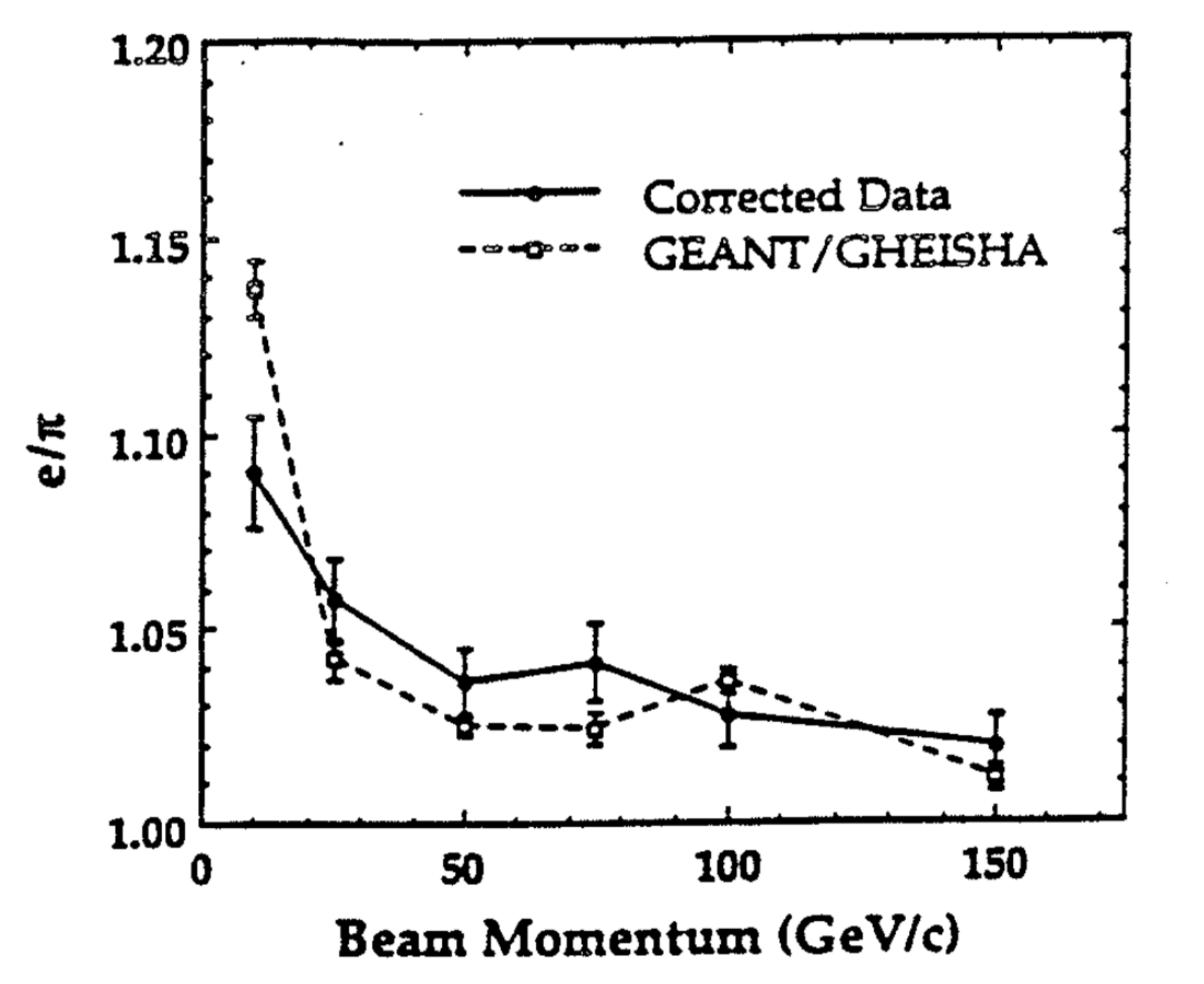}
\caption{$e/\pi$ energy response ratio versus beam energy as measured in the 
D0 Liquid Argon-Uranium calorimeter test beam experiment that took place in 
1991~\cite{d0testbeam}.}
\label{d0eoverpi}
\end{figure}

The situation was radically 
different for the Tevatron experiments in the early nineties when computers 
were slower, simulation included poor approximations in exchange 
for time performance, and less advanced remote communication technology 
among scientists made it 
significantly more challenging to establish an international work program to 
understand and optimize the physics of GEANT3. Furthermore, the Tevatron 
test beam programs were very limited in scope and data-taking capabilities, 
in detriment of the MC validation exercise. 
Fig.~\ref{d0eoverpi} shows the electron-to-pion energy response 
ratio, $e/\pi$, versus
beam energy as measured in the D0 Liquid Argon-Uranium calorimeter 
test beam experiment that took place in 1991~\cite{d0testbeam}. 
In the case of CMS, statistical uncertainties in simulated 
data are negligible and the agreement is much better than for D0 in the energy 
range covered by the experiments. Furthermore, the CMS comparison extends to 
energies as low as 1~GeV while the D0 measurement and comparison stops at 
10~GeV. In CMS, the modeling of the momentum dependence of the single particle 
response is excellent, while in D0 $e/\pi$ flattens out much faster in MC than 
in data. 

\section{Applications of Simulation to HEP Collider Experiments} \label{usesofsimu}

There are many applications of simulation to HEP collider experiments.
One area is the analysis of the experimental data collected by
the detectors and the interpretation of the resulting physics measurements
in the light of theoretical predictions.
Another use of simulation is in studies to design and optimize detectors for 
best physics performance. Simulation is also a critical tool utilized  
to develop calibration methods and reconstruction algorithms, as well as to
preform stress-testing of the computing infrastructure. 

\subsection{Simulation in Data Analysis} \label{DataAnalysis}

Until recently, pure Geant-based simulation applications were rarely used to 
make a direct ``MC truth'' extraction of 
calibration factors, particle identification and 
reconstruction efficiencies, or backgrounds for 
particle searches. Experiments used well-tuned ParSim options instead, 
such as QFL or the GEANT3-GFLASH tools developed in CDF.
Pure Geant MC samples were either not accurate enough due to
approximations to gain speed, they were statistically limited, or both.
The situation changed significantly in the last few years 
when the availability of large samples of significantly more 
realistic simulated events became the norm in 
HEP experiments. As a result, MC-driven methods are being used
with increasing frequency and confidence,
as long as they are based on simulation code that has been thoroughly 
validated within systematic uncertainties, using thin 
target, test beam, and in situ experimental data. 
Data-driven methods, based on physics laws applied to real data, are still at 
the core of the derivation of calibration and correction factors applied to 
data measurements, while closure tests are essential to test the validity and 
precision of the methods. Closure tests are based on the comparison between
the detector-level MC data, treated as if it were real data, and the Monte
Carlo truth information associated with the particles of an event before they 
interacted with the detector.

\subsubsection{Data-driven Methods} \label{DataDriven}

Data-driven methods are analysis techniques that use real experimental data, 
detector properties, and physics laws to perform detector calibration and 
alignment, estimate backgrounds in particle searches and, in general, 
determine correction factors applied to physics measurements. Simulation 
plays an essential role in the process of developing the methods, in the 
demonstration of their prediction power and the mitigation of biases, and in 
the derivation of the associated systematic uncertainties. A few examples of 
these data-driven techniques are described in the following paragraphs.

\paragraph{Object Balance for Jet Energy Calibration} \label{ObjectBalance}   

Quark- and gluon-initiated jets are the most common physics objects in
hadron collider experiments. The observed energy of jets in HEP detectors 
needs to be calibrated 
with a scale factor, which depends on the jet type and kinematics, and
includes corrections for electronic noise, additional hard interactions
in the same beam-beam crossing, 
detector response, and reconstruction algorithm effects.
In a collider experiment, the response correction may be derived using 
conservation of momentum in the transverse plane, and the fact that the 
energy response and 
resolution are much better for electromagnetically interacting physics objects,
such as photons or electrons, than they are for jets. 
The relatively small energy 
calibration factors for photons are similar to those for electrons, 
which are typically derived from 
$Z \rightarrow e^{+}e^{-}$ samples. Once 
the photon scale is adjusted, jets may be calibrated using “$p_T$ balancing”, 
that is transverse momentum conservation in each event. 
To increase sample statistics and the 
accuracy of the measurement, jets in forward $\eta$ regions may be calibrated 
from di-jet events with one jet in the central region ($\eta$ close to 0). 
As a bonus, the jet 
energy or $p_T$ resolution may be derived from the width of the 
asymmetry distribution, 
$A=(p^{\rm jet}_{T_1}-p^{\rm jet}_{T_2})/(p^{\rm jet}_{T_1}+p^{\rm jet}_{T_2})$. 
This is the approach used by the D0 experiment~\cite{d0jes1,d0jes2}, 
while CDF used QFL and its successor, the ParSim approach based on GEANT3 and 
GFLASH~\cite{cdfjes}. The CMS experiment uses a hybrid approach where the MC 
truth prediction of the jet energy scale is adjusted by small 
factors derived from the above-mentioned 
data-driven techniques~\cite{cmsjes}, in order to take into account
the differences of jet energy scale in data and simulated 
events. ATLAS uses a similar calibration scheme based 
on both MC-driven and data-driven 
techniques~\cite{atlasjes}.

\begin{figure}[htbp]
  \centering
  \begin{minipage}{.75\textwidth}
    \centering
    \includegraphics[width=\textwidth]{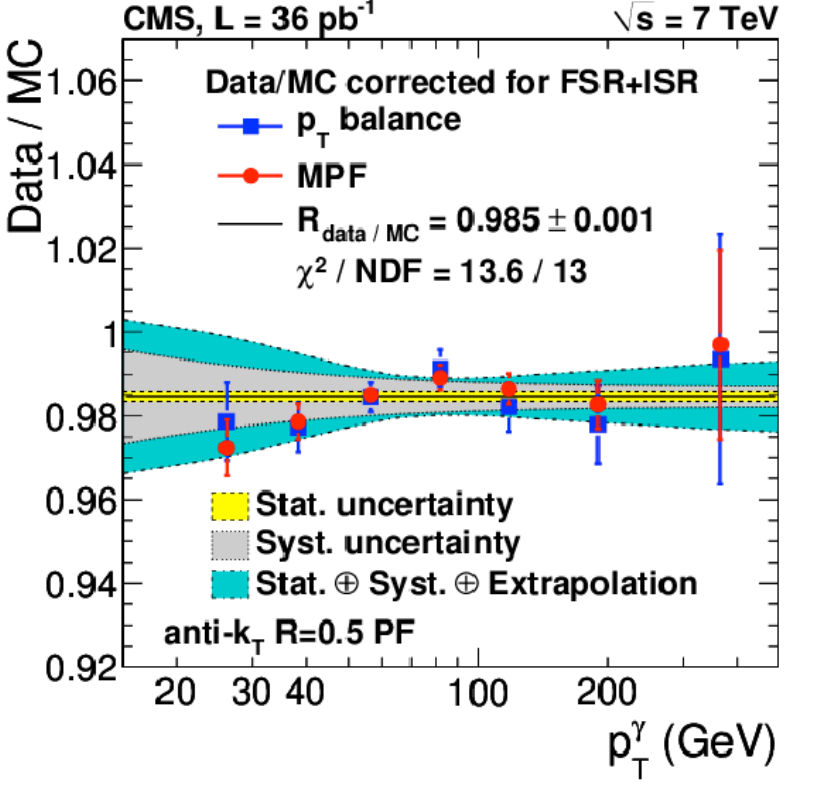}
  \end{minipage}
  \begin{minipage}{.8\textwidth}
    \centering
    \includegraphics[width=\textwidth]{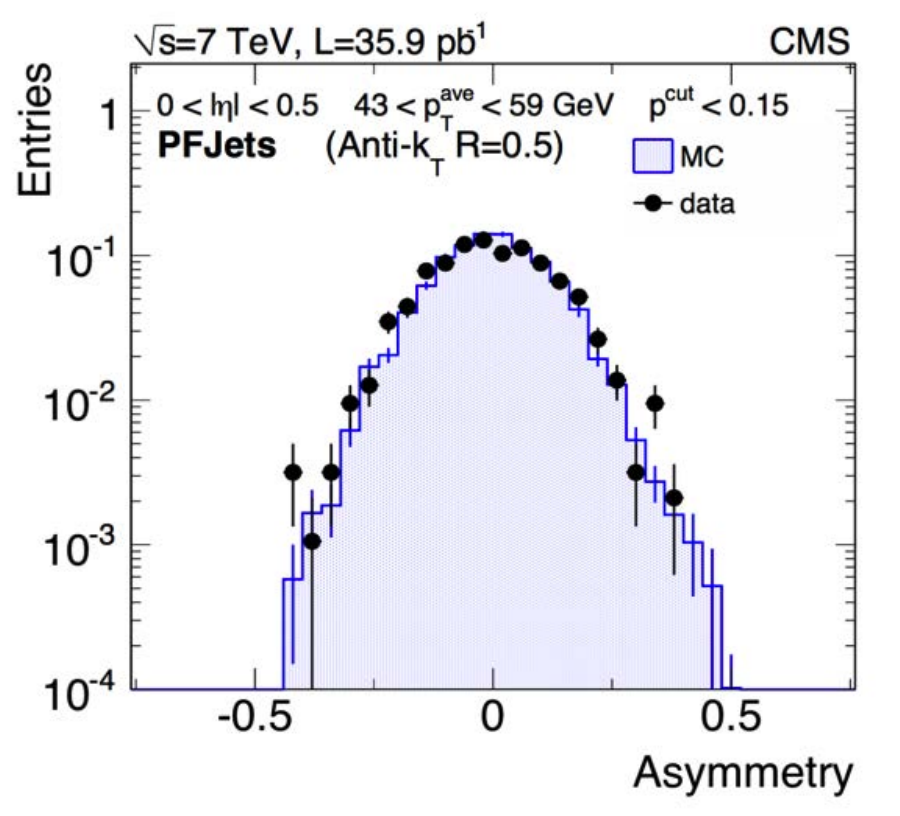}
  \end{minipage}
\caption{Top: CMS data-to-MC ratio of the jet energy response as a function of 
jet $p_T$, determined from two different data-driven methods~\cite{cmsjes}: 
$p_T$ balancing and 
Missing $p_T$ Fraction. Bottom: Asymmetry distribution, 
$A=(p^{\rm jet}_{T_1}-p^{\rm jet}_{T_2})/(p^{\rm jet}_{T_1}+p^{\rm jet}_{T_2})$,
for a CMS sample of 
di-jet events from where the jet $p_T$ resolution is measured. The
asymmetry variable measurement shows statistical errors only.}
\label{cmsjetresponse}
\end{figure}

Fig.~\ref{cmsjetresponse} (top) shows the CMS data-to-MC ratio of the jet 
energy response as a 
function of jet $p_T$, determined from two different data-driven methods: 
$p_T$ balancing (solid squares) and Missing $p_T$ Fraction (solid circles). 
The Missing $p_T$ Fraction Method (MPF) is a variation of $p_T$ balancing, 
that uses the projection of the event transverse momentum imbalance vector 
onto the direction of the 
photon to estimate the response of the hadronic recoil. 
The message contained in Fig.~\ref{cmsjetresponse} is that the normalization 
factor between the jet energy response measured in data and modeled in the
CMS Full Simulation is approximately ${\rm SF_R}=$0.985 with an uncertainty 
of less
than 2$\%$. Moreover, the ratio shows that the $p_T$ dependence is flat
to within the small uncertainties represented by the error bands. 
Although this ratio also depends on the jet flavor
and pseudorapidity, the data-to-MC normalization or ``scale'' factors
are in all cases small enough to allow CMS to follow the approach of extracting
the jet energy response directly from MC truth information. In other words,
the jet energy response applied as a correction to the jet energy in 
real collider data, $R_{\rm jet}$, is calculated as 
${\rm SF_R}\times (p^{\rm reco}_{T}/p^{\rm part}_{T})$, where $p^{\rm reco}_{T}$ is 
the detector
level jet $p_T$ obtained from the reconstruction software algorithms applied to
MC events, and $p^{\rm part}_{T}$ is the jet true $p_T$, with ``true $p_T$'' 
referring to the $p_T$ of the particle-level jet, after 
fragmentation and hadronization, before it hit the detector.
The ${\rm SF_R}$ factor puts MC and data in the same footing, by shifting 
$R_{\rm jet}$ in MC to model what was measured in data.
This approach is significantly more accurate because, since the same 
data-driven method is applied to 
both MC and data, the uncertainty in the ratio is much smaller than that
in the numerator and denominator, given that 
most uncertainty components are correlated and cancel. 
It is not the jet response what is measured but how different the
measurements are in data and MC.

Fig.~\ref{cmsjetresponse} (bottom) shows the asymmetry 
distribution $A$ for 
a CMS sample of di-jet events from where the jet $p_T$ resolution is measured. 
The agreement between MC and data is excellent except in the non-Gaussian 
tails of the distribution, which are very difficult to model in MC because 
they come from non-linear contributions to the detector response. Although
the CMS calorimenter system (ECAL+HCAL) is undercompensating, with an 
$e/h>2$ ($e$, $h$ are the response to the energy deposited by
an incident hadron through electromagnetic and nuclear interactions
respectively), non-linear behavior is
corrected to a large extent during calibration and through the use of 
tracking information by the particle flow algorithm~\cite{pflow}. As in the 
case of the jet energy response, MC truth resolutions, properly adjusted
with data-to-MC scale factors, are utilized in 
data analysis.

\begin{figure}[htbp]
  \centering
  \begin{minipage}{.7\textwidth}
    \centering
    \includegraphics[width=\textwidth]{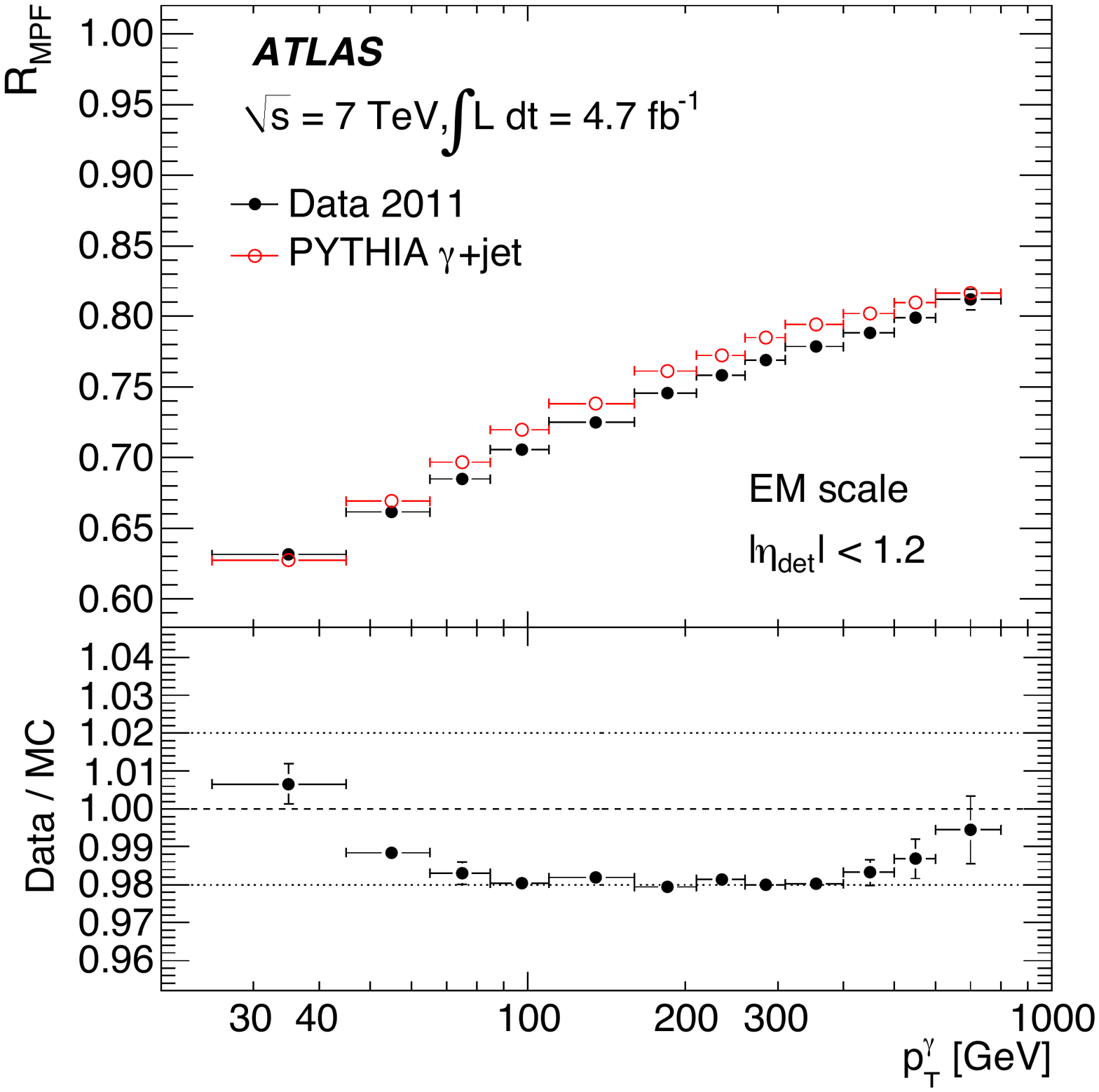}
    \vspace{-4.5cm}
  \end{minipage}
  \begin{minipage}{.8\textwidth}
    \centering
    \includegraphics[width=\textwidth]{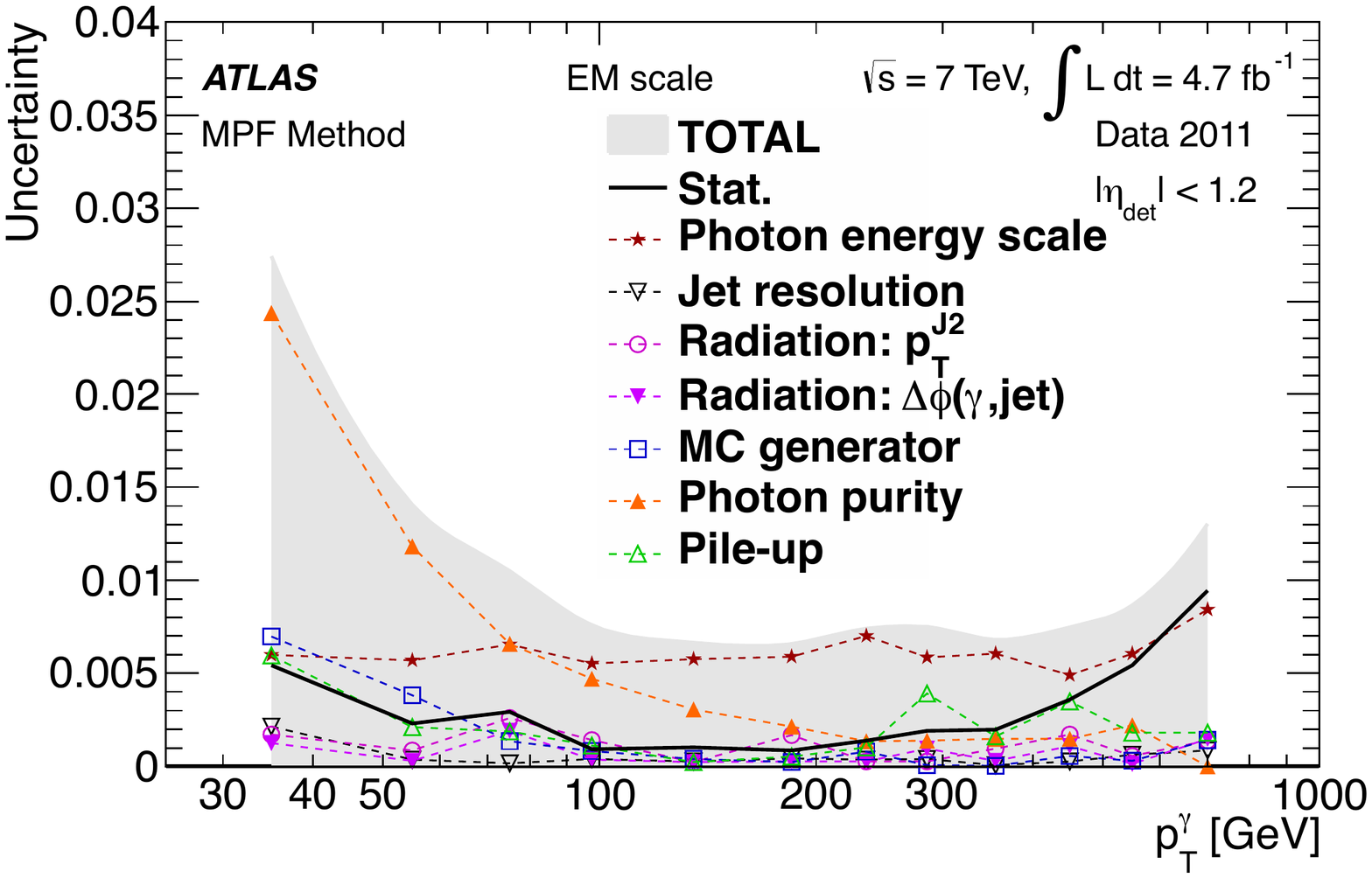}
  \end{minipage}
\vspace{-3.0cm}
\caption{Top: Jet response measured in ATLAS with the MPF method applied to 
photon plus jets events. Photons are calibrated to account
for the EM scale, derived from a $Z \rightarrow e^{+}e^{-}$ data sample.
The data-to-MC response ratio is shown in the bottom inset of the figure. 
Only the statistical uncertainties are shown. Bottom: Systematic uncertainties 
on the MPF response data-to-MC ratio, with photons calibrated to the
electromagnetic (EM) scale. See Ref.~\cite{atlasjesplots}.}
\label{atlasjesresponse}
\end{figure}

\begin{figure}[htbp]
  \centering
  \begin{minipage}{.8\textwidth}
    \centering
    \includegraphics[width=\textwidth]{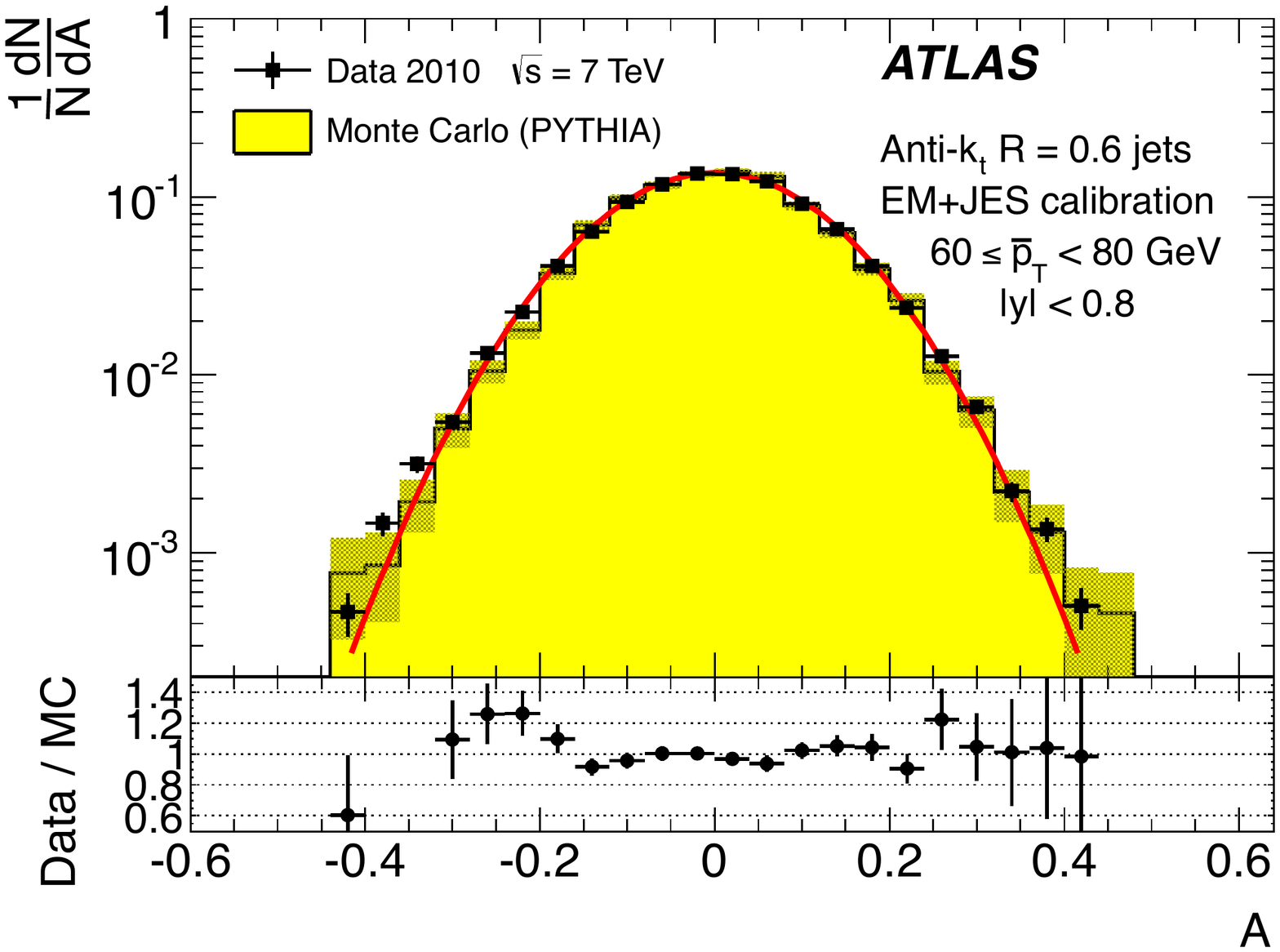}
    \vspace{-6.5cm}
  \end{minipage}
  \begin{minipage}{.85\textwidth}
    \centering
    \includegraphics[width=\textwidth]{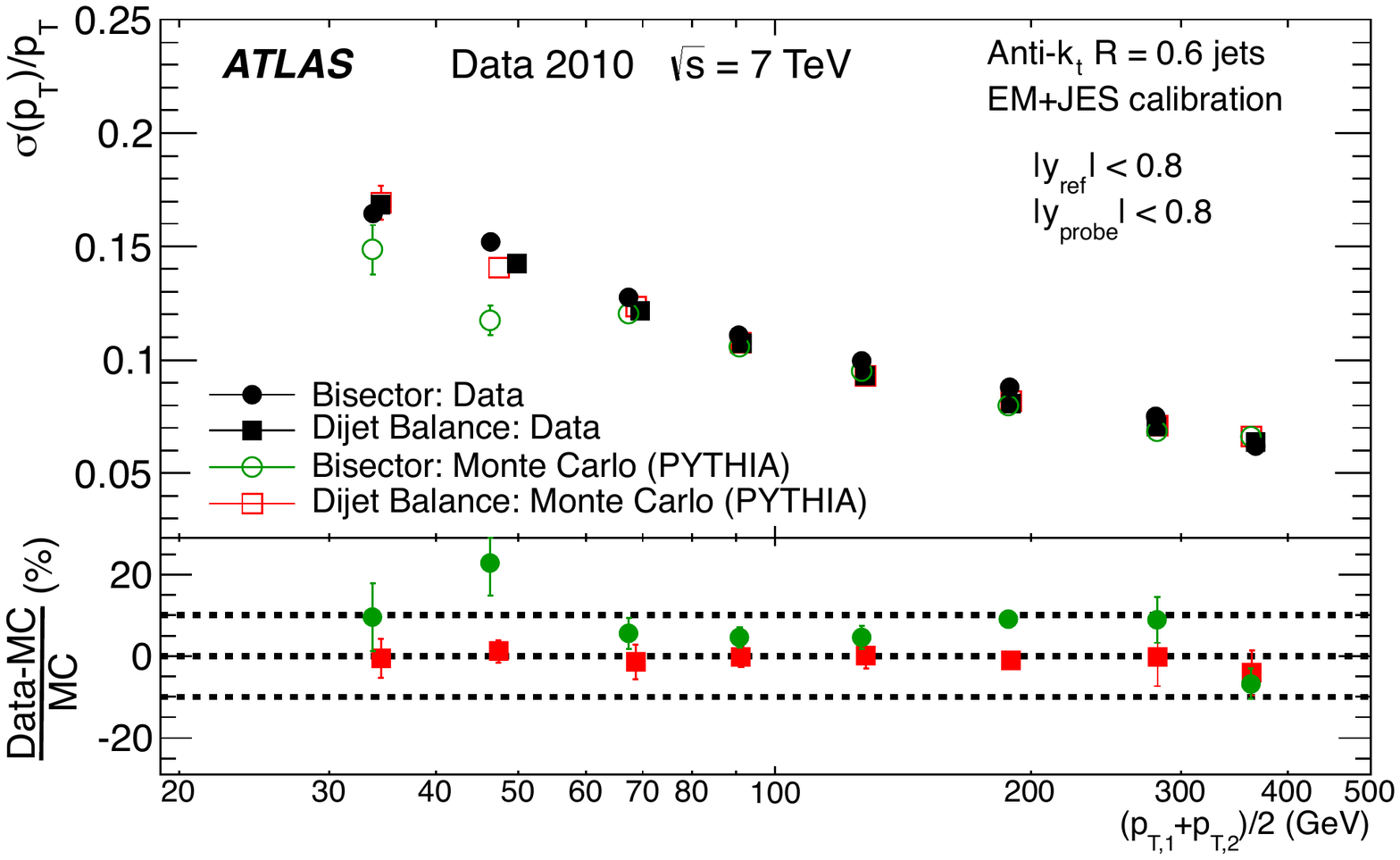}
  \end{minipage}
\vspace{-3.5cm}
\caption{Top: Asymmetry distribution, 
$A=(p^{\rm jet}_{T_1}-p^{\rm jet}_{T_2})/(p^{\rm jet}_{T_1}+p^{\rm jet}_{T_2})$, for an
ATLAS sample of di-jet events.
Bottom: ATLAS jet $p_T$ resolutions measured using the di-jet balance and
bisector methods. In both plots, error bars are statistical and the lower
panels show data-to-MC ratios to illustrate the level of agreement between
simulation and data. See Ref.~\cite{atlasjerplots}.}
\label{atlasjetresolution}
\end{figure}

As in CMS, one of the methods utilized by ATLAS to perform the jet 
calibration is MPF. Fig.~\ref{atlasjesresponse} (top) shows the measured MPF 
response in a photon plus jets sample, as a function of $p_{T}^{\gamma}$ in the
central pseudorapidity region~\cite{atlasjesplots}. 
Also for ATLAS, the MC models the
MPF response very accurately over the whole kinematic range of interest, 
deviating slightly from a 0.98 flat data-to-MC ratio in the lowest and highest
extremes of the range. The shaded band in Fig.~\ref{atlasjesresponse} (bottom) 
shows the total uncertainty in the data-to-MC MPF response ratio, which is
less than 1\% for $p_{T}^{\gamma}>70$~GeV.
The excellent data-to-MC agreement for the asymmetry distribution, $A$, is
illustrated in Fig.~\ref{atlasjetresolution} (top) for the ATLAS 
experiment~\cite{atlasjerplots}. Although calorimeter jets are used in this
example, the ATLAS asymmetry distribution is reasonably well described by a
Gaussian function, since the $e/h=1.37$ value for the ATLAS 
calorimeter system does
not deviate that much from perfect compensation, $e/h=1$.
Still, the modeling of the residual non-linear behavior contributing to the
tails is difficult and both the data measurement and the MC prediction 
deviates from a perfect Gaussian distribution in the tails. 
Fig.~\ref{atlasjetresolution} (bottom) shows
the measured jet energy resolution, $\sigma(p_T)/p_T$, versus the average
jet $p_T$ of the two jets in the di-jet sample used for the derivation. The
agreement between MC and data for the di-jet balance method is impressive, while
the agreement when using the bisector method is within 10\%. The bisector
method~\cite{atlasjerplots} is a variant of the di-jet balance technique, 
and measures the
variance of the $p_T$ balance vector projected along an orthogonal coordinate
system in the transverse plane, where one of the axes is chosen in the 
direction that bisects the azimuthal angle formed by the two leading jets.

\paragraph{Control Samples for Background Estimation} \label{ControlSamples}

One essential aspect of a search for a
new particle, or a characterization of a known particle, is the 
selection of a signal-enhanced sample, where the particle under study 
represents the signal, and all other physics processes resulting in similar
final states or detector signatures represent the backgrounds.
Control Samples (CS's) or control regions (CR's) are background-only samples, 
or regions of phase space, used to estimate the background contributions in 
a signal region (SR) from a combination of measurements in the CR, event 
properties, and physics laws. For example, it may be known that a given 
functional form fits well a Standard Model process which is a background to a 
beyond-the-Standard-Model (BSM) signal under study. To estimate the
background in the SR, the function may be fit to the data in the 
CR and then extrapolated to the SR to make the prediction. 

\begin{figure}[htbp]
  \centering
  \begin{minipage}{.9\textwidth}
    \centering
    \includegraphics[width=\textwidth]{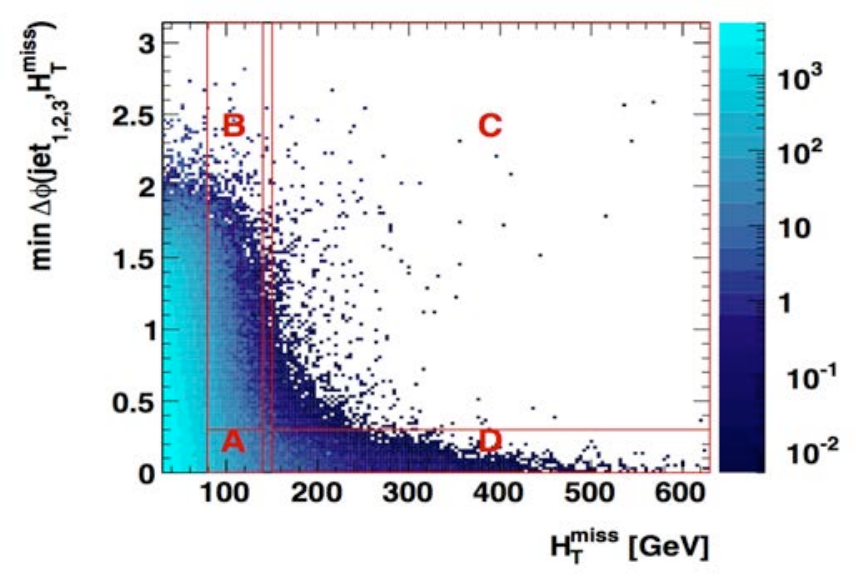}
  \end{minipage}
  \begin{minipage}{.9\textwidth}
    \centering
    \includegraphics[width=\textwidth]{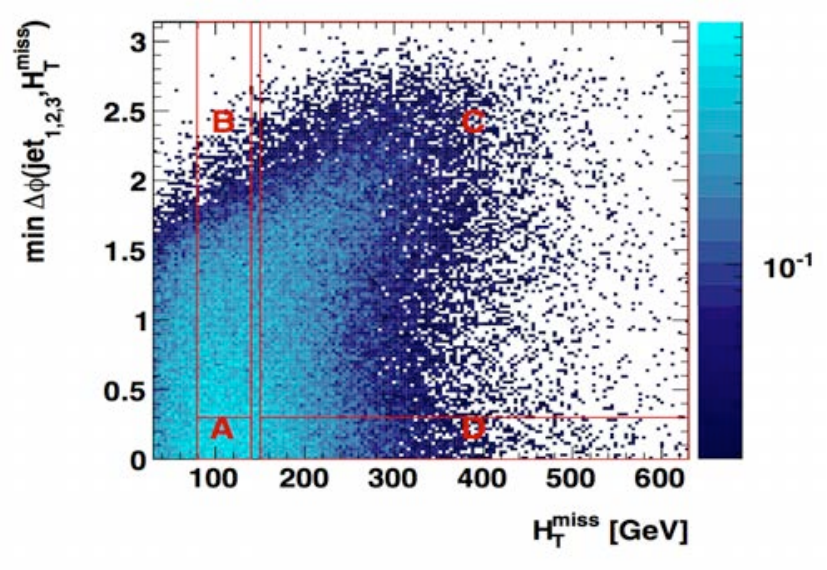}
  \end{minipage}
\caption{Illustration of the ABCD or factorization method for multi-jets 
(QCD) background estimation. The minimum azimuthal angular distance between 
the three leading jets in the event,
${\rm min}\, \Delta \phi({\rm jet}_{1,2,3},H^{\rm miss}_{T})$, is shown as a 
function of the event $H^{\rm miss}_{T}$, for a QCD 
background only sample (top) and a SUSY signal sample (bottom). It is 
apparent that the A, B, and D regions are dominated by the background, while 
the C region is background depleted and signal enhanced~\cite{tschum}.}
\label{abcd}
\end{figure}

Fig.~\ref{abcd} illustrates how MC samples are used to 
establish the boundaries of CR's and SR's for use in the 
data-driven prediction of the QCD background to multi-jet final states in a 
CMS Supersymmetry (SUSY)~\cite{susy} search~\cite{ra2}. SUSY is a theory
based on a symmetry that relates bosons and fermions, offering a solution
to the Higgs boson hierarchy problem, and predicting the unification of 
gauge couplings and a dark matter candidate.  
Fig.~\ref{abcd} shows the minimum azimuthal angular distance
${\rm min}\, \Delta \phi({\rm jet}_{1,2,3},H^{\rm miss}_{T})$ between 
the three leading jets in the event and the event $\vec{H}^{\rm miss}_{T}$, 
defined as the the negative vector sum of the $p_T$'s of 
all jets in the event~\cite{tschum}. This angle is plotted as a function of 
$H^{\rm miss}_{T}$, the absolute value of the vector. For QCD background events, 
${\rm min}\, \Delta \phi({\rm jet}_{1,2,3},H^{\rm miss}_{T})$ is small and 
$H^{\rm miss}_{T}$ low, since the $p_T$ 
imbalance comes from detector response and resolution effects, and tends to
be aligned with the two approximately back-to-back 
leading jets. For certain SUSY models, the two leading jets and two weakly 
interacting BSM stable particles, called neutralinos ($\tilde{\chi}^0_1$), 
tend to be in opposite 
hemispheres of the transverse 
plane, generating high missing $p_T$ in the event as well as a large 
${\rm min}\, \Delta \phi({\rm jet}_{1,2,3},H^{\rm miss}_{T})$. 
The data-driven technique illustrated in Fig.~\ref{abcd} with a
MC sample is known as the factorization or ABCD method. It consists
of identifying three CR's (A, B, and D) and a SR (C), 
which corresponds to events with a large angular distance and high 
$H^{\rm miss}_{T}$. 
The background 
in C may be estimated as $N_C=(N_B/N_A)\times N_D$ , if the variables are 
uncorrelated and as $N_C=f(H^{\rm miss}_{T})\times N_D$  if they 
are correlated, with $f(H^{\rm miss}_{T})$ determined from an
extrapolation of a fit of an appropriate function to data in the CR. 

\paragraph{Tag-and-Probe Method for Efficiency and Fake Rates} \label{TagandProbe}

The principle of the tag-and-probe method, when applied to measure particle
reconstruction and identification efficiencies, is to use the a priori 
knowledge of the identity of a reconstructed physics object (Tagged Object),
for example a known resonance, and ask the question on the fraction of 
the times a given Probe Object is identified by a reconstruction and 
identification algorithm correctly. The method is also used to measure 
trigger efficiencies, or the probability for a hardware or software 
trigger system to flag an event it is designed to identify from all
collisions occurring in the experiment. Particle isolation efficiency is
the probability for a particle to pass a requirement of isolation with
respect to other particles in the event, and fake rate is the probability
for a particle to be miss-identified as a different particle by a software
algorithm. Both may also be measured from real data using the 
tag-and-probe method.  
For example, the probability that an electron 
is miss-identified as a photon, the $e \rightarrow \gamma$ fake rate, 
may be derived from three di-object samples 
where each of the two objects has been identified either as an electron or a 
photon: $e^{+}e^{-}$, $e^{+/-}\gamma$,$\gamma \gamma$. 
The di-object invariant mass distribution will show a 
distinct peak centered at the mass value of the $Z$ boson. In average, 
once the non-$Z$ 
di-electron continuous and monotonically decreasing background is subtracted, 
all the events within the $m_Z \sim 91.2$~GeV peak are bound to be 
electrons, no 
matter whether they have been identified as such or as $e\gamma$, or 
$\gamma\gamma$, because the decay rate of $Z$ 
bosons to photons is negligible. In the example, the electrons 
are the tagged objects and the photons the probe objects. The photons 
are fake photons, in reality electrons miss-identified as photons. Thus the   
$e\rightarrow \gamma$ fake rate is calculated as
$f_{e \rightarrow \gamma}=(N^{\gamma}_{ee}+N^{\gamma}_{e\gamma}+N^{\gamma}_{\gamma \gamma})/N^{em}_{TOT}$,  where $N^{\gamma}_{ee}$, $N^{\gamma}_{e\gamma}$,
$N^{\gamma}_{\gamma \gamma}$ are the number of 
photons in the $e^{+}e^{-}$, $e^{+/-}\gamma$, and $\gamma \gamma$ samples 
and $N^{em}_{TOT}$ is the total number of electromagnetic interacting 
objects, either $e$ or $\gamma$. 
Another application of tag-and-probe is the determination of 
the $\tau$-lepton 
reconstruction plus identification efficiency from a di-lepton sample.
In this case, one 
lepton is identified as a muon and the other lepton as a 
hadronically-decaying $\tau$ ($\tau_{\rm had}$) using the standard 
identification selections. 
When computing the di-lepton invariant mass distribution, a $Z$ 
boson peak is visible, populated with the events where the $\mu$ 
comes from a 
leptonically-decaying $\tau$ ($\tau_{\rm lep}$). Here, the 
$\tau_{\rm had}$ is the 
tagged object and the 
muon the probe object. The $\tau$ reconstruction plus identification 
efficiency is calculated as
$\varepsilon^{\mathrm{reco+id}}_{\tau}=N^{\rm evt}_{\rm pass}/(N^{\rm evt}_{\rm pass}+N^{\rm evt}_{\rm fail})$, where $N^{\rm evt}_{\rm pass}$, $N^{\rm evt}_{\rm fail}$ are the number of 
events passing and failing the 
requirement that there are two $\tau$ 
leptons. 
Following the same procedure described for the jet energy corrections, 
experiments typically extract object 
identification efficiencies directly from MC truth predictions and adjust them 
with scale factors computed as ratios between the data-driven efficiencies 
obtained from real data and MC samples, respectively.

\begin{figure}[htbp]
  \centering
  \begin{minipage}{.65\textwidth}
    \centering
    \includegraphics[width=\textwidth]{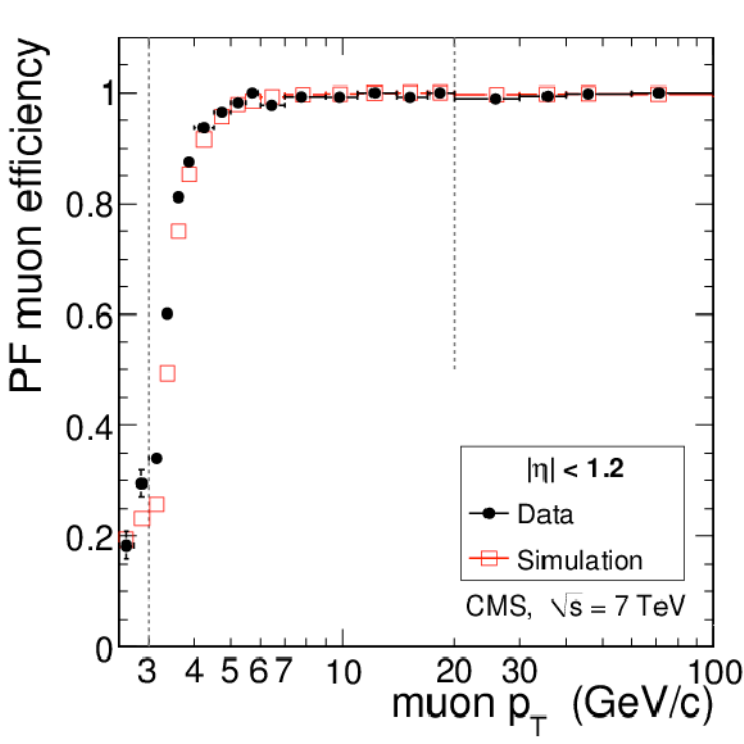}
  \end{minipage}
  \begin{minipage}{.7\textwidth}
    \centering
    \includegraphics[width=\textwidth]{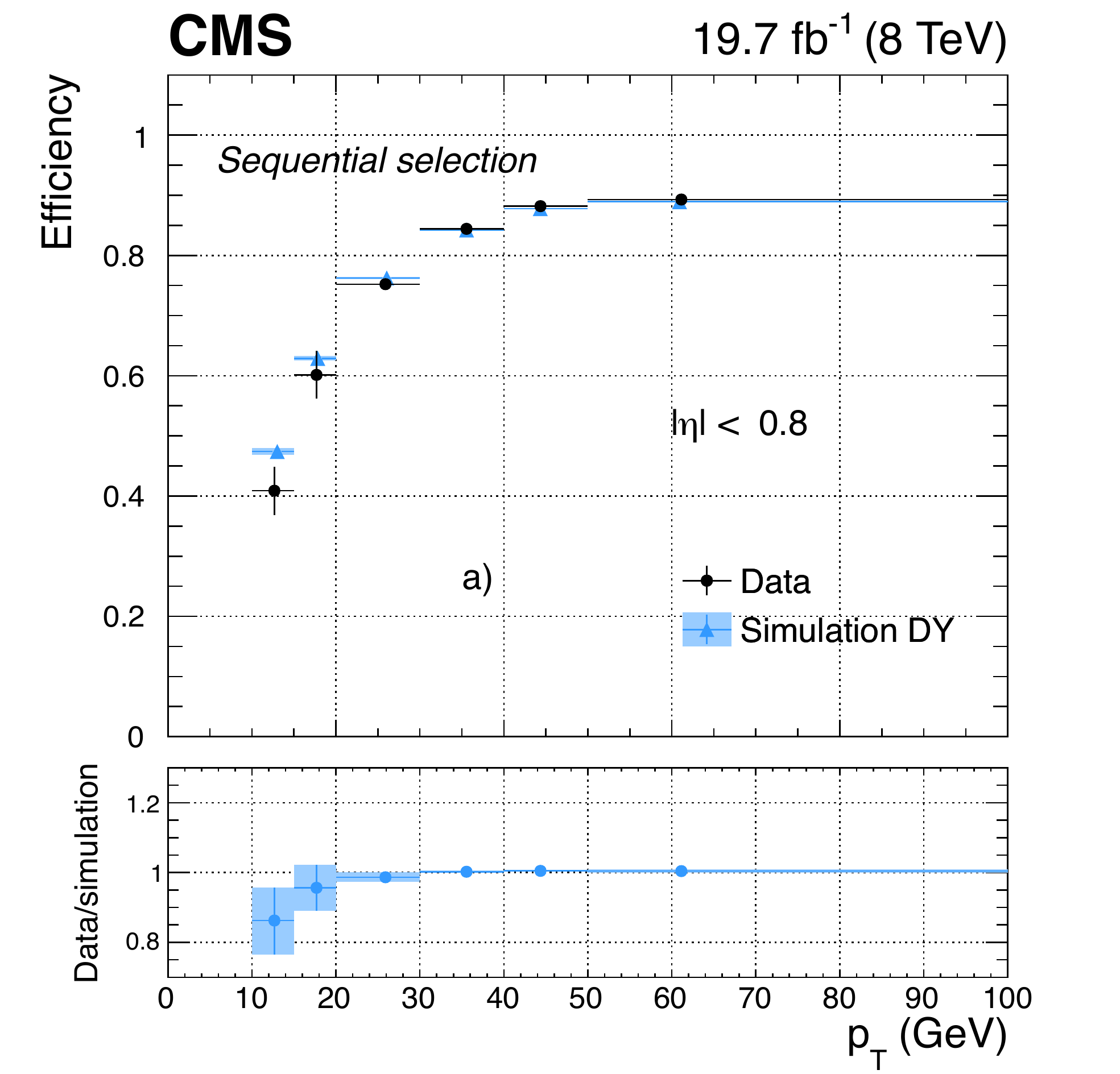}
  \end{minipage}
\caption{Data-to-MC comparison of the CMS 
muon reconstruction plus identification efficiency, 
$\varepsilon_{\mu}^{\mathrm{reco+id}}$~\cite{muonpog} (top),
and the electron identification efficiency for medium electrons, 
$\varepsilon_{e}^{\mathrm{id}}$~\cite{electronpog} (bottom),
measured 
with the tag-and-probe method. The lower panel displays the data-to-MC
scale factors used to correct the electron MC prediction for use in 
data analysis.
The uncertainties shown in the muon efficiency plot are statistical only, and
the uncertainty band in the electron efficiency plot represents 
statistical and systematic errors on the method added in quadrature.}
\label{efficiencies}
\end{figure}

Fig.~\ref{efficiencies} shows a data-to-MC comparison of the CMS 
muon reconstruction plus identification efficiency~\cite{muonpog},
$\varepsilon_{\mu}^{\mathrm{reco+id}}$, and the 
electron identification efficiency for medium electrons~\cite{electronpog},
$\varepsilon_{e}^{\mathrm{id}}$, measured 
with the tag-and-probe method. In this particular case, the method is referred 
to as tight-and-loose, because it utilizes
$Z \rightarrow \mu^{+}\mu^{-}$, $J/\psi \rightarrow \mu^{+}\mu^{-}$, and  
$Z \rightarrow e^{+}e^{-}$ samples 
to measure the efficiencies, where the tagged lepton is selected with a 
stringent (tight) criteria and the probed lepton with a relaxed (loose) 
criteria. Efficiencies are not defined the same way for all physics objects 
and in all experiments, and a good understanding of the definition details is
important for their correct utilization in physics analysis. For example, the
CMS muon efficiency shown in Fig.~\ref{efficiencies} refers to reconstruction
plus identification (or muon sample selection) efficiency. It is the
conditional probability of identifying a muon with a looser or tighter
selection criteria given that a track in the tracker system exists. The CMS
electron efficiency in Fig.~\ref{efficiencies} accounts for identification 
only, and must be multiplied by the reconstruction only efficiency to yield
the combined reconstruction plus identification efficiency.    

\begin{figure}[htbp]
  \centering
  \begin{minipage}{.75\textwidth}
    \centering
    \includegraphics[width=\textwidth]{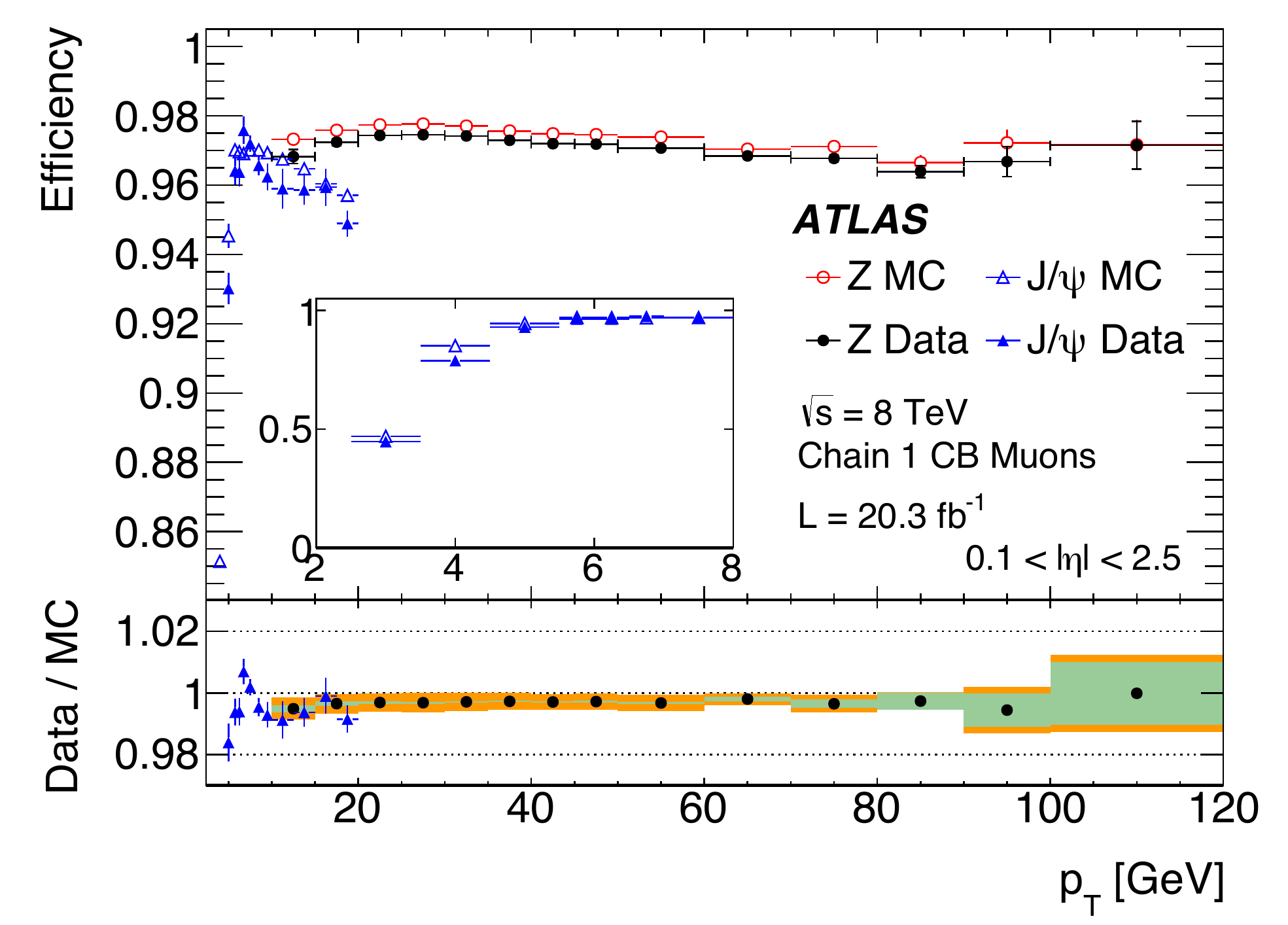}
  \end{minipage}
  \begin{minipage}{.7\textwidth}
    \centering
    \includegraphics[width=\textwidth]{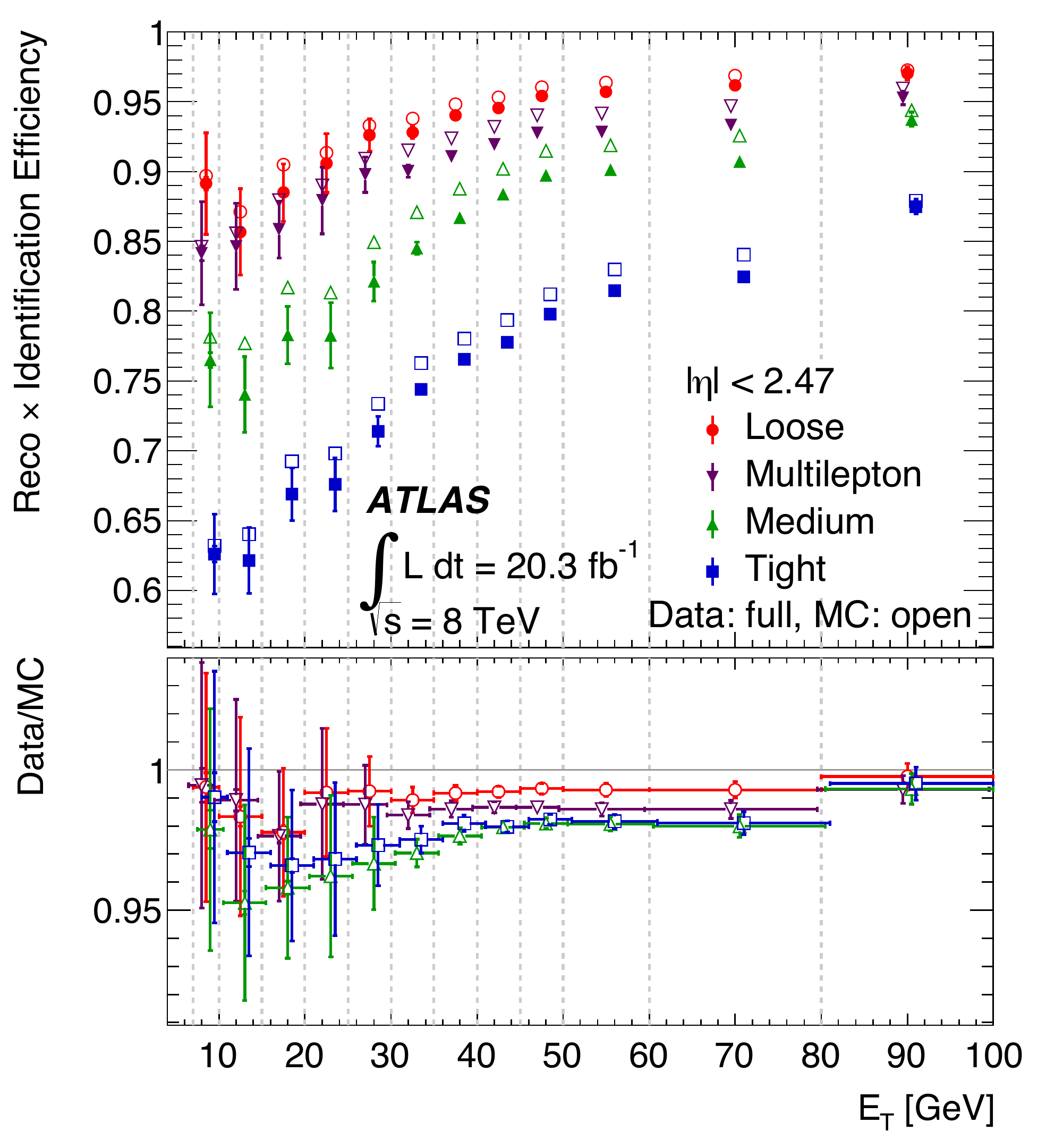}
  \end{minipage}
\caption{Data-to-MC comparison of the ATLAS 
muon~\cite{atlasmuon} (top) and electron~\cite{atlaselectron} (bottom)  
reconstruction plus identification efficiencies,
$\varepsilon_{\mu/e}^{\mathrm{reco+id}}$, measured 
with the tag-and-probe method. The lower panels show the data-to-MC
scale factors used to correct the MC predictions for use in data analysis.
The green uncertainty band depicts statistical uncertainties only, while the
orange band also includes the systematic uncertainties. In the case of the
electrons, the inner error bars are statistical and the outer bars depict
the total error.}
\label{efficienciesatlas}
\end{figure}

ATLAS also uses the tag-and-probe method to measure
$\varepsilon^{\mathrm{reco+id}}_{\mu}$~\cite{atlasmuon} and 
$\varepsilon^{\rm reco+id}_{e}$~\cite{atlaselectron} in
$Z \rightarrow \mu^{+}\mu^{-}$, $J/\psi \rightarrow \mu^{+}\mu^{-}$, 
and  $Z \rightarrow e^{+}e^{-}$ samples. The muon efficiency in 
Fig.~\ref{efficienciesatlas} is the conditional probability of reconstructing
a muon that successfully combines inner detector and muon system 
information (CB), given that a 
track is found in the inner detector. The ATLAS electron efficiency in
Fig.~\ref{efficienciesatlas} is the product of the reconstruction and
identification efficiencies.

Except in the case of the CMS muon efficiency, all other plots in 
Figs.~\ref{efficiencies},~\ref{efficienciesatlas} show an 
inset with the data-to-MC ratio and total uncertainty, which 
translates into the scale factors utilized to adjust the MC truth efficiencies 
used in physics analysis. For muons with $p_T>5$~GeV, the agreement between 
simulation and data is excellent for both experiments, 
while for electrons ATLAS shows a small disagreement of the 
order of 1-2\% depending on the selection criteria, larger for $E_T< 40$~GeV,
which is due to mismodeling of the shower shape in the forward calorimeter.
In the case of CMS, once the electron reconstruction efficiency, which is
not shown and varies in the 90-97\% range for 15-100~GeV electrons,  
is multiplied
by the medium electron identification efficiency, 
the results are similar to ATLAS's. However, CMS's coverage is restricted to 
the central $\eta$ region in this particular
plot, which shows excellent data-to-MC agreement above 30~GeV, and a trend
towards MC overestimation en the 10-30~GeV range.

\subsubsection{Closure Tests} \label{closuretests}

While the use of accurate simulation increases the chances that the
software developed with it performs out-of-the-box in real experiments, 
closure 
tests are fundamental tools to demonstrate that a given data-driven 
method to measure calibration factors or efficiencies works as
advertised, and without biases outside of the quoted uncertainties. In other 
words, a method not closing indicates the need to go back to the
drawing board and understand the biases in the measurement procedure that are 
responsible for the lack of closure. 
Closure tests are only useful if the simulated samples accurately 
model the details of real data because, otherwise, some effects from the 
measuring procedure may be missed. The basic principle of closure tests is the 
possibility to compare MC truth values to data-driven measurements, 
with the former calculated directly from detector-level-to-particle-level
information, and the latter derived from methods applied to 
detector-level-only information. For example, the MC truth jet energy response 
is determined from the ratio between the energy of detector-level
reconstructed jets and the MC truth jet energy calculated as the 
sum of energies of all the final state particles 
identified as part of the jet before they hit the detector. If the jet energy 
response measured from detector-level MC information using data-driven 
methods, such as $\gamma$-jet and di-jet balance, 
is consistent within uncertainties with the MC 
truth jet energy response, then the method 
closes. Typically, closure tests take the form of 
detector-level-to-particle-level ratios for an observable of interest as
a function of variables such as transverse momentum or pseudorapidity. For
a method to close, the ratio has to be consistent with unity within the 
quoted uncertainties. 

The lack of 
high-quality, high statistics MC samples in the D0 experiment was one of the 
causes of delay in the publication of a number of physics measurements. 
For example, the D0 Run 1 
jet papers to validate QCD predictions were only published in the late 
1990's and early 2000's, once the jet energy was calibrated to 
a $\sim 3\%$ accuracy level.
The challenge to uncover the biases of the data-driven methods used
to measure the jet energy scale was at the core of the publication delay.
The difficulty arose from the lack of MC samples that
modeled response linearity and shower shapes to the necessary
level of accuracy.

\begin{figure}[htbp]
\centering
\vspace{-3cm}
\includegraphics[width=\textwidth]{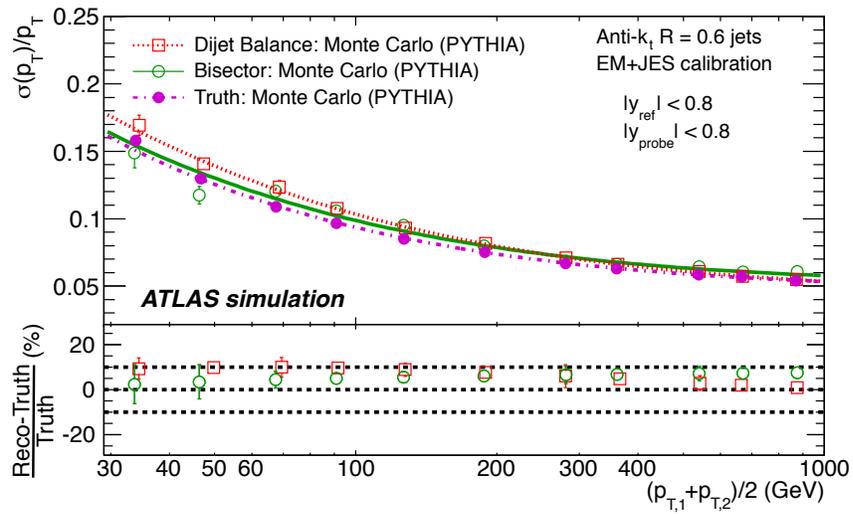}
\vspace{-4.0cm}
\caption{Illustration of a closure test~\cite{atlasjerplots}.
Comparison between the MC truth
jet $p_T$ resolution and the results obtained from the bisector and
di-jet balance methods applied to detector level MC as if it were data.
The lower panel shows the percentage difference, obtained from the fits.
The errors shown are only statistical.}
\label{atlasclosure}
\end{figure}

The concept of closure test is illustrated in Fig.~\ref{atlasclosure}, which
demonstrate the di-jet balance and bisector methods to measure jet
energy resolutions in ATLAS~\cite{atlasjerplots}. 
The MC truth resolution is shown in
full circles, and the measured resolutions extracted from MC detector
level samples using the di-jet balance and bisector techniques are shown
in open squares and circles respectively. The lower panel demonstrates a
better than 10\% accuracy level and indicates the methods are biased to 
slightly overestimate the resolutions.

\begin{figure}[htbp]
  \centering
  \begin{minipage}{.75\textwidth}
    \centering
    \includegraphics[width=\textwidth]{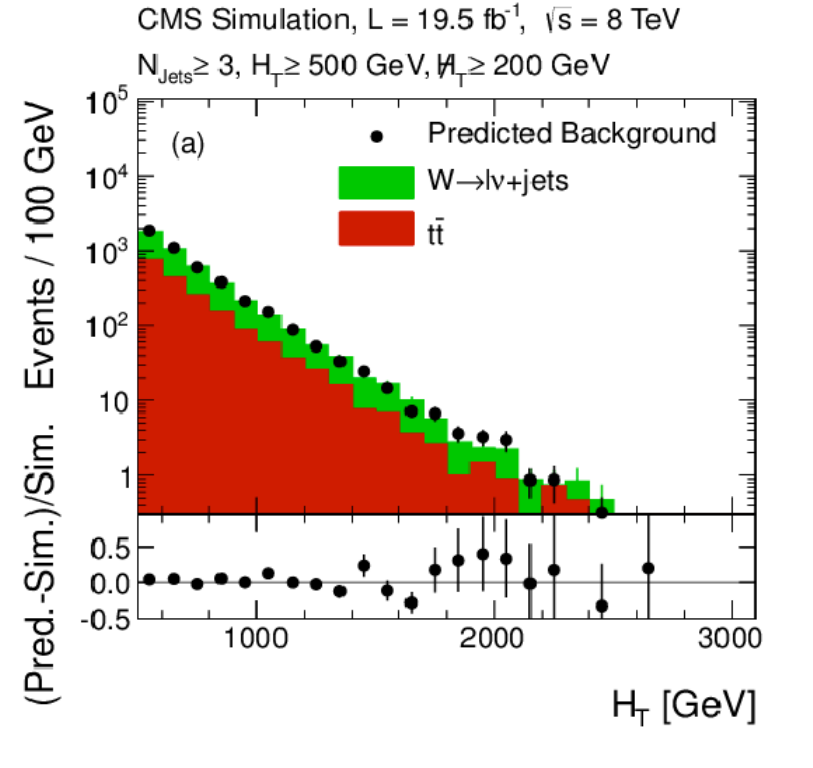}
  \end{minipage}
  \begin{minipage}{.75\textwidth}
    \centering
    \includegraphics[width=\textwidth]{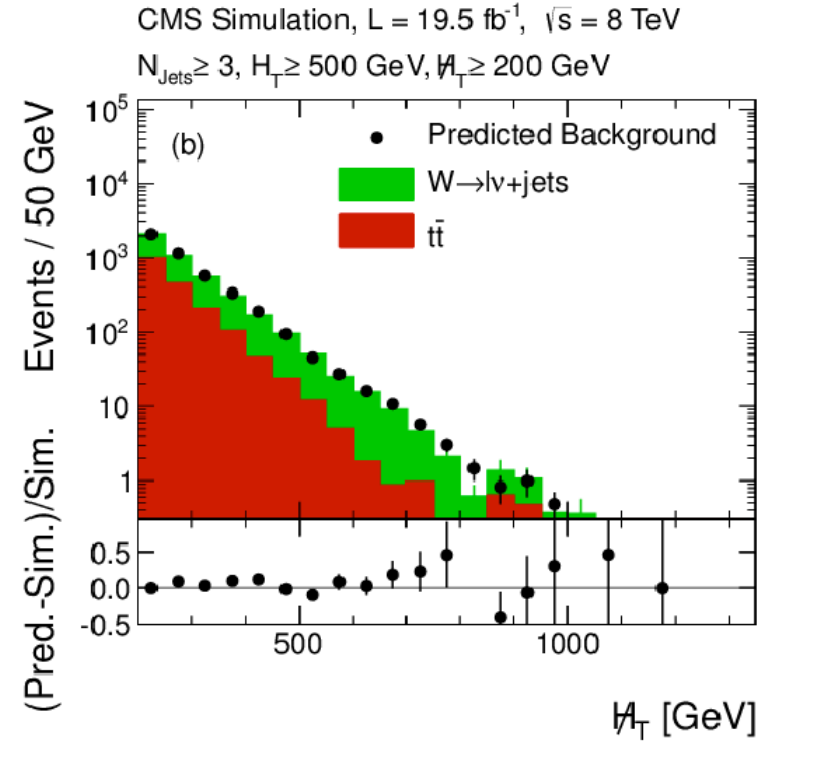}
  \end{minipage}
\caption{Illustration of a closure test. 
Data-driven $t\overline{t}$+jet and $W$+jets backgrounds (circles) 
are compared with the MC truth estimate (histograms) 
for the $H_T$ and $H^{\rm miss}_{T}$ observables in a CMS multi-jets SUSY 
search~\cite{susymultijets}.}
\label{closure1}
\end{figure}

\begin{figure}[htbp]
\centering
\includegraphics[width=0.75\linewidth]{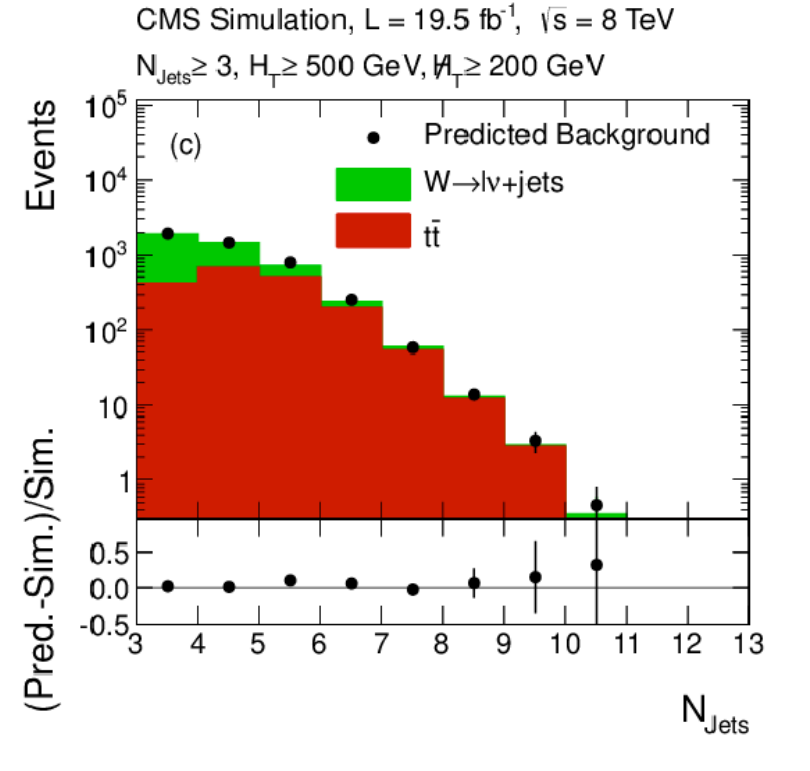}
\caption{Illustration of a closure test. 
Data-driven $t\overline{t}$+jet and $W$+jets backgrounds (circles) 
are compared with the MC truth estimation (histograms) 
for the jet multiplicity distribution in a CMS multi-jets SUSY 
search~\cite{susymultijets}.}
\label{closure2}
\end{figure}

For the same CMS muti-jets SUSY final state introduced in 
Sec.~\ref{ControlSamples}, 
Figs.~\ref{closure1},~\ref{closure2} illustrate MC closure 
tests of a data-driven method to predict the SM background coming from 
$t \overline{t}$+jets and $W$+jets events~\cite{susymultijets}. 
The background arises from events that mistakenly pass the 
$p^{\rm lep}_{T}> 10$~GeV veto cut on the 
$p_T$ of isolated leptons, introduced to remove events with high $p_T$
isolated leptons from the
signal sample. These mistakes occur when the lepton escapes 
detection due reconstruction and identification inefficiencies. 
The data-driven method is based on the 
selection of a one lepton plus jets control sample by inverting the lepton veto 
requirement. In such a sample, 97\% of the events are either 
$t \overline{t}$+jets or $W$+jets. 
Once the number of events in the control region is 
normalized by a factor accounting for reconstruction and identification 
efficiencies,
$(1/\varepsilon_{\rm iso})[(1-\varepsilon_{\rm iso})/\varepsilon_{\mathrm{reco+id}}]$, and the sample is restricted to the signal region, also referred to 
as the search region, the control sample predicts the number of 
electroweak background events in the search region.
Figs.~\ref{closure1},~\ref{closure2} show a 
comparison between the predicted background (circles) and the MC truth 
estimated background (histograms) for $H_T$ (scalar sum of the jet $p_T$'s),
$H^{\rm miss}_{T}$, and jet 
multiplicity. The excellent closure within statistical uncertainties for all 
three observables indicates that the method predicts the background
with high accuracy and any potential biases are under control, within
the quoted uncertainties.
Had there been deviations of the ratios from unity, 
outside statistical and systematic uncertainties, potential
sources of biases on the method would have been further investigated and 
eventually removed, most probably at 
the cost of additional systematic uncertainties.

\subsection{Simulation in Detector Design and Optimization} \label{designoptimization}

HEP collider detectors consist of devises based on 
diverse technologies specialized in observing and characterizing the 
different types of particles that result from high-energy collisions. 
A typical HEP collider detector includes tracking modules to measure the 
interaction vertices and the tracks of charged particles, calorimeters to 
measure energy depositions as a result of electromagnetic and hadronic 
showers, wire chambers to detect high energy muons, and magnets for particle 
identification and momentum measurement. For event reconstruction, 
modern experiments follow a holistic approach, using all sub-detector 
components to reconstruct each particle individually by means of 
complex software algorithms. This is the case of the particle flow 
technique used by CMS~\cite{pflow}. 

To design a HEP detector, different technologies and physical 
characteristics are modeled and optimized in simulation for best physics 
performance. For example, the efficiency and precision of particle
tracking algorithms in a silicon 
detector typically improves by increasing the pixel and strip density, the 
number of layers, and the angular coverage, as well as by minimizing the 
amount of material a particle traverses. Muon detection improves with the 
wire chamber density, number of layers in the radial direction, and angular 
coverage. More powerful or weaker magnets allow for more compact or 
larger designs, 
with a range of momentum resolutions. An ideal calorimeter provides full solid 
angle coverage and hermeticity, and improves its performance with higher 
transverse granularity, longitudinal segmentation, and materials that yield 
Gaussian and narrow response functions. These parameters are 
varied in the simulation and the final design is selected using a
cost-benefit equation that considers monetary cost versus detector
physics performance.
 
Monte Carlo simulation campaigns for detector design and optimization 
produce millions of events generated with different 
detector scenarios that vary in technology options and physical parameters. 
Goals range from making a case for a given 
detector configuration, to 
optimizing a design for maximum physics output, or investigating the physics 
impact of detector de-scoping options driven by budgetary constraints. 
Nowadays, these simulation efforts are an absolute requirement for every 
HEP experiment seeking approval from funding agencies. 
Interestingly, Geant4 plays a dual role in the process. Firstly, it helps
select the optimal design that does the physics job for the available 
budget. Secondly, Geant4 influences 
the detector design by adding its own software and computing constraints to 
the decision process. In other words, the optimal geometric designs are often 
too difficult to model with Geant4 or very expensive in computing resources. 
While Geant4 evolves to support experiments 
with more features and speed, detector configurations also adapt to play to 
the strengths of the Geant4 simulation toolkit.  

\begin{figure}[htbp]
  \centering
  \begin{minipage}{.8\textwidth}
    \centering
    \includegraphics[width=\textwidth]{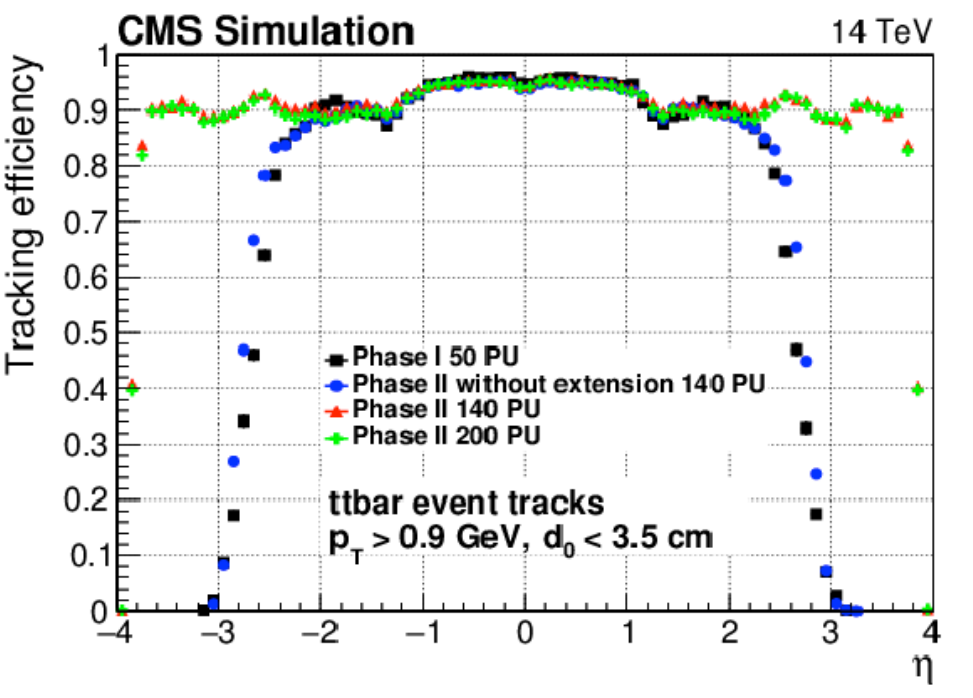}
  \end{minipage}
  \begin{minipage}{.8\textwidth}
    \centering
    \includegraphics[width=\textwidth]{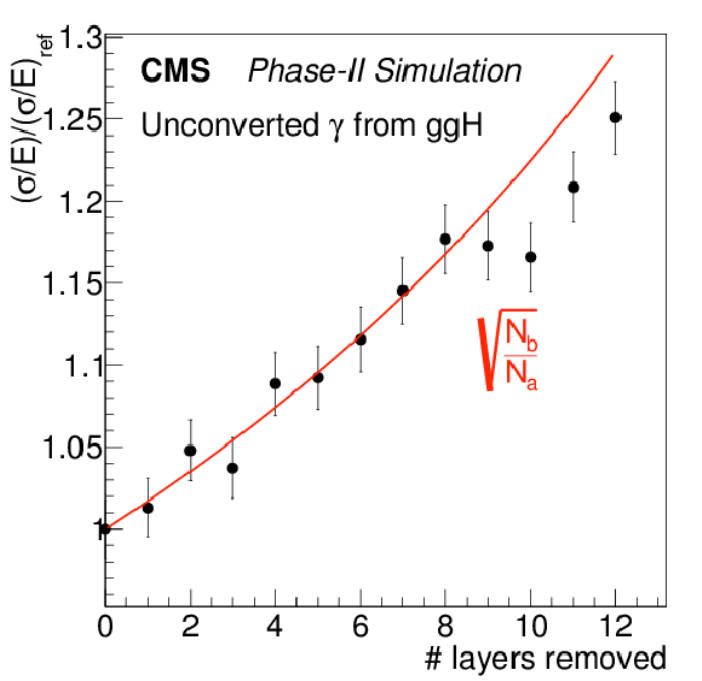}
  \end{minipage}
\caption{Top: Predicted CMS tracking efficiency versus $\eta$ for a range of 
luminosity scenarios (number of pile-up events), and two 
detector upgrade options, with and without extended $\eta$ 
coverage~\cite{cmsupgrade}.
Bottom: Relative degradation in photon energy resolution in gluon fusion Higgs 
events 
as a function of the number of layers removed from the proposed CMS endcap 
calorimeter~\cite{cmsupgrade}.}
\label{detectordesign1}
\end{figure}

\begin{figure}[htbp]
  \centering
  \begin{minipage}{.85\textwidth}
    \centering
    \includegraphics[width=\textwidth]{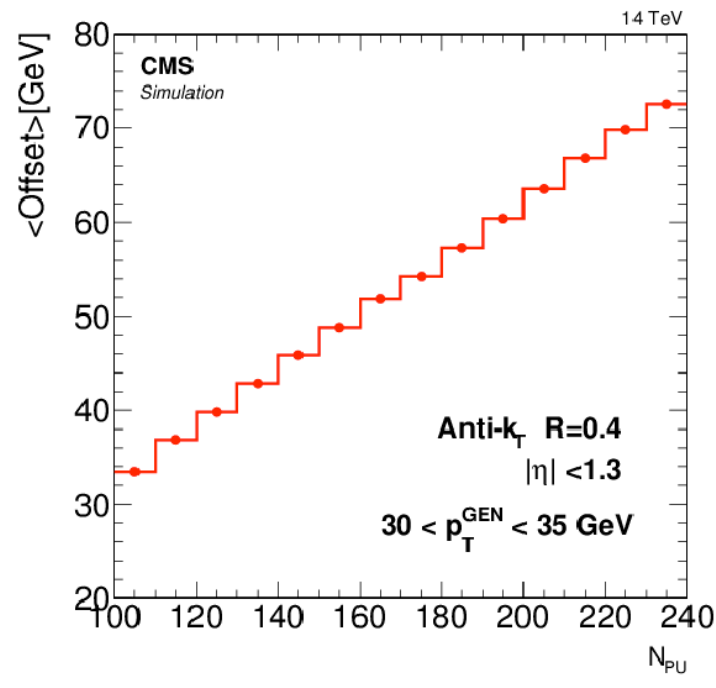}
  \end{minipage}
  \begin{minipage}{.85\textwidth}
    \centering
    \includegraphics[width=\textwidth]{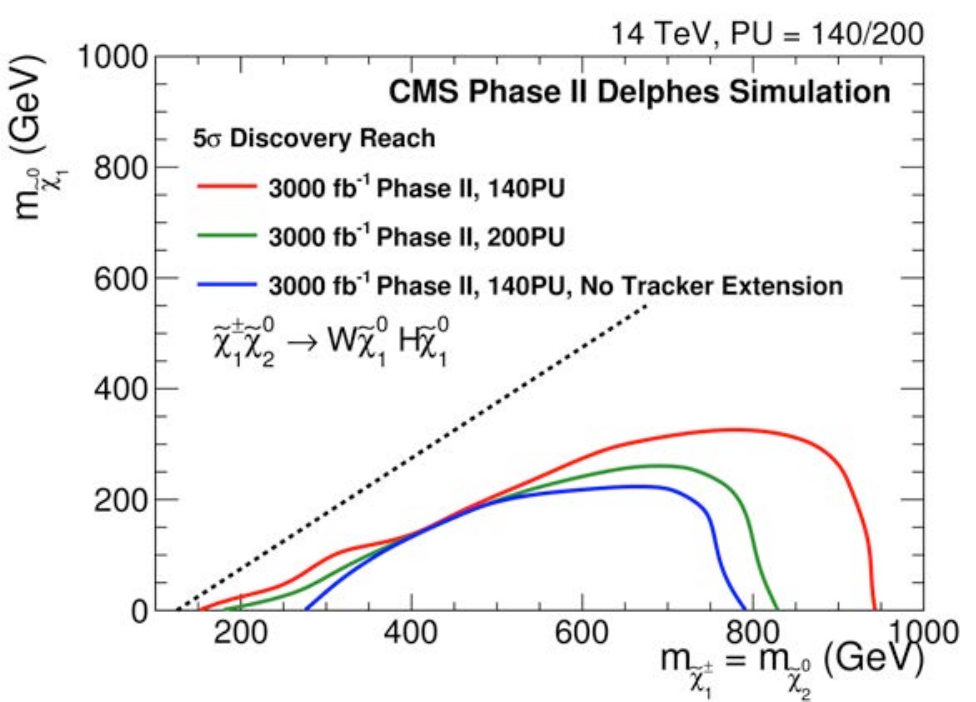}
  \end{minipage}
\caption{Top: Average energy from pile-up events contributing to a 
reconstructed jet as the number of pile-up events increases with 
luminosity~\cite{cmsupgrade}. Bottom: Effect of luminosity and impact of a 
proposed CMS tracker extension in the sensitivity to SUSY particle 
production~\cite{cmsupgrade,delphes}.}
\label{detectordesign2}
\end{figure}

Figs.~\ref{detectordesign1},~\ref{detectordesign2} illustrate the use of 
simulation for design studies in the 
CMS experiment~\cite{cmsupgrade}. Performance tests evaluate basic detector 
level observables, such as track efficiencies, photon, 
electron, muon and jet resolutions, as well as the potential precision 
of cross section or mass measurements, or the discovery reach for new 
particles. As an example, Fig.~\ref{detectordesign1} (top) shows the 
predicted CMS tracking efficiency versus pseudorapidity for various tracker
design options and accelerator performance parameter values associated with the
high-luminosity LHC (HL-LHC) run scheduled to start in 2026. Efficiency is
studied for different accelerator performance scenarios, expressed in terms
of instantaneous luminosity, $L=\frac{1}{\sigma}\frac{dN}{dt}$, where
$dN/dt$ is the number of events produced as the result of the hard
collision and $\sigma$ is the interaction cross section. $L$ depends on
detector parameters such as the number of particles in a bunch within the beam,
and the size of the beams. 
At the LHC, as $L$ increases, the probability of multiple 
proton-proton (pp) interactions per crossing with low momentum transfer 
increases. 
These spurious 
interactions pile-up and overlap with the high-$p_T$ event of interest 
which fired the physics trigger. For different pile-up
(PU) scenarios, measured in terms of the number of spurious pp interactions, 
Fig.~\ref{detectordesign1} (top) shows the tracking 
efficiency for the 2017 detector (Phase I detector, black squares) and the 
proposed 2026 detector (Phase II detector, blue, red, 
green symbols) with and without a tracker upgrade that would extend the 
$\eta$ coverage to 3.8. 
The addition would extend the coverage in the region near the beam pipe, 
improving the tracking efficiency and reducing the fake rate (not shown), 
thus allowing to suppress more efficiently 
any spurious contribution from pile-up events to the high-$p_T$ event
under study. Fig.~\ref{detectordesign1} (bottom) 
shows the relative 
degradation in photon energy resolution, measured in gluon fusion Higgs events, 
as a function of the number of layers removed from the proposed CMS endcap 
calorimeter. $N_{b/a}$ is the number of layers before and after removal. This 
degradation affects measurements with the Higgs boson decaying to two photons 
or four electrons. 

Fig.~\ref{detectordesign2} (top) shows how much energy from 
pile-up events 
contribute on average to a reconstructed jet as the number of pile-up events 
increases with luminosity. This spurious contribution changes the jet 
multiplicity of the event, distorts the jet energy response, and 
degrades jet energy and 
missing transverse energy resolutions. Missing transverse energy is a measure 
of the event momentum imbalance in the transverse plane and will be defined 
and discussed properly in Sec.~\ref{misstransmom}.
Fig.~\ref{detectordesign2} 
(bottom) shows, using the Delphes parametrized simulation 
framework ~\cite{delphes}, the 
impact of the tracker extension and the number of pile-up events on the 
sensitivity of the proposed detector to SUSY particle production. The
study investigates a model with electroweak production of a chargino-neutralino
pair, $\tilde{\chi}^{\pm}_{1} \tilde{\chi}^{0}_{2}$, decaying to $WH$ and two
stable neutralinos,  $\tilde{\chi}^{0}_{1}$, the latter being 
the lightest stable particle (LSP) predicted by the model. 
For an integrated luminosity of 
3000~fb$^{-1}$, that is the integral of the instantaneous luminosity over the 
time covering the full data set expected to be collected by the CMS experiment
during the HL-LHC run, the sensitivity is explored for three different
scenarios: 140 PU events with and without a tracker extension, and 200~PU
events with a tracker extension. The limits
on the chargino mass are very sensitive to an increase in the number of
PU events, although a fraction of the sensitivity is
recovered with the tracker extension. 

\begin{figure}[htbp]
  \centering
  \begin{minipage}{1.0\textwidth}
    \centering
\vspace{-4cm}
    \includegraphics[width=\textwidth]{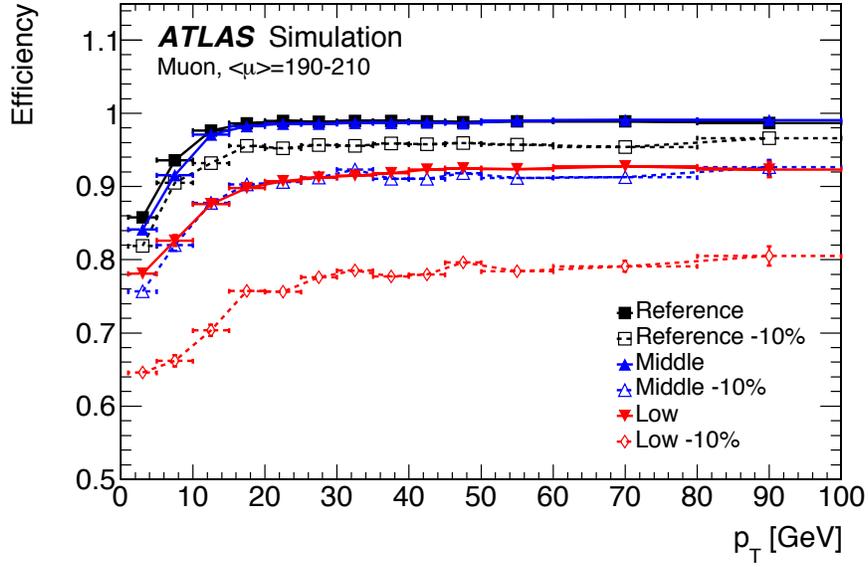}
\vspace{-6cm}
  \end{minipage}
  \begin{minipage}{0.9\textwidth}
    \centering
    \includegraphics[width=\textwidth]{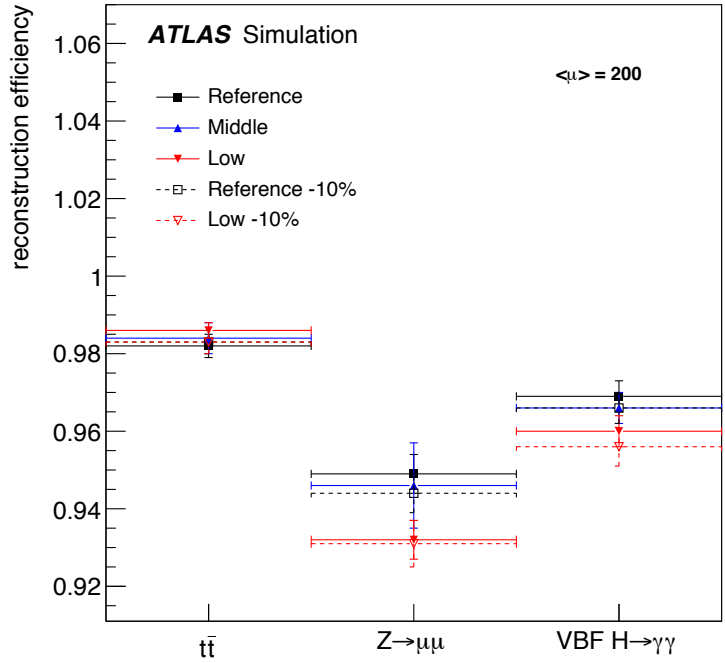}
\vspace{-3cm}
  \end{minipage}
\caption{Top: Momentum dependence of the 
muon reconstruction plus identification efficiency for
different ATLAS detector upgrade scenarios and 200 PU 
events~\cite{atlasupgrade}.
Bottom: Primary vertex reconstruction efficiency in 
$t\overline{t}$, $Z \rightarrow \mu^{+}\mu^{-}$, and Vector Boson Fusion (VBF)
$H\rightarrow \gamma \gamma$ events for the three ATLAS 
detector upgrade scenarios and 200 PU events~\cite{atlasupgrade}.}
\label{atlasupgrade1}
\end{figure}

\begin{figure}[htbp]
  \centering
  \begin{minipage}{1.0\textwidth}
    \centering
\vspace{-4cm}
    \includegraphics[width=\textwidth]{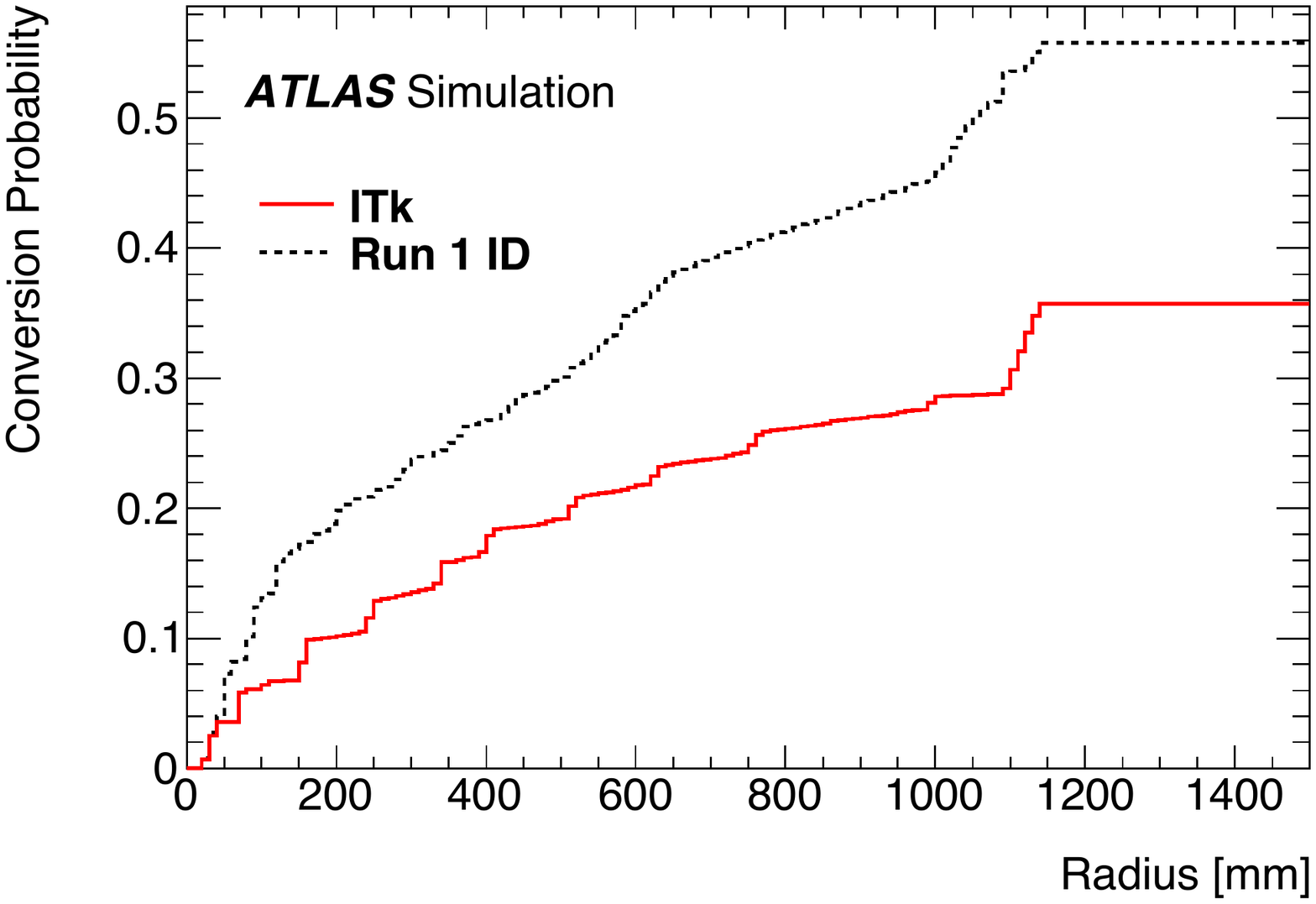}
\vspace{-8.0cm}
  \end{minipage}
  \begin{minipage}{1.0\textwidth}
    \centering
    \includegraphics[width=\textwidth]{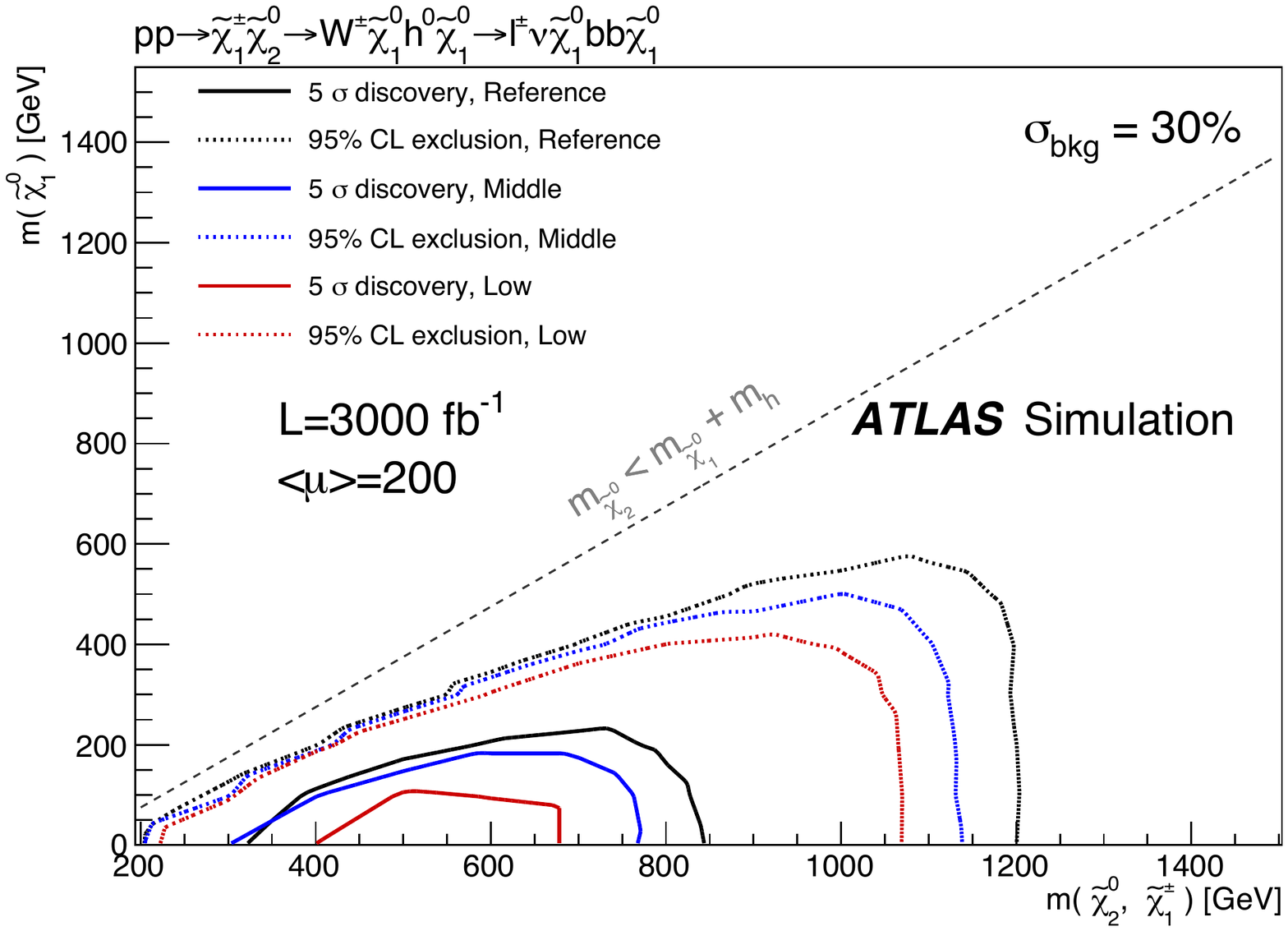}
\vspace{-4cm}
  \end{minipage}
\caption{Top: Photon conversion cumulative probability as a function of
the distance from the interaction vertex (radius) for the Run 1 and HL-LHC
ATLAS detectors~\cite{atlasupgrade}. 
Bottom: Sensitivity to the different detector upgrade scenarios of an 
ATLAS SUSY search for the production of a chargino-neutralino pair 
$\tilde{\chi}^{\pm}_{1} \tilde{\chi}^{0}_{2}$ in the case of 200 PU 
events~\cite{atlasupgrade}.}
\label{atlasupgrade2}
\end{figure}

The ATLAS Collaboration also performed various MC studies to optimize its
detector design for the HL-LHC era. Ref.~\cite{atlasupgrade} contains a
description of the upgrade options under consideration, which include
extensions to the tracker and pixel detectors to cover pseudorapidity 
ranges of $|\eta|<4$ (Reference scenario),  $|\eta|<3.2$ (Middle scenario),
or $|\eta|<2.7$ (Low scenario). The upgrade also includes improvements 
to the trigger system, detector electronics, and 
forward calorimetry (Reference scenario). The results presented here are based
on simulated events produced with the ATLAS Geant4-based simulation
application.
Fig.~\ref{atlasupgrade1} (top) shows the momentum dependence of the 
muon reconstruction plus identification efficiency for
different ATLAS detector upgrade scenarios and 200 PU events. 
It is apparent that the detector descoping
from the Reference scenario to the Low scenario would cost the 
experiment 10\% in
muon efficiency. The primary vertex reconstruction efficiency in ATLAS for 
$t\overline{t}$, $Z \rightarrow \mu^{+}\mu^{-}$, and Vector Boson Fusion (VBF)
$H\rightarrow \gamma \gamma$ events is shown in Fig.~\ref{atlasupgrade1} 
(bottom) in the case of 200 PU events. 
The lesson learned is that vertexing performance does not 
depend strongly on the tracker layout, and it varies with physics process.
While the efficiency does not change for $t\overline{t}$, it goes down
by 1\%(2\%) for VBF $H\rightarrow \gamma \gamma$ 
($Z \rightarrow \mu^{+}\mu^{-}$) events when switching from the
Reference to the Low scenario. 
Fig.~\ref{atlasupgrade2} (top) illustrates on the
reduction in the photon conversion cumulative probability as a function of
the distance from the interaction vertex (radius) when the ATLAS 
Inner Tracker (ITk)
is upgraded. The ATLAS SUSY search shown in Fig.~\ref{atlasupgrade2} (bottom)
is for the production of a chargino-neutralino pair 
$\tilde{\chi}^{\pm}_{1} \tilde{\chi}^{0}_{2}$, which 
decays into a $W$, a SM-like Higgs boson, and LSP (neutralinos). 
The final state consists of two jets, one isolated lepton, two b-jets, and 
large missing transverse momentum coming from the weakly interacting
neutralino, $\tilde{\chi}^{0}_{1}$. The study is performed for 200 PU events and
show a 200~GeV improvement
in the limit to the chargino/neutralino mass for low LPS masses, when
switching from the Low to the Reference detector scenario. 

\subsection{Simulation in Software and Computing Design and Testing} \label{softwaredesign}

Simulation is also an essential tool to develop each element of the workflow 
and dataflow associated with data handling in large HEP experiments. At the
LHC, the Worldwide LHC Computing Grid (WLCG)~\cite{cerngrid} is used to 
process, store and
analyze the data collected or generated by the experiments.
The WLCG is composed of four levels or ``Tiers'': 0, 1, 2, 3. The difference
between Tiers is in the services they provide, whether they host raw data,
and how well they are interconnected.
The Tier 0 is located at CERN in Geneva, Switzerland and at the Wigner
Research Centre for Physics in Budapest, Hungary. All data passes through the 
two Tier 0 sites, which are 
connected by two dedicated 100~Gbit/s data links and provide less than 
20\% of the compute capacity. The main role of the Tier 0 is
to safe-keep the raw data, perform a first pass reconstruction, reprocess 
data when the LHC is not running, and distribute
the raw and reconstructed data to the Tier 1 centers.
The Tier 1 consists of 13 computing centers with large storage capacity
distributed all over the world. They are responsible for the safe-keeping
of different shares of all raw and reconstructed data, as well as for 
performing large-scale reprocessing and storing the 
associated output. The Tier 1 centers distribute data to the Tier 2 
centers and store a share of the simulation output produced by the Tier 2's.
A dedicated high-bandwidth network, consisting of 10~Gbit/s 
optical-fiber links, connect CERN to most of 
the Tier 1 centers around the world.
The approximately 160 Tier 2 centers are typically located at research 
institutions outside CERN
and provide data storage capacity and computing power for
simulated event production
and reconstruction, as well as for data analysis tasks. 
Tier 3 computing resources are not part of the WLCG and 
refer to local clusters in universities, other scientific institutes, or 
even individual PC's, that 
scientists use to access the WLCG resources.

\begin{figure}[htbp]
\centering
\includegraphics[width=0.9\linewidth]{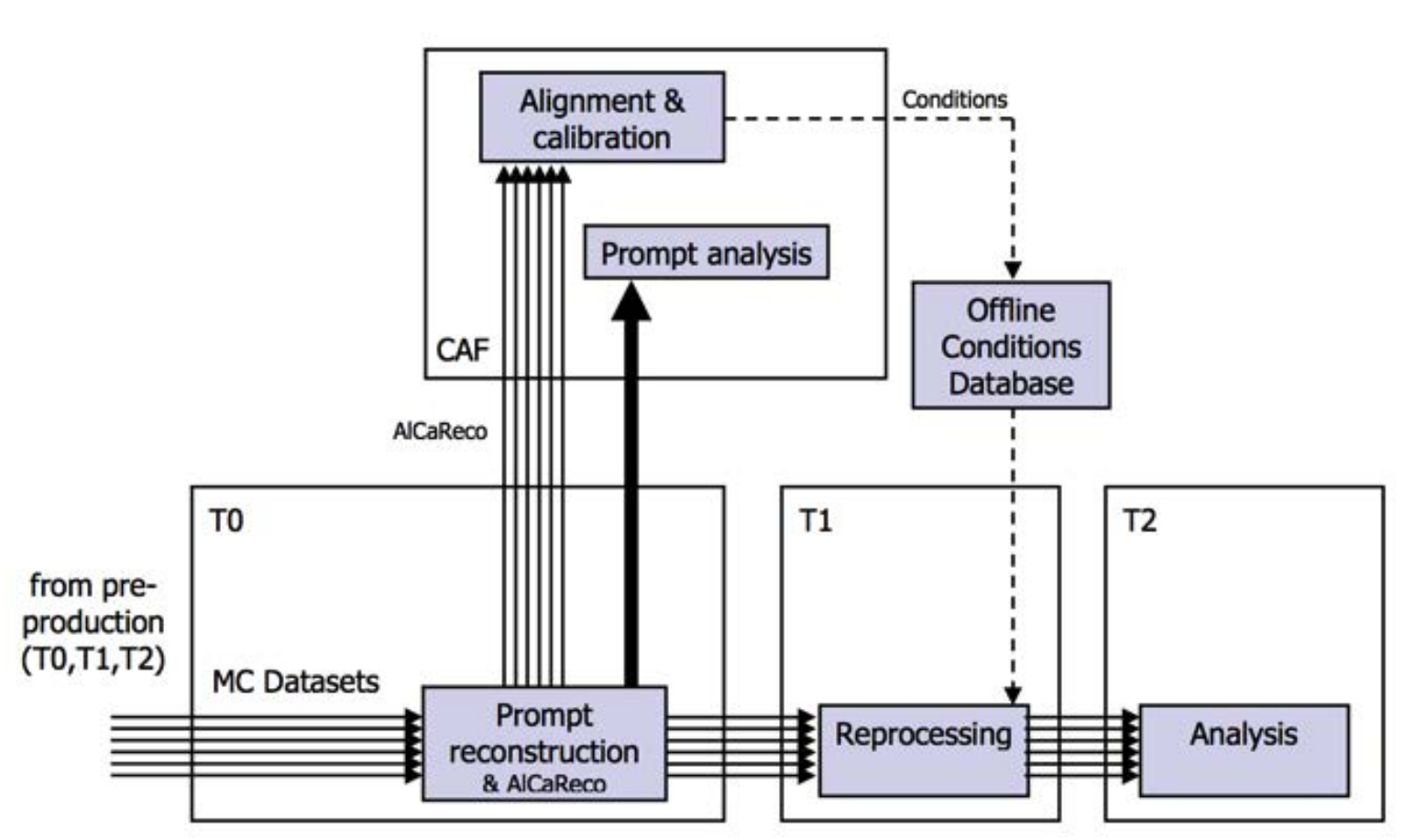}
\caption{Workflow for the CMS 2008 Software, Computing, and Analysis 
challenge (CSA)~\cite{csa2008}.}
\label{csaworkflow}
\end{figure}

In CMS, the combined procedure of data acquisition, processing, transfer,
and storage using WLCG resources was tested in a series of computing, 
software and analysis (CSA) challenges. The tests included components such as 
the preparation of large simulated data-sets, prompt reconstruction at the 
Tier 0 center, the distribution of output files to Tier 1 centers for 
re-reconstruction and skimming, calibration jobs on alignment and calibration 
data-sets, and physics analysis in Tier 2 centers. 
In a series of exercises in 2006, 2007, 2008 
(Run 1) and 2014 (Run 2), the computing system was stress tested at 
25\%, 50\%, and 100\% capacity. In preparation for Run 1, 
150 million simulated events were 
produced, realistic trigger rates were modeled, and reconstruction and physics 
analysis performed in real time for event samples representing an integrated 
luminosity in excess of a quarter of the total delivered in 2010. For 
illustration, the workflow for the CMS 2008 CSA challenge\cite{csa2008}
is shown in Fig.~\ref{csaworkflow}. The ``pre-production'' samples are
simulated data modeling the real raw data acquired by the detector and
filtered by the trigger according to the same physics requirements coded
in the actual trigger system. This step was performed in various Tier 0, 1
and 2 computing centers, and the resulting output data copied to the Tier 0 
center at CERN, where prompt reconstruction followed. Next, the MC data-sets 
utilized for calibration and alignment, the ``AlCaReco'' files, were produced 
and transferred to the CERN Analysis Facility (CAF). Then, the calibration and 
alignment constants were derived at the CAF and transferred to the conditions 
database. The data was reprocessed in the Tier 1 center, re-reconstructed as 
is typical in the experiments to correct mistakes in the first pass. Finally,
the physics analysis was performed in the Tier 2 centers.

The realism of these rehearsals in 21$^{st}$ century experiments has allowed 
them to reach data taking with an unprecedented degree of preparedness. Event 
and file sizes, memory and CPU time consumption, detector geometry description 
and alignment, particle showering in the detector material, electronics, 
calibration procedures, prompt reconstruction, data transfer between computing 
processing centers were tested so accurately and realistically with
MC samples, that 
experiments as complex as ATLAS and CMS did not meet major surprises during 
start-up, with most components working as predicted, within design 
specifications and, basically, out of the box. 

\section{Simulation of Collider Physics Observables for Particles and Events} \label{physperf}

The level of agreement between the MC predictions of physics
observables and 
the corresponding data measurements are a test of 
the accuracy of the simulation software. 
This section starts with a discussion on the impact of the detector
geometry and materials modeling on the simulation of photons, electrons, 
and muons. It follows with data-to-MC comparisons for b jet identification
variables, a set of $W/Z$+jets observables, and missing transverse
energy distributions and resolutions. The impact of simulation in the 
precision of jet cross
section measurements and publication timeline is presented at the 
end as a case study.

\subsection{Geometry and Material Modeling Effects on Photon, Electron, and Muon Simulation} \label{geometry}

Accurate simulation of electrons and photons necessitates a
very detailed description of the material and thickness of the tracker system 
components. 
Typically, trackers are highly segmented to provide 
efficient particle identification and precise measurements of particle 
trajectories and momentum in the presence of a magnetic field. In addition,
these detectors must be thin and light to minimize interactions
before the particles reach the calorimeters. In an ideal detector, 
electrons, photons, and hadrons would traverse the 
tracker unperturbed and experience their first destructive 
interaction and subsequent 
shower in the calorimeters, the detector components designed to measure 
energy. In real detectors, most significantly in the case of silicon
trackers, particles do interact and disappear (photon conversion) or 
loose a large fraction of their total 
energy while traversing the detector material (charged particles). 
This is a price that most modern experiments are willing
to pay in exchange for the more precise position and momentum measurements,
faster readout, and better radiation tolerance
offered by silicon-based detectors. For example in CMS, 
photons have a 70\% probability to convert 
into electrons within the silicon tracker volume, a difficult challenge to
overcome given the key role that photons play in Higgs measurements 
($H \rightarrow \gamma \gamma$), direct photon strong production studies, 
and BSM searches. Simulation is a useful tool
to understand the impact of tracking detector material on physics measurements, 
keep the systematic uncertainties under control, and deliver competitive 
results. The necessary condition is that the tracker materials,
shapes, and thicknesses are described with precision in the geometry code, and
that photon-nucleus interactions, 
photon conversions and energy loss, as well as multiple scattering are 
accurately modeled in Geant4. 

\begin{figure}[htbp]
  \centering
  \begin{minipage}{.75\textwidth}
    \centering
    \includegraphics[width=\textwidth]{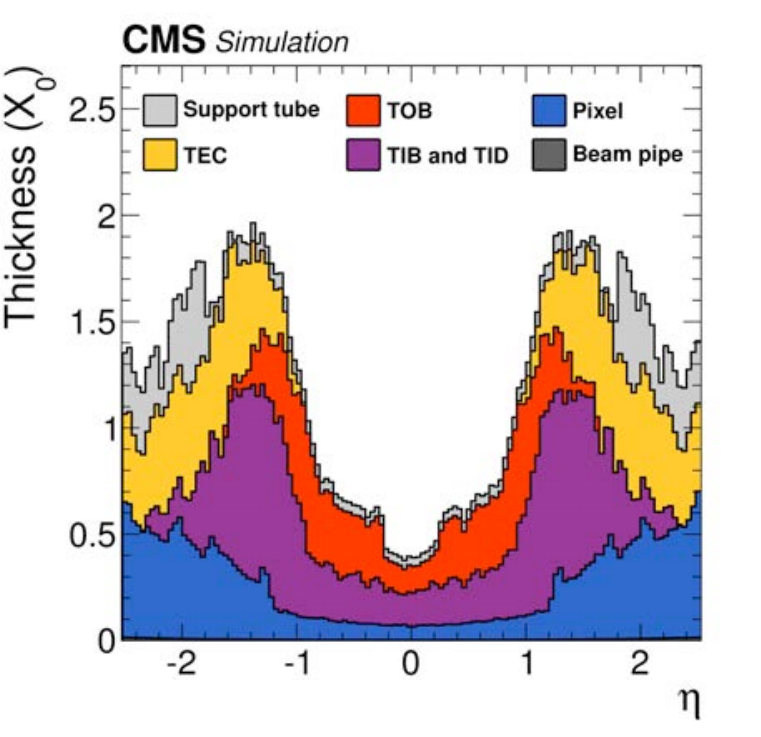}
  \end{minipage}
  \begin{minipage}{.70\textwidth}
    \centering
    \includegraphics[width=\textwidth]{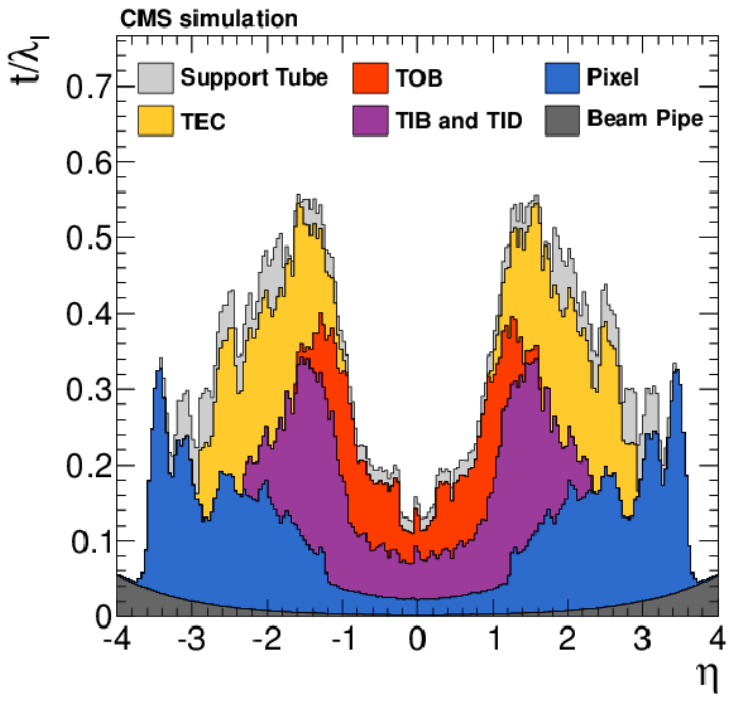}
  \end{minipage}
\caption{Total thickness of the CMS tracker material, in units of radiation 
lengths $X_0$ (top) and interaction lengths $\lambda_I$ (bottom), that a 
particle produced at the center of the detector would traverse as it moves 
along different pseudorapidity directions in the $\eta<2.5$ acceptance region. 
The contribution to the total material of each of the subsystems that 
comprise the CMS tracker is shown separately.}
\label{material}
\end{figure}

\begin{figure}[htbp]
\centering
\includegraphics[width=1.0\linewidth]{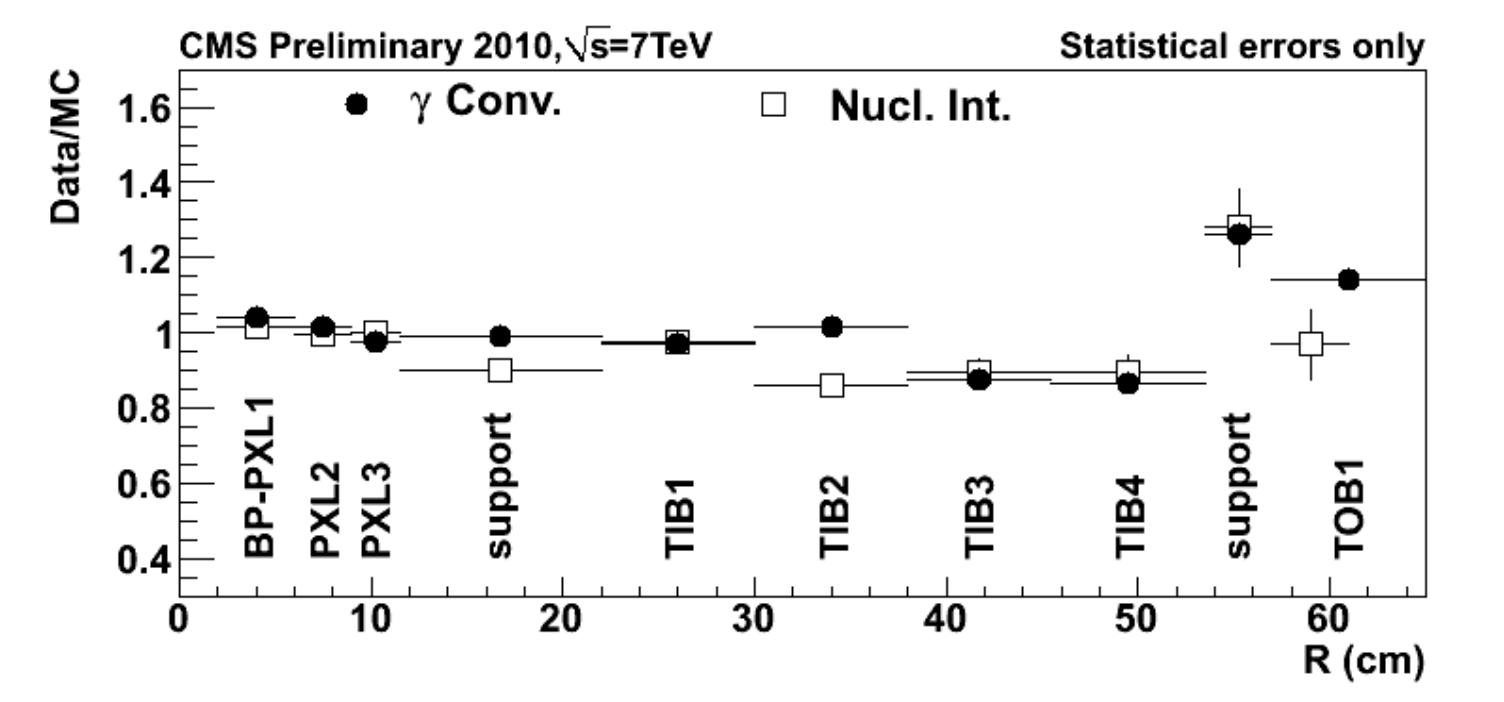}
\caption{Data-to-MC ratio of the fraction of photons undergoing conversions
(full circles) and nuclear interactions (open circles) in the CMS tracker
volume. Discrepancies between MC and data can be related directly to 
deficiencies in the detector geometry modeling of the tracker~\cite{trackmat}. 
The ratio is plotted as a function of the radial distance ($R$) from the 
center of the detector,
which is correlated with different sub-detector components.}
\label{conversions}
\end{figure}

In the CMS simulation software, the implementation of the shapes and materials 
of the tracker geometry elements (350,000 volumes) was 
followed by careful validation to achieve accuracy. 
Fig.~\ref{material} shows the total thickness of the CMS tracker material
in units of radiation lengths $X_0$ (top) and interaction lengths $\lambda_I$ 
(bottom) that a particle produced at the center of the detector would traverse 
as it moves along different pseudorapidity directions in the 
$\eta < 2.5$ acceptance region. The contribution to the total material of each 
of the subsystems that comprise the CMS tracker is given separately: the pixel 
tracker, the strip tracker which consists of the tracker endcap (TEC), the 
tracker outer barrel (TOB), the tracker inner barrel (TIB), and the tracker 
inner disks (TID), the support tube that surrounds the tracker, and the beam 
pipe~\cite{trackperf}. Fig.~\ref{conversions} presents the data-to-MC ratio of 
the 
fraction of photons undergoing conversions and nuclear interactions as a 
function of the radial distance ($R$) from the center of the detector, which is
correlated with different sub-detector components. This ratio is computed
from data-driven measurements of the conversion and nuclear interaction
probabilities respectively, demonstrate 
agreement between data and MC within 15\%, and may be used as scaling factors
to correct the MC before using it in physics analysis. 
Discrepancies observed in Fig.~\ref{conversions} can be directly
related to deficiencies in the detector geometry modeling~\cite{trackmat}.

\begin{figure}[htbp]
  \centering
  \begin{minipage}{.9\textwidth}
    \centering
    \includegraphics[width=\textwidth]{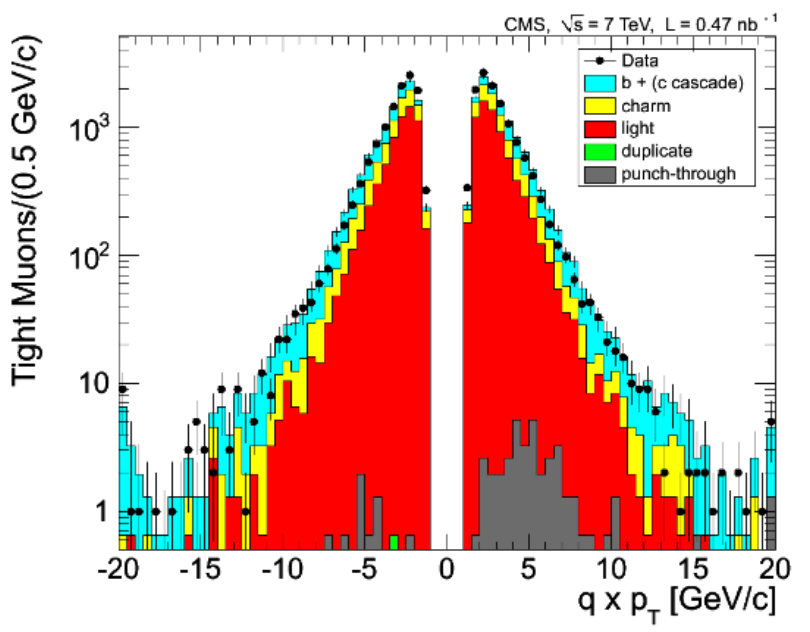}
  \end{minipage}
  \begin{minipage}{.9\textwidth}
    \centering
    \includegraphics[width=\textwidth]{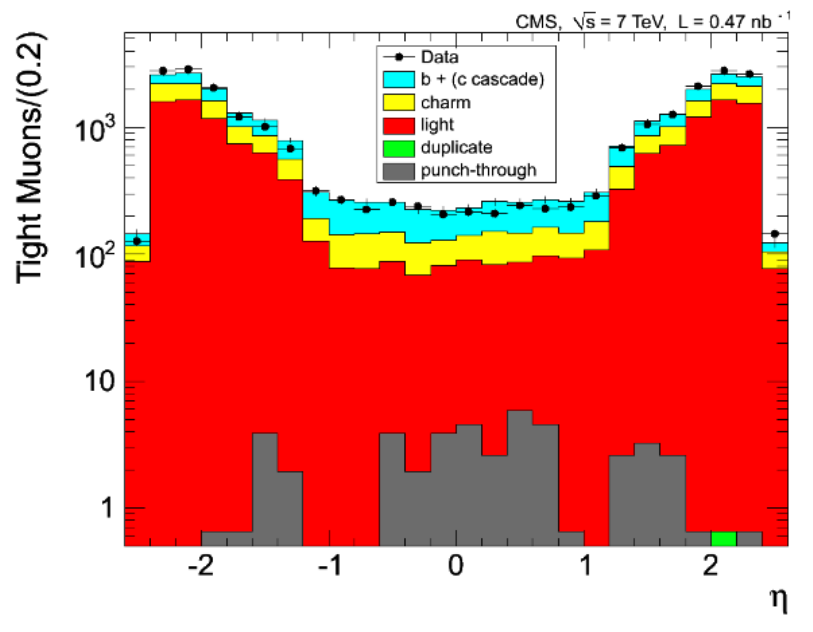}
  \end{minipage}
\caption{Muon $q\times$ $p_T$ ($q$ is the charge) and $\eta$ distributions 
for a CMS sample of muons selected by the zero-bias trigger. Data is
presented in full circles and simulation in histograms~\cite{muonpog}. 
The error bars indicate the statistical uncertainty.}
\label{muons}
\end{figure}

Muons are also particularly sensitive
to the modeling of the detector geometry and material, because they interact
very little with matter, and therefore traverse all detector sub-systems in a 
collider experiment.
Fig.~\ref{muons} shows the $q\times p_T$ and $\eta$ distributions, where
$q$ is the muon charge, 
for CMS muons selected from zero-bias data~\cite{muonpog}. Zero-bias refers 
to a sample of events collected from random proton bunch crossings without any 
specific trigger requirement. The sub-sample of all muons contained in the
zero-bias sample includes the contributions of prompt muons from 
$W$ and $Z$ decays, muons from heavy flavor decays 
(b- and c-quarks or $\tau$-leptons), light hadrons ($\pi$, $K$) or decays of 
particles produced in nuclear interactions, and muons from hadrons that 
penetrate the detector beyond the limits of the calorimeters. 
In Fig.~\ref{muons}, the inclusive muon sample selected in data is compared
with the sum of the MC predictions for each of the above-mentioned processes. 
The excellent agreement in the kinematic regions where data and MC are 
compared, $p_T=1-20$~GeV and $|\eta|<2.6$, is remarkable given that the pixel, 
tracker, and muon systems are all used in muon reconstruction, 
involving a diversity of technologies, shapes,
and materials, as well as abrupt transitions between sub-detector systems. 

\begin{figure}[htbp]
  \centering
\vspace{-3cm} 
 \begin{minipage}{.9\textwidth}
    \centering
    \includegraphics[width=\textwidth]{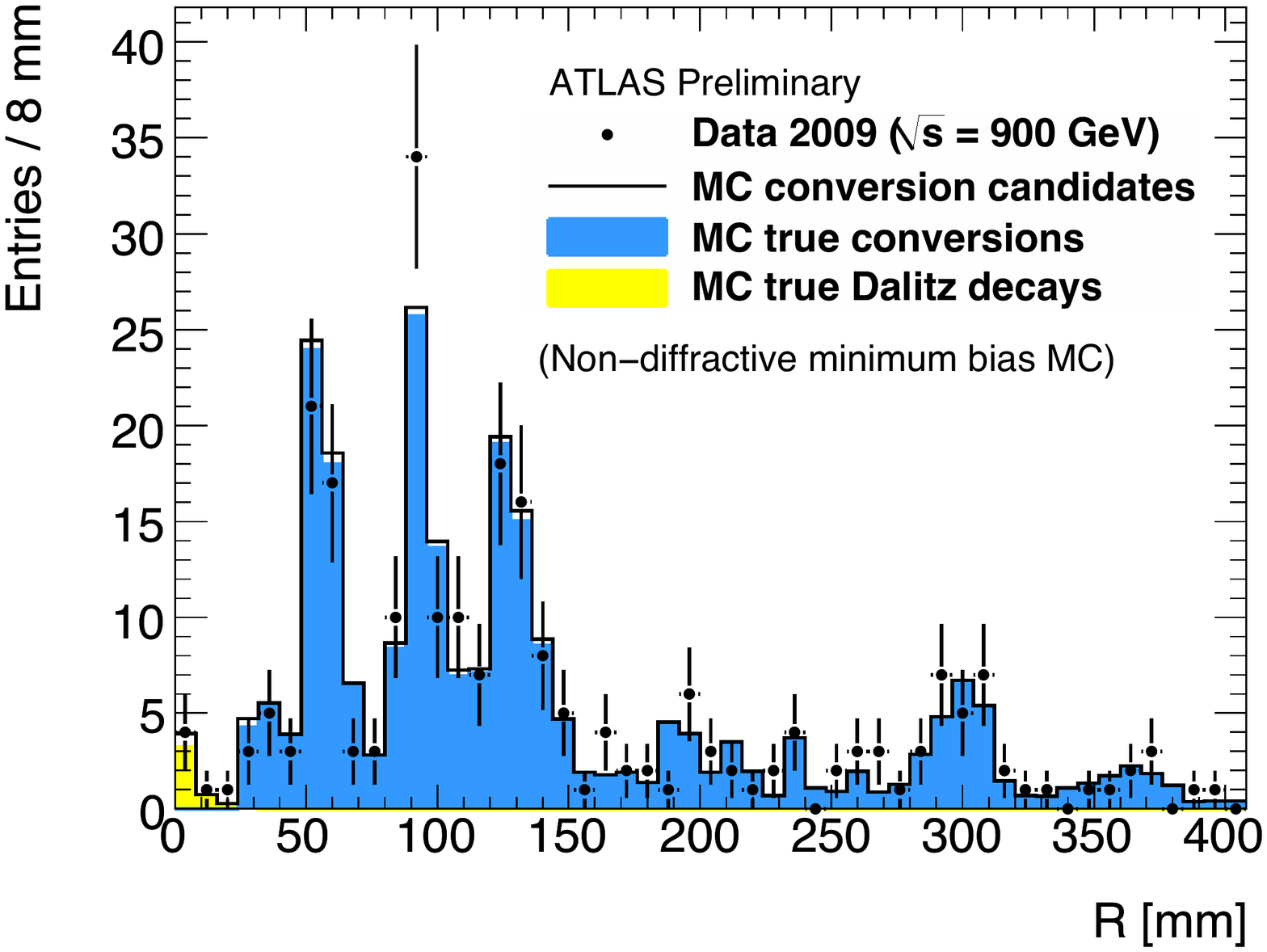}
\vspace{-6.5cm} 
  \end{minipage}
  \begin{minipage}{.9\textwidth}
    \centering
    \includegraphics[width=\textwidth]{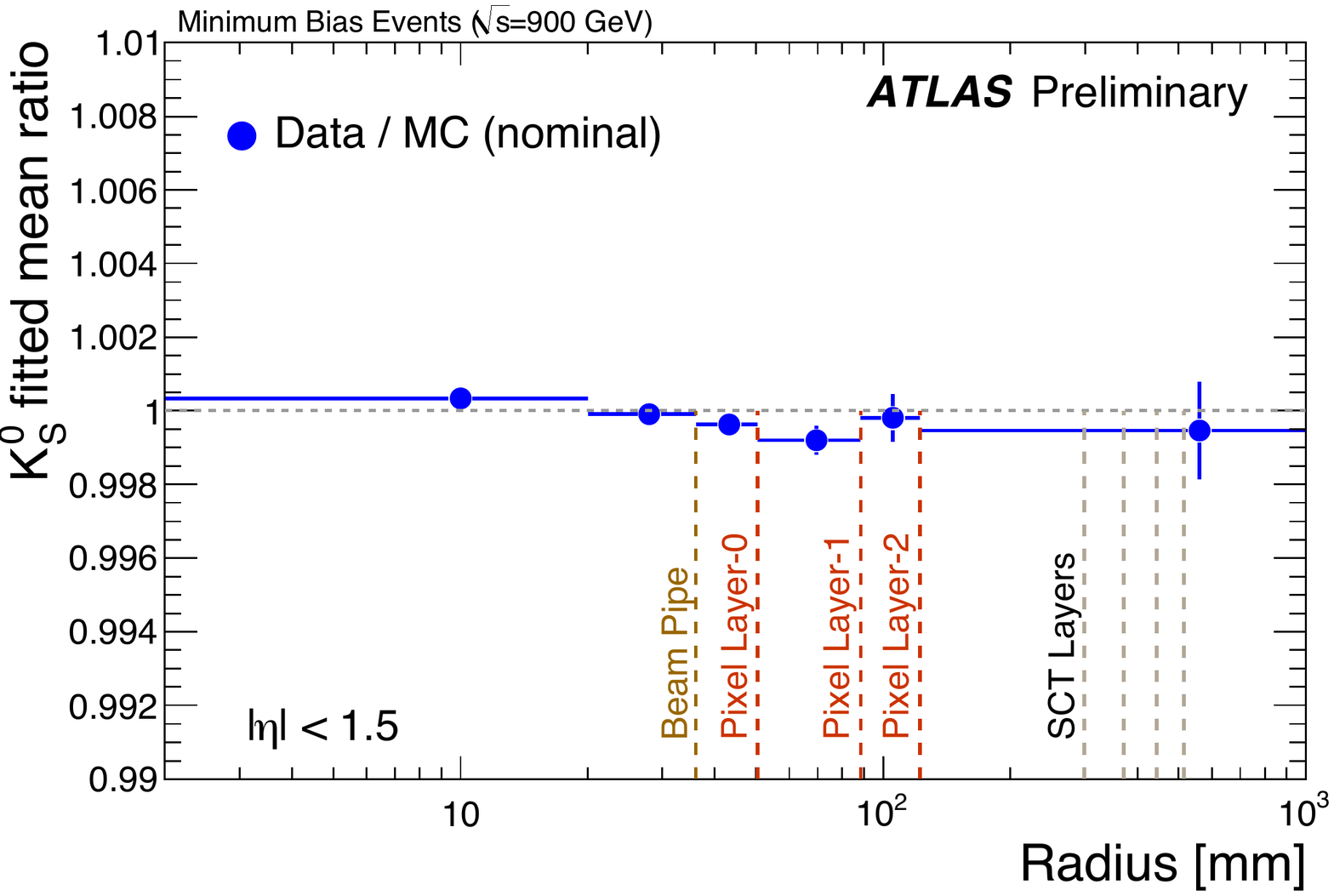}
\vspace{-4cm} 
  \end{minipage}
\caption{Top: ATLAS distributions of photon conversion
vertices in the radial direction starting from the detector center are
shown for collider data, conversion candidates, obtained using the
same analysis method applied to the data, true photon conversions, and
true Dalitz decays of neutral mesons~\cite{atlasphotconv}.
Bottom: Data-to-MC ratio of the ATLAS measured $K_{s}^{0}$ mass as a function 
of the radial distance to the center of the detector, with dashed lines 
marking the boundaries of the sub-detector systems~\cite{atlaskshort}.}
\label{atlasgeom}
\end{figure}

As in the case of CMS, the ATLAS detector also includes a silicon-based 
inner detector for vertex and track reconstruction, which extends to a 
radius of 1.15~m, and is 7~m in length along the beam pipe. The development
and validation of simulation code to model the detector shape, thickness, and 
materials was therefore an activity of utmost importance.
Fig.~\ref{atlasgeom} (top) shows the distribution of photon conversion
vertices in the radial direction starting from the detector 
center~\cite{atlasphotconv}. 
Full circles represent the collider data measurement, while the solid
line shows the distribution of conversion candidates obtained using the
same analysis method applied to the data.
The histogram shows the true MC distribution for the conversions (blue) and
the Dalitz decays of neutral mesons (yellow).
The good agreement between MC and data, although based on limited statistics,
is a measure of the excellent modeling of the material distribution in the MC.
A second example of material modeling validation is shown in 
Fig.~\ref{atlasgeom} (bottom)~\cite{atlaskshort} and consists of
reconstructing particles with well known masses and 
lifetimes from detector tracks. Flaws in the material
modeling of the detector would result in incorrect compensation for 
effects of energy loss and multiple scattering on the tracks, 
resulting in biases to the reconstructed tracks momenta, 
which propagate to the reconstructed mass. In the case of the $K_{s}^{0}$,
which decays with a proper length of $c\tau \sim 2.7$~cm, it is possible
to study the detector material modeling accuracy as a function of the
radial position of the decay vertex.
The sample used in the study consists of a 
selection of oppositely charged track pairs with $p_T>100$~MeV. $K_{s}^{0}$ 
candidates were reconstructed with a fit to the pairs satisfying a selection 
criteria. Fig.~\ref{atlasgeom} (bottom) shows the data-to-MC ratio
of the measured $K_{s}^{0}$ mass as a function of the radial distance to the
center of the detector, with dashed lines marking the boundaries of the
sub-detector systems. Once again, the high level of agreement of the 
fitted $K_{s}^{0}$ in
MC and data is a measure of the excellent modeling of the ATLAS
silicon tracker material distribution and thickness in the 
simulation.

\subsection{Modeling of Particle and Event Properties and Kinematics} \label{physicsobjects}

This section includes a number of data-to-MC comparisons from ATLAS and CMS
focused on 
event or particle properties and kinematics. 
Examples are presented for photons, electrons, 
muons, and jets from light and heavy quarks. These particles are 
observed and reconstructed as physics objects in the detector, and 
constitute the basic ingredients of every measurement. Excellent understanding 
of their kinematic distributions, as 
well as their reconstruction and identification efficiencies is a first step 
for any experiment to deliver robust physics measurements of high quality and 
precision.

\subsubsection{Tagging of Heavy Quarks}

The ability to model b-jet reconstruction and identification is an
important simulation benchmark. In hadron colliders, the 
identification of jets originating from b quarks is critical for both SM 
measurements and BSM searches, given that top quarks decay into a b jet and a 
$W$ boson, and flavor is intimately tied with the Electroweak Symmetry 
Breaking mechanism (EWSB). Furthermore, SUSY and EWSB are related via the 
hierarchy problem. Thus b-jet identification is a key component of the 
event selection criteria developed for BSM searches, and accurate modeling of 
b-jets and b-tagging related variables are essential to understand data 
selection efficiencies and simulate the signal samples.

\begin{figure}[htbp]
\centering
\includegraphics[width=0.70\linewidth]{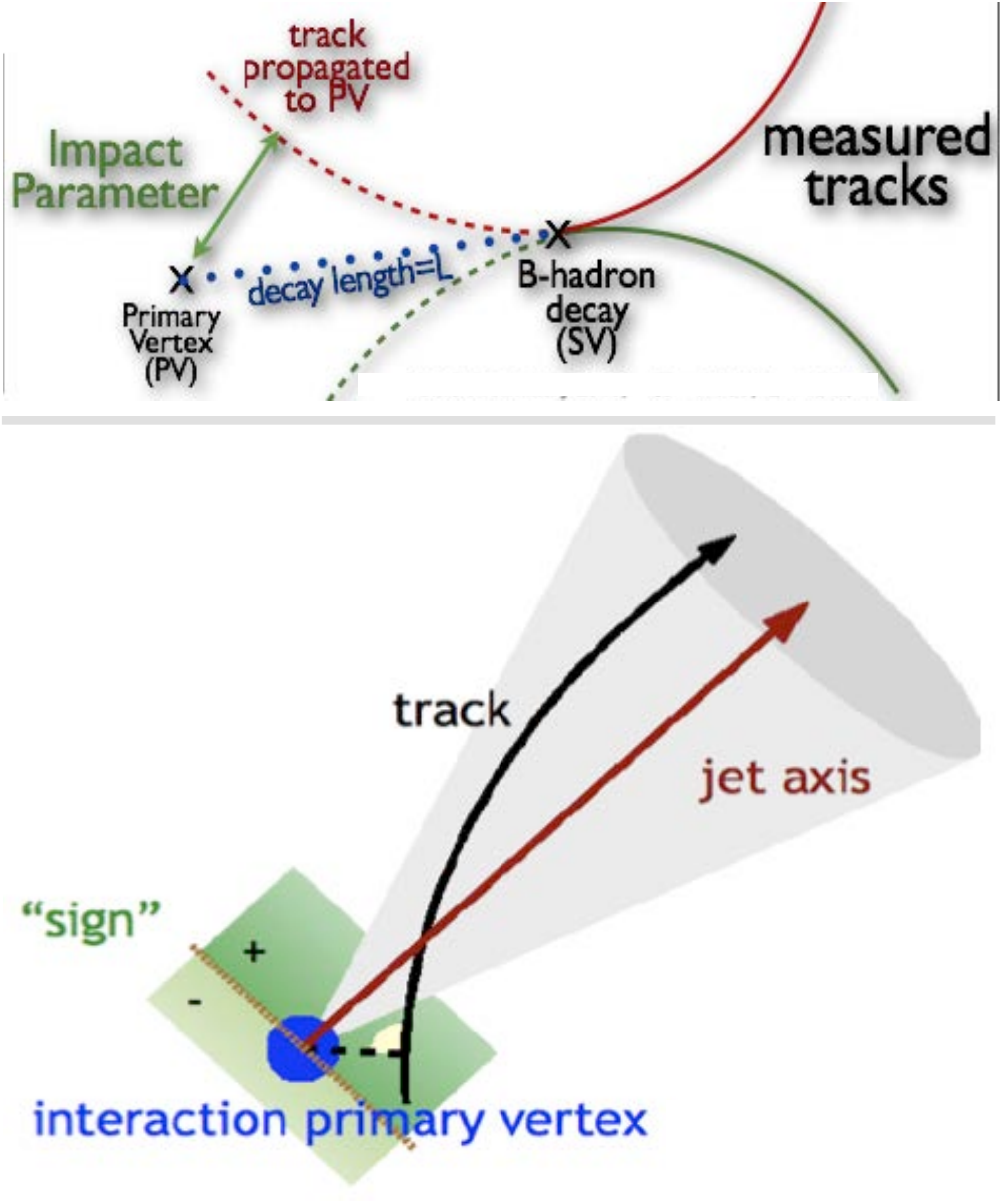}
\caption{Top: Illustration of a b jet, including tracks 
originating from the secondary vertex (SV) and bending outwards due
to the effect of the solenoidal magnetic field. The impact parameter is 
indicated as the distance between the primary vertex (PV) and the 
back-propagated tracks. Bottom: Sign convention for the impact parameter.}
\label{btaggsketch}
\end{figure}

The b-jet identification procedure, or b-tagging~\cite{btaggingperf}, depends
on variables and requirements such as the 
impact parameters of charged-particle tracks in a jet, 
the properties of reconstructed 
decay vertices in the jet, and the presence or absence of a lepton 
within a jet. The 3-Dimensional 
Impact Parameter (3D IP) 
is defined as the point of closest approach 
between a track and the event primary vertex (PV). The impact parameter has 
the same sign as the scalar product of the vector pointing from the primary 
vertex to the point of closest approach with the jet direction. 
In an ideal detector, tracks 
originating from the decay of long-lived particles such as b quarks traveling 
along the jet axis would have positive IP values, while the impact 
parameters of light-flavor quarks coming from the PV would be still be 
positive but close to zero. 
However, in a real detector, both negative and positive values are possible 
due to resolution effects. 
While distributions are significantly asymmetric
for b-quarks, they are almost symmetric for light quarks, with a deviation
towards a small positive mean value 
due to contributions of secondary vertices from
particles decaying within the light
jets, such as kaons and lambdas. Fig.~\ref{btaggsketch} (top) shows
two tracks originating from the secondary vertex (SV) and bending outwards due
to the effect of the solenoidal magnetic field. The impact parameter is 
indicated as the distance between the primary vertex (PV) and the 
back-propagated tracks. Fig.~\ref{btaggsketch} (bottom) illustrates the
sign convention for the impact parameter. 

\begin{figure}[htbp]
\centering
\includegraphics[width=0.9\linewidth]{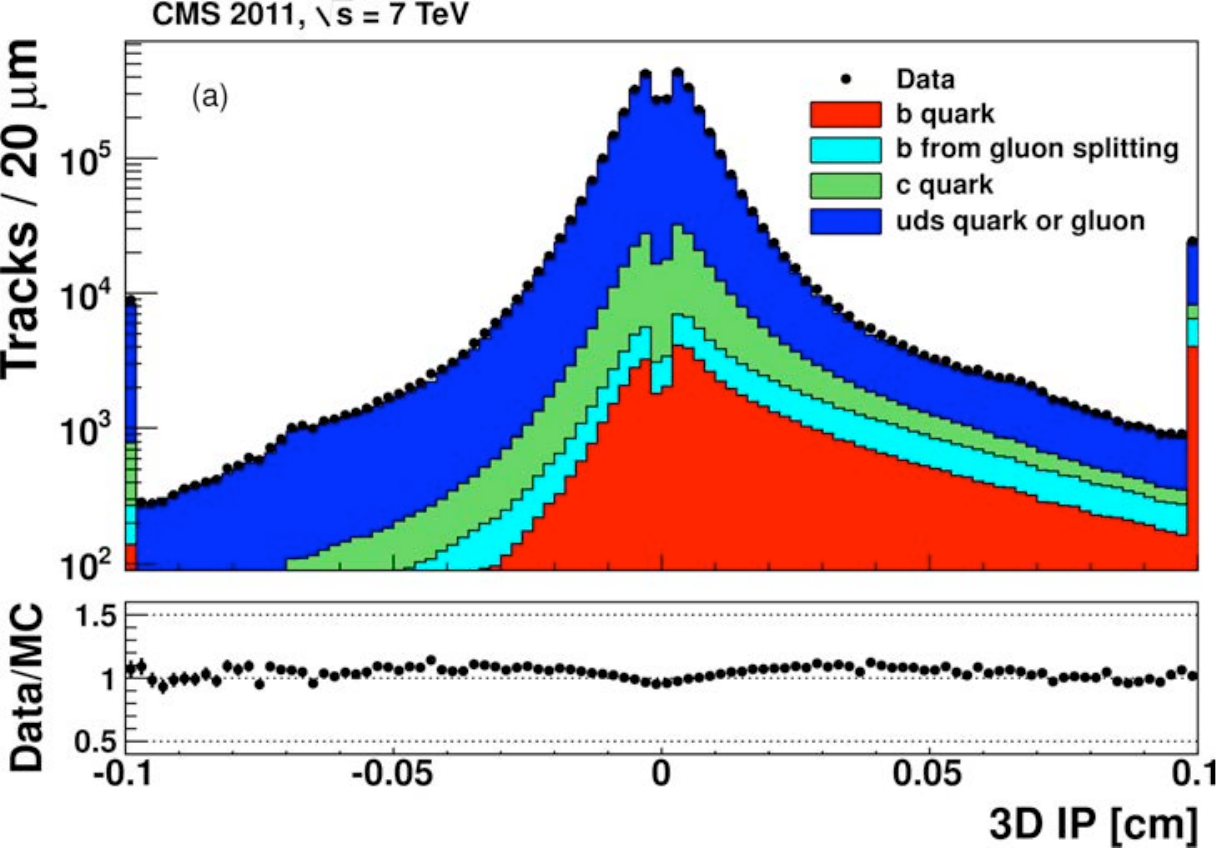}
\caption{MC and data 3D Impact Parameter distributions for tracks in jets 
selected in a CMS di-jet trigger sample~\cite{btaggingperf}. Data is
shown in full circles and the MC predictions in colored histograms.
Underflow and overflow are added to 
the first and last bins, respectively.}
\label{impactparameter}
\end{figure}

\begin{figure}[htbp]
\centering
\includegraphics[width=1.0\linewidth]{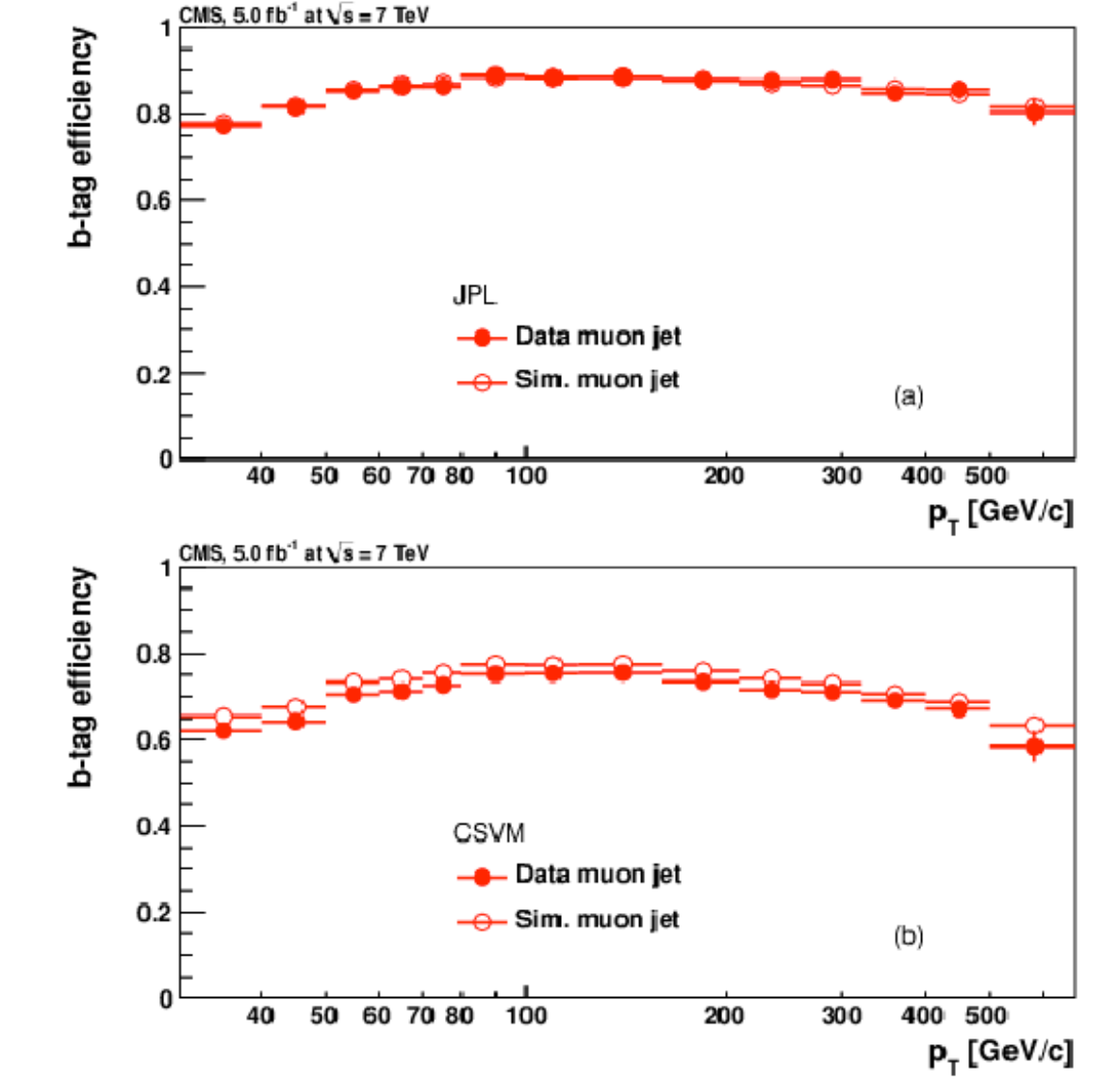}
\caption{Efficiencies for the identification of b-jets in CMS are measured for 
the JPL (top) and the CSVM (bottom) 
b-tagging algorithms using a sample of jets with
muons~\cite{btaggingperf}. 
Full and open circles correspond to data and simulation, 
respectively.}
\label{btaggingeff}
\end{figure}

The 3D IP distribution for tracks in jets 
selected in a CMS di-jet trigger sample is presented in 
Fig.~\ref{impactparameter}~\cite{btaggingperf}. 
Data is compared with MC 
predictions for all the parton flavors contributing to the inclusive 
di-jet sample. As expected, while the distributions for tracks in heavy-flavor 
jets are significantly asymmetric with a positive mean value, the distribution 
for light quarks and gluons is almost symmetric.
The excellent agreement between 3D IP distributions in data and MC, within
less than 10\%, is a 
precondition to the development of accurate data driven methods to 
measure b-tagging efficiencies. These efficiencies, 
shown in Fig.~\ref{btaggingeff}, are derived from a sample of jets with muons 
for two different b-tagging algorithms, known by their acronyms JPL and 
CSVM~\cite{btaggingperf}. The data-to-MC ratios
of b-tagging efficiencies obtained from these plots are 
used to adjust the MC truth predictions for use in physics measurements.

\begin{figure}[htbp]
\centering
\vspace{-1cm}
\includegraphics[width=0.8\linewidth]{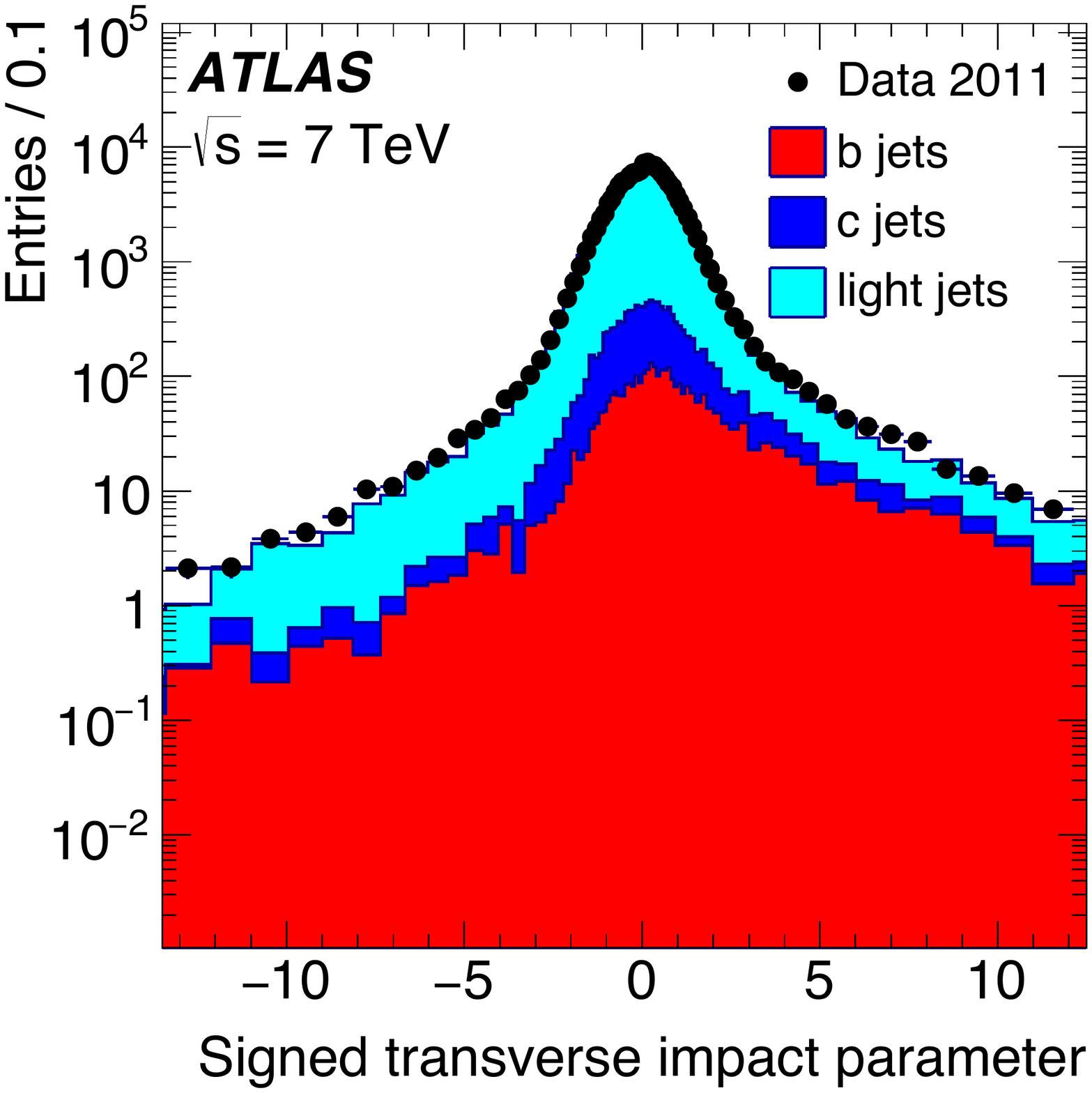}
\vspace{-2cm}
\caption{Signed transverse impact parameter
significance distribution measured in an ATLAS di-jet sample, compared to a MC
distribution~\cite{btaggingperfatlas}.}
\label{btaggingIPatlas}
\end{figure}

\begin{figure}[htbp]
\centering
\vspace{-3cm}
\includegraphics[width=0.9\linewidth]{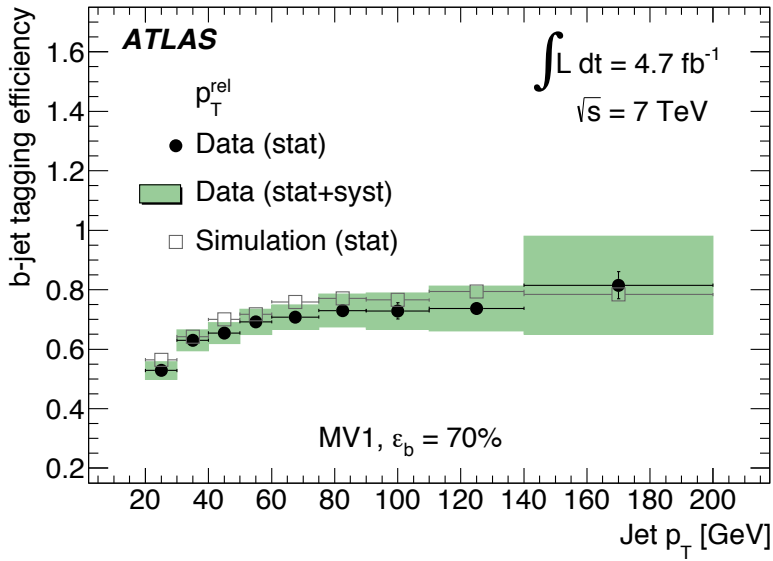}
\vspace{-3.5cm}
\caption{ATLAS's b-tagging efficiency versus jet $p_T$  for
a the MV1 neural network tagger. Data is shown in full circles and
MC in open squares.
Error bars are statistical and the band corresponds to the systematic
uncertainty in the method~\cite{btaggingperfatlas}.}
\vspace{0.0cm}
\label{btaggingeffatlas}
\end{figure}

In the context of b-tagging studies~\cite{btaggingperfatlas}, 
ATLAS defines the signed transverse impact parameter significance as
$S_{d_0} \equiv d_0/\sigma_{d_0}$, where $\sigma_{d_0}$ is the uncertainty on the 
reconstructed transverse impact parameter $d_0$, with $d_0$ the
$r-\phi$ projection of the distance of closest approach of the track to the PV.
Fig.~\ref{btaggingIPatlas} shows the signed transverse impact parameter
significance distribution measured in an ATLAS di-jet sample compared to a MC
distribution. The overall agreement is good except in the tails of the
distribution, which are more difficult to model. The b-tagging efficiency
as a function of the jet $p_T$ is shown in Fig~\ref{btaggingeffatlas} for
a neural network tagger known by its acronym MV1. As in the case of CMS, the
MC derivation of b-tagging efficiencies using data-driven methods is within
less than 5\% of the equivalent measurement in data, and the difference is
accounted for in physics measurements through scale factors computed as the
ratio of the values represented by the full circles over those in open squares.
Differences between CMS and ATLAS efficiencies when comparing 
Fig.~\ref{btaggingeff} with 
Fig.~\ref{btaggingeffatlas} are not relevant because taggers are 
typically tuned to
different efficiency operating points depending on the fake (or mis-tag) rate 
tolerance for a particular physics measurement. The mis-tag probability for
light-parton jets to be mis-identified as b jets is measured from data in
the ATLAS and CMS experiments using ``negative taggers''. These inverted
tagging algorithms select non-b jets using the same variables and
techniques as the b-tagging algorithms. 
An accurate determination of the mis-tag rate is important because, since 
the cross section for
light jets is much larger than for b jets, even a low rate of 
``false positives'' (mis-tagged jets) affects the b-jet sample purity
in a significant way. Simulating mis-tag rates is tricky because the
contributing jets originate in the tails of the IP distributions, which
are not trivial to model. For a mis-tag rate tolerance in the 0.01-0.03 range, 
CMS reports a $p_T$ dependence of the
data-to-MC mis-tag rate scale factors of about 20\%~\cite{btaggingperf}, 
while ATLAS reports factors of
2-3 for a tolerance in the 0.002-0.005 range~\cite{btaggingperfatlas}.

\subsubsection{$W$, $Z$ and Photon Event Distributions}

Gauge bosons such as the $W$, the $Z$ and the photon, 
are at the core of SM measurements and contribute 
backgrounds to most BSM searches. Event topologies and kinematic distributions
for $W/Z/\gamma$+jets events must therefore be modeled with high accuracy.
Although physics generators are the limiting factor in the case of events
with heavy flavor and many jets, the focus
in this section will be on the detector modeling, clarifying when 
generators play a significant role.

$t \overline{t}$+jets, 
$W$+jets, $Z$+jets backgrounds 
contribute at different levels to SUSY searches with jets, leptons, or photons 
in the final state. Although simulation is typically not used as the main tool 
to predict these backgrounds, MC samples are used to design and develop 
data-driven methods for background estimation and to perform the associated 
closure tests. For instance, kinematic distributions of final state particles 
measured in data, inspire physically motivated families of functional forms 
which 
also fit well the simulated spectra in both the control and signal regions
and are ultimately used to predict the backgrounds in SR's from 
extrapolations of fits to data in CR's. This is a common practice in many
cases where the electroweak (EWK) processes are known to be accurately modeled 
by physics generators. The use of MC samples to assist on the derivation of 
EWK backgrounds for final states with high jet 
multiplicity and heavy-flavor jets is more challenging, because of limitations 
of the physics generators rather than those of detector modeling. In other
words, the standard machinery of the Pythia~\cite{pythia1,pythia2} event 
generator is based on leading-level matrix elements combined with parton 
showers. From matrix elements calculations, the MadGraph~\cite{madgraph} 
event generator produces events based on processes modeled to LO accuracy 
for any user-defined Lagrangian, and to the NLO accuracy for QCD corrections 
to SM processes. Matrix elements at the tree-level and one-loop-level can also 
be generated. 
Consequently, predictions for final states with high particle multiplicity 
and heavy-flavors are either inaccurate, or computationally expensive once
loop-level calculations are included.
Exceptionally, in the
case of rare SM processes that contribute sub-dominant backgrounds, such as
$t \overline{t}V$, $t \overline{t}H$, $VH$ in same-sign leptonic BSM searches, 
backgrounds are predicted directly from 
MC truth information. The cost of this approach is a large uncertainty on a 
small fraction of the total background, which ultimately does not affect
the sensitivity of the analysis. 

\begin{figure}[htbp]
  \centering
  \begin{minipage}{.45\textwidth}
    \centering
    \includegraphics[width=\textwidth]{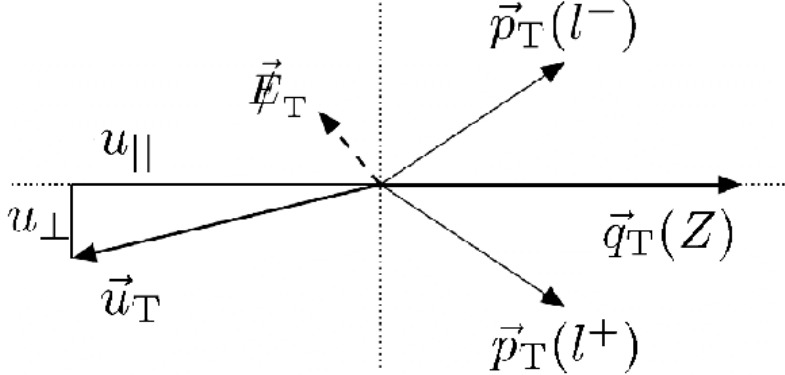}
  \end{minipage}
  \begin{minipage}{.45\textwidth}
    \centering
    \includegraphics[width=\textwidth]{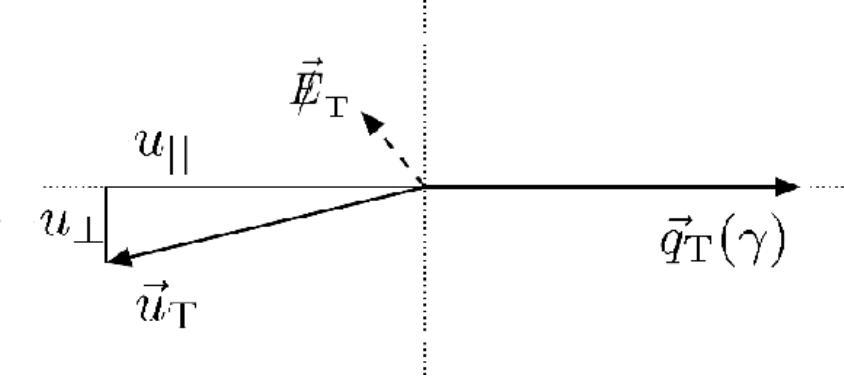}
  \end{minipage}
\caption{Kinematics of $Z$+jets and 
$\gamma$+jets events in the transverse plane. The events consist 
of either a $Z$ decaying to leptons or a $\gamma$ recoiling against jets that 
balance the transverse momentum, $\vec{q}_T$, of the gauge boson. 
The total transverse momentum of the hadronic recoil is indicated by the 
vector $\vec{u}_T$, while $u_{\perp}$ and $u_{\parallel}$ are the components 
perpendicular and parallel to the gauge boson. The \met vector is a measure
of the $p_T$ imbalance in the event.}
\label{zgammakine}
\end{figure}

Fig.~\ref{zgammakine} describes the kinematics in the transverse plane 
of $Z$+jets and $\gamma$+jets collider events, which are used to 
illustrate the data-to-MC agreement of quantities involving gauge boson
production. 
These events consist of either a $Z$ boson decaying to leptons or a $\gamma$ 
recoiling against jets that balance the transverse momentum, 
$\vec{q}_T$, of the gauge boson. The total transverse momentum of the 
hadronic recoil is indicated in Fig.~\ref{zgammakine} by the vector
$\vec{u}_T$, while $u_{\perp}$ and $u_{\parallel}$ are the components 
perpendicular
and parallel to the gauge boson. The \met vector is a measure of the $p_T$
imbalance in the event and will be discussed in detail in 
Sec.~\ref{misstransmom}. 

\begin{figure}[htbp]
  \centering
  \begin{minipage}{.70\textwidth}
    \centering
    \includegraphics[width=\textwidth]{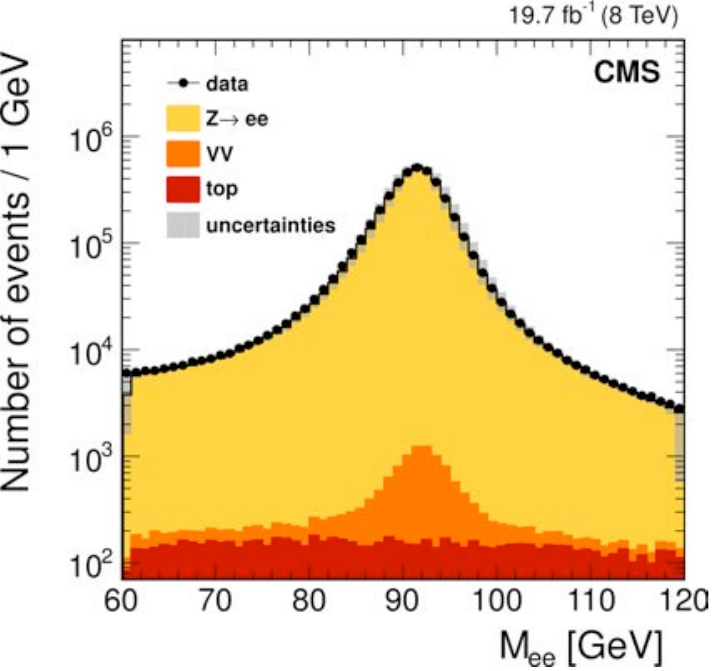}
  \end{minipage}
  \begin{minipage}{.75\textwidth}
    \centering
    \includegraphics[width=\textwidth]{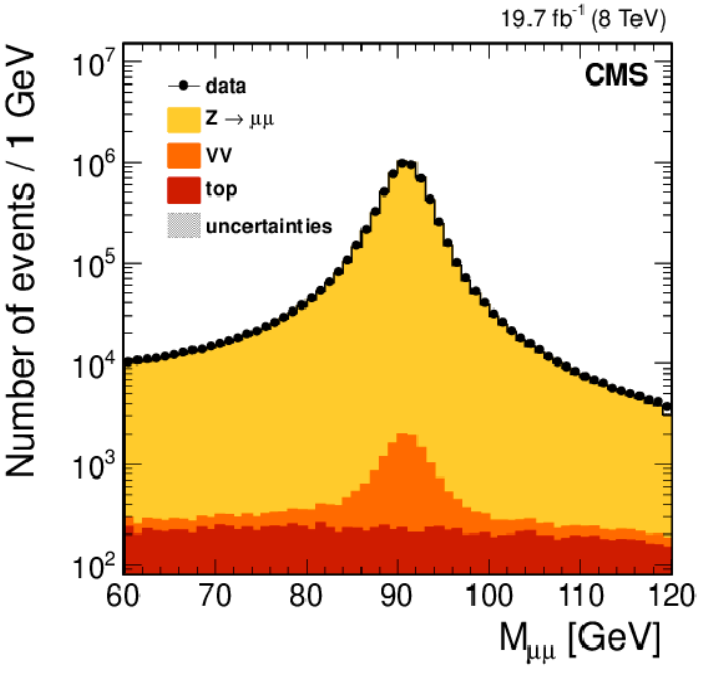}
  \end{minipage}
\caption{Di-lepton invariant mass distributions in CMS events passing the 
$Z \rightarrow \mu^{+}\mu^{-}$ and $Z \rightarrow e^{+}e^{-}$ 
selections~\cite{metperf}. The
full circles indicate the data measurement while the histograms are
the MC predictions.
The $VV$ contribution corresponds to processes with two electroweak bosons in 
the final state.}
\label{zmass}
\end{figure}

\begin{figure}[htbp]
  \centering
  \begin{minipage}{.65\textwidth}
    \centering
    \includegraphics[width=\textwidth]{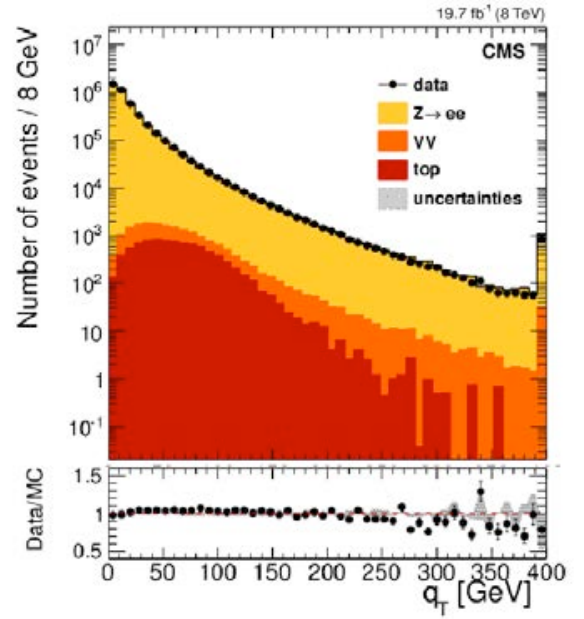}
  \end{minipage}
  \begin{minipage}{.6\textwidth}
    \centering
    \includegraphics[width=\textwidth]{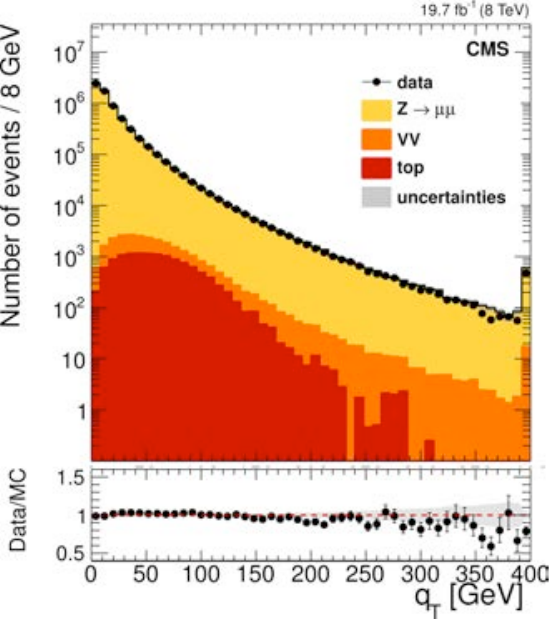}
  \end{minipage}
\caption{$Z$ transverse momentum, $q_T$, distributions in CMS
$Z \rightarrow \mu^{+}\mu^{-}$ and $Z \rightarrow e^{+}e^{-}$ 
events~\cite{metperf}. 
The full circles indicate the data 
measurement while the histograms are the MC predictions.
The lower panels show the data-to-MC ratio, 
including the statistical uncertainties in both data and simulation, and the 
gray error band displays the systematic uncertainty of the measurement.}
\label{zmomentum1}
\end{figure}

\begin{figure}[htbp]
\centering
\includegraphics[width=0.75\linewidth]{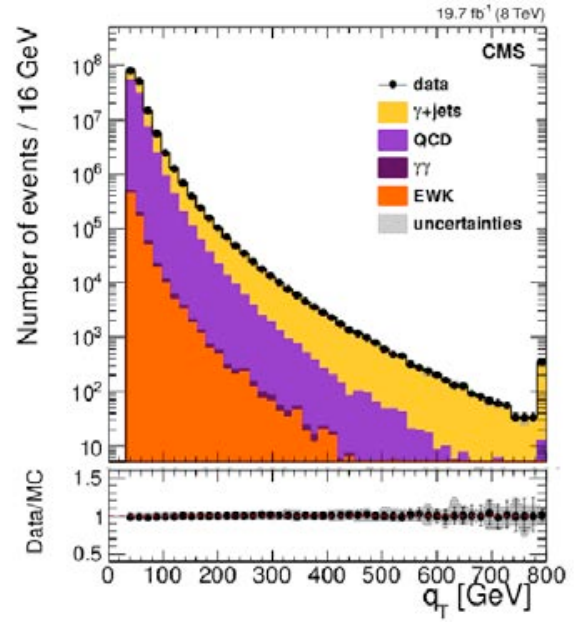}
\caption{$\gamma$ transverse momentum, $q_T$, distributions in CMS
direct-photon events~\cite{metperf}. The full circles indicate the data 
measurement while the histograms are the MC predictions.
The lower panels show the data-to-MC ratio, 
including the statistical uncertainties in both data and simulation, and the 
gray error band displays the systematic uncertainty of the measurement.}
\label{zmomentum2}
\end{figure}

Figs.~\ref{zmass},~\ref{zmomentum1},~\ref{zmomentum2} show the CMS di-lepton 
($\mu^{+}\mu^{-}$ and $e^{+}e^{-}$) invariant mass distributions for the 
$Z$+jets samples, and the $\vec{q}$ distributions for the $Z$+jets and 
$\gamma$+jets samples~\cite{metperf}. 
The full circles correspond to the number of events in the data samples, 
while the histograms represent the MC predictions for the number of events 
contributed by all the physics processes with final states passing the 
selection criteria. The 
agreement between the data measurements and the MC predictions is 
excellent, except in the range of
$q_T>200$~GeV for the $Z$+jets sample, where the MC overestimates the data
by a difference that grows linearly with $q_T$.

\begin{figure}[htbp]
  \centering
\vspace{-3cm}
  \begin{minipage}{.75\textwidth}
    \centering
    \includegraphics[width=\textwidth]{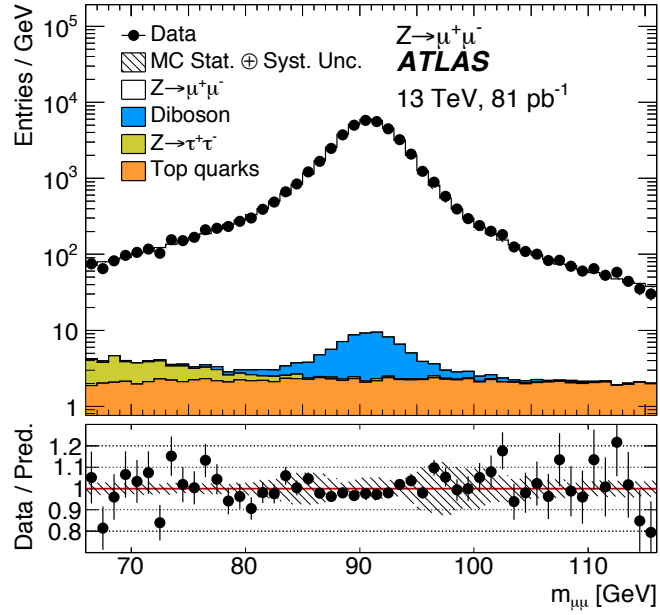}
\vspace{-3cm}
  \end{minipage}
  \begin{minipage}{.75\textwidth}
    \centering
    \includegraphics[width=\textwidth]{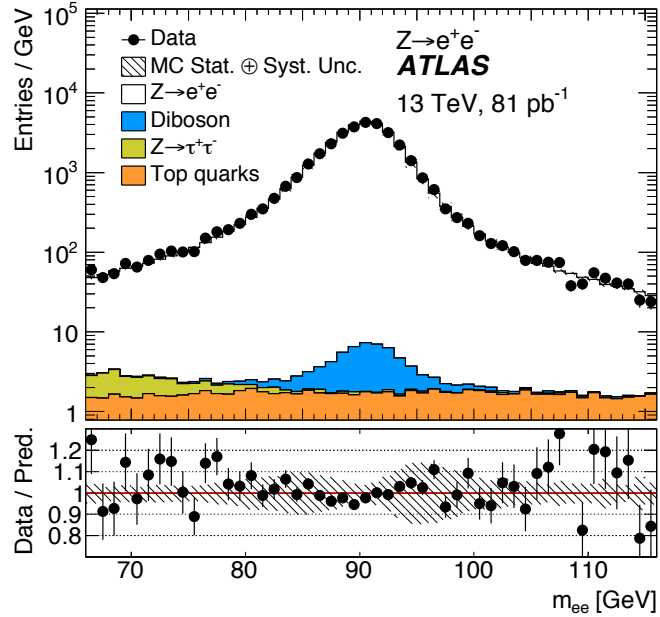}
\vspace{-2cm}
  \end{minipage}
\caption{Di-lepton invariant mass 
distributions in an ATLAS $Z$+jets sample for the  
$\mu^{+}\mu^{-}$ and $e^{+}e^{-}$ decay channels~\cite{wzprodatlas}.
The full circles correspond to the data, while the histograms 
represent the MC predictions for all the physics processes with final states 
and kinematics passing the selection criteria. Error bars are statistical
and the shaded band shows the systematic uncertainty in the data 
measurement.}
\label{zmassatlas}
\end{figure}

\begin{figure}[htbp]
  \centering
\vspace{-3cm}
  \begin{minipage}{.75\textwidth}
    \centering
    \includegraphics[width=\textwidth]{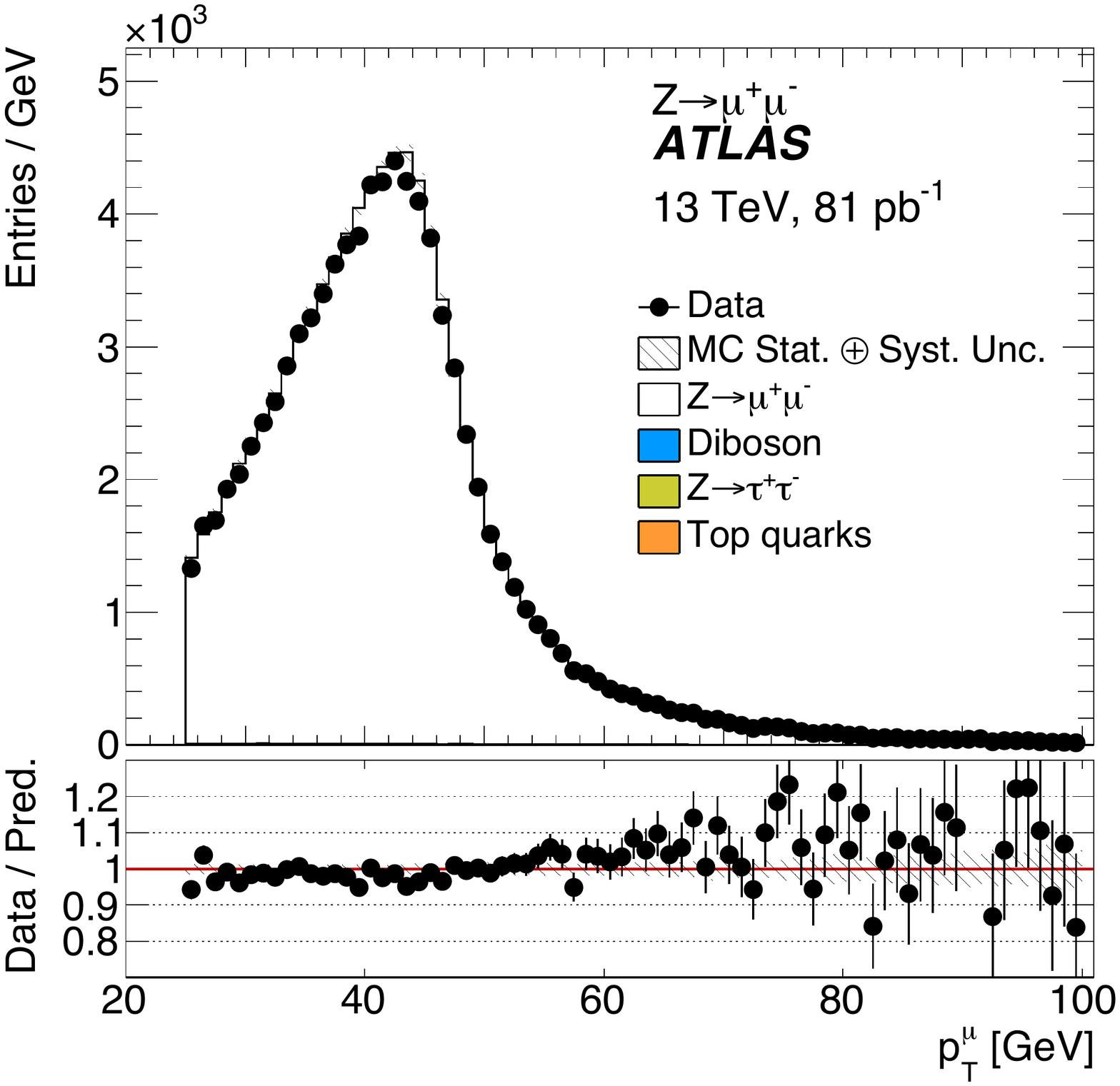}
\vspace{-3cm}
  \end{minipage}
  \begin{minipage}{.75\textwidth}
   \centering
    \includegraphics[width=\textwidth]{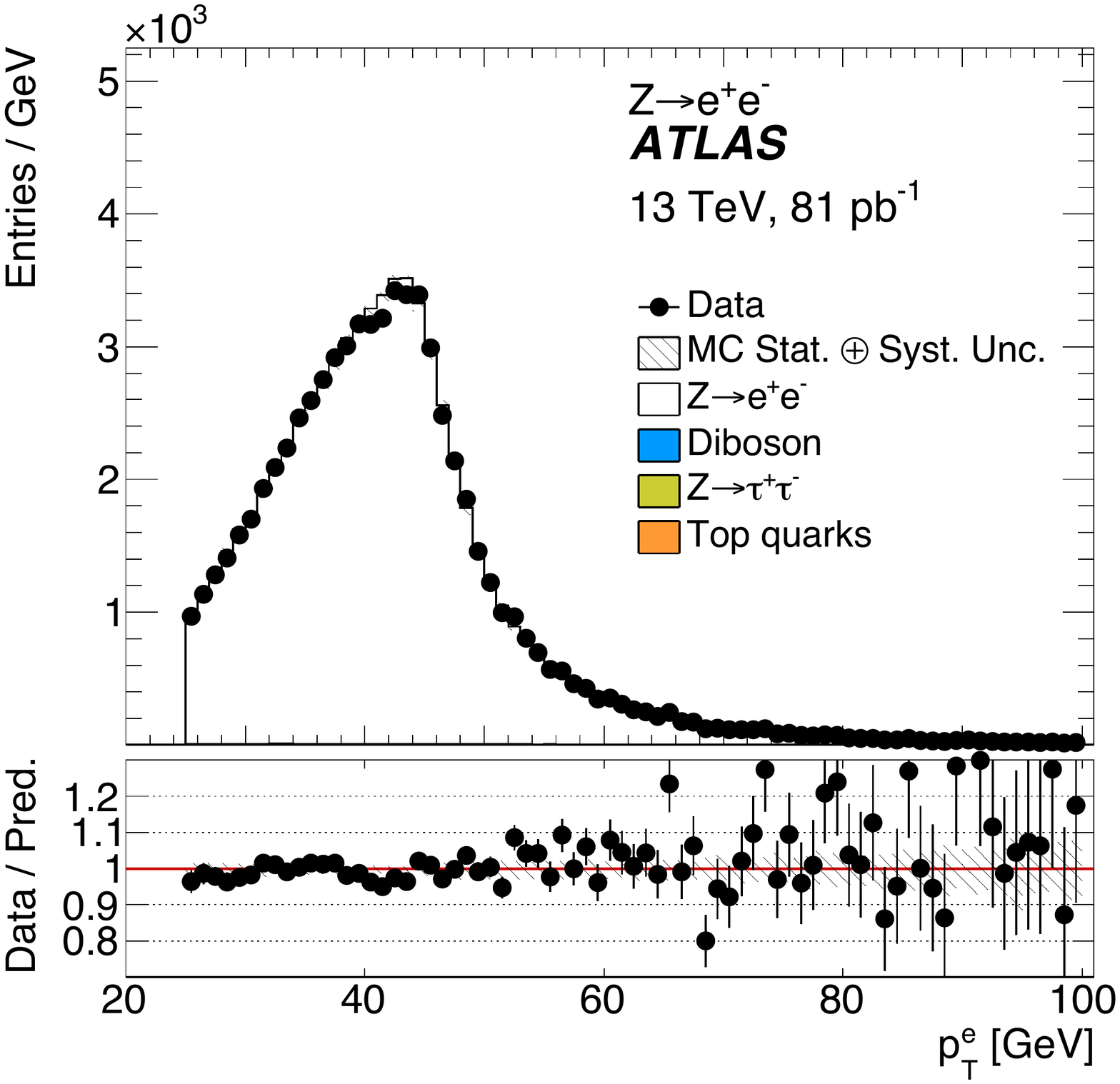}
\vspace{-2cm}
  \end{minipage}
\caption{Muon and electron transverse momentum distributions in 
an ATLAS $Z$+jets 
sample for the $\mu^{+}\mu^{-}$ and $e^{+}e^{-}$ decay channels~\cite{wzprodatlas}.
The full circles correspond to the data, while the histograms 
represent the MC predictions for all the physics processes with final states 
and kinematics passing the selection criteria. Error bars are statistical
and the shaded band shows the systematic uncertainty in the data 
measurement.}
\label{zmomentumatlas}
\end{figure}

\begin{figure}[htbp]
  \centering
\vspace{-3cm}  
\begin{minipage}{.70\textwidth}
    \centering
    \includegraphics[width=\textwidth]{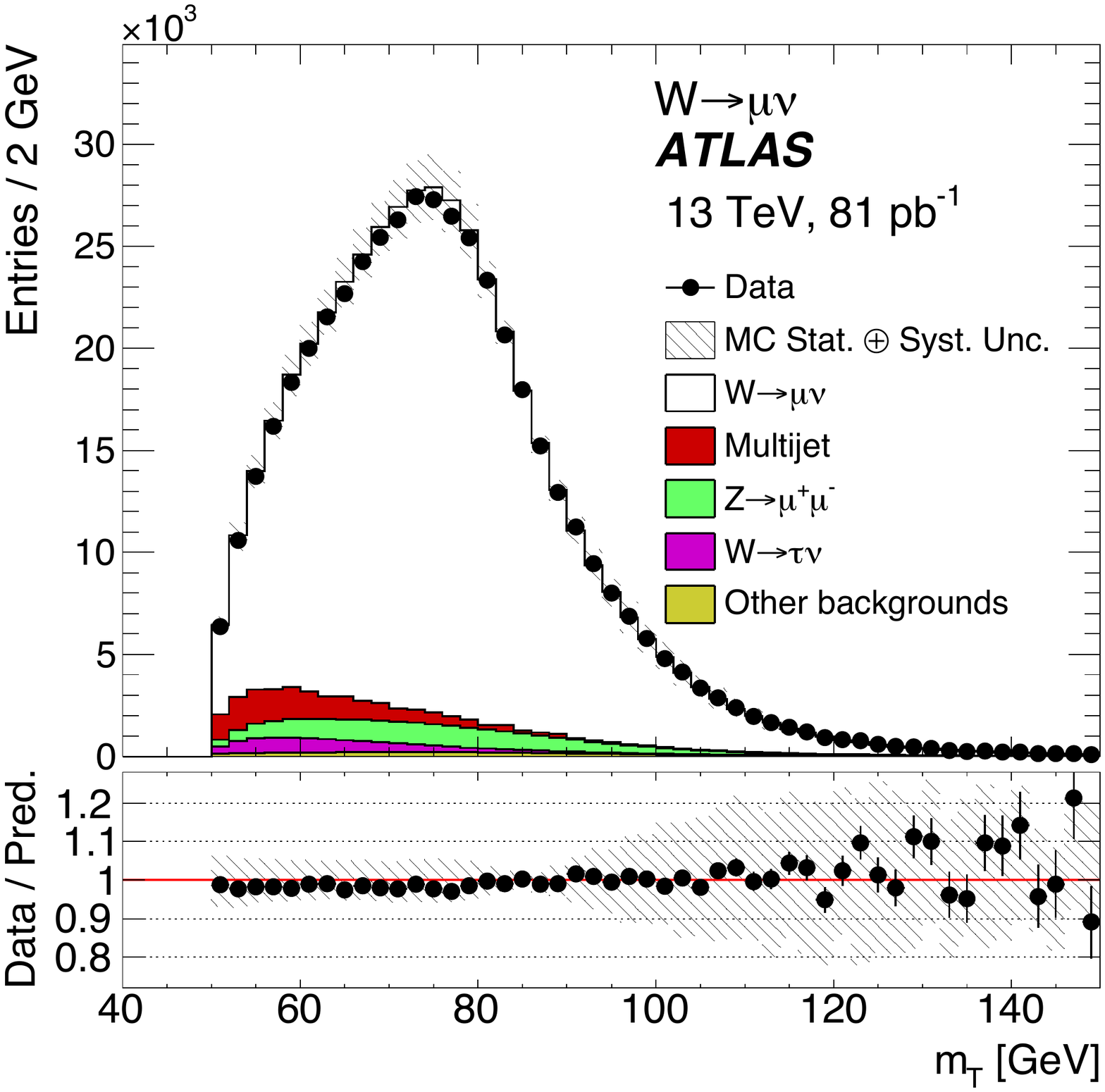}
\vspace{-3cm}
  \end{minipage}
  \begin{minipage}{.75\textwidth}
    \centering
    \includegraphics[width=\textwidth]{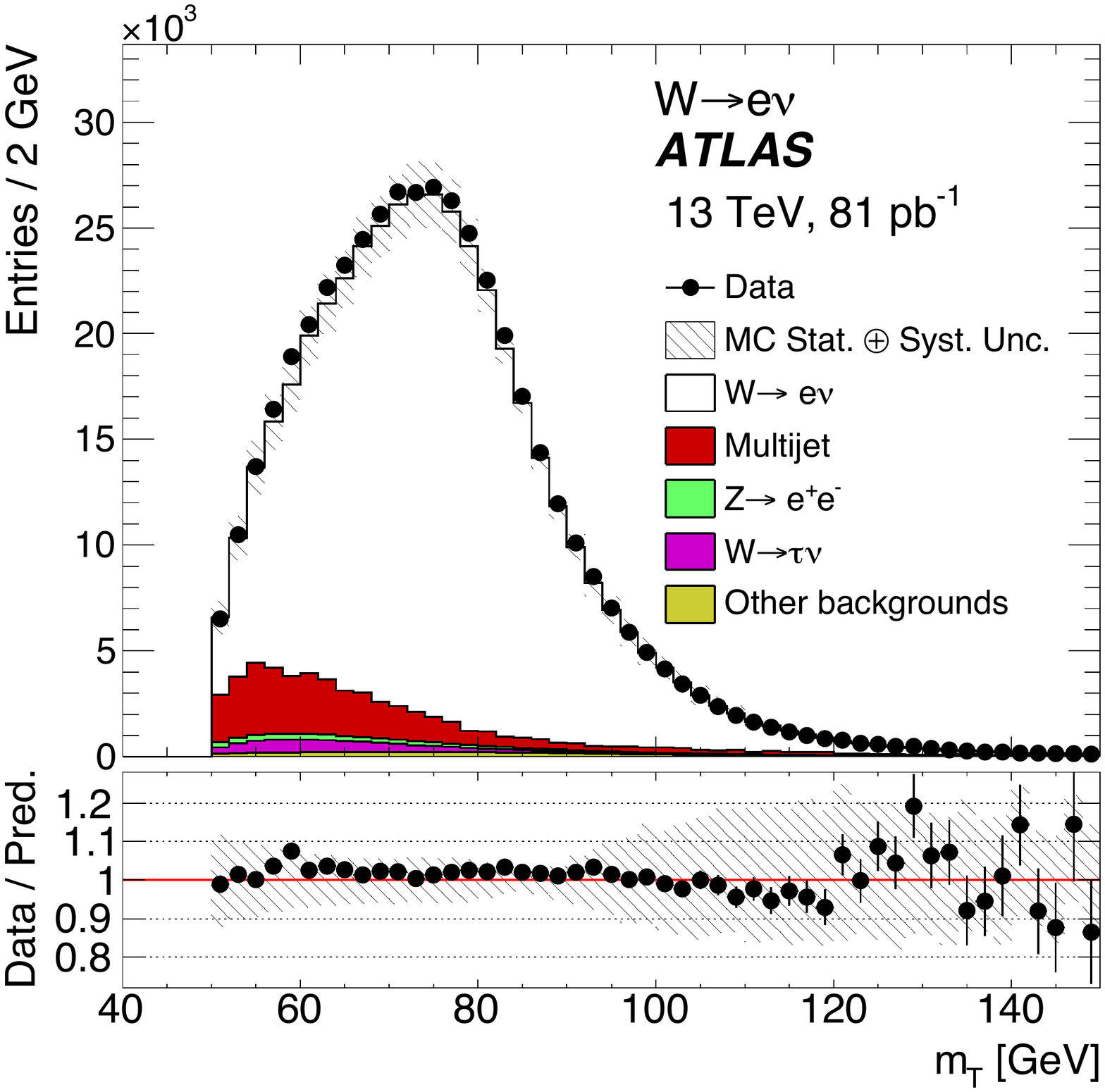}
\vspace{-2cm}
  \end{minipage}
\caption{Transverse mass distribution, $M_T=\sqrt{2p_T^lp_T^{\nu}[1-{\rm cos}(\phi_l-\phi_{\nu})]}$, in an ATLAS $W$+jets sample for the  
$\mu\nu$ and $e\nu$ decay channels~\cite{wzprodatlas}.
The full circles correspond to the data, while the histograms 
represent the MC predictions for all the physics processes with final states 
and kinematics passing the selection criteria. Error bars are statistical
and the shaded band shows the systematic uncertainty in the data 
measurement.}
\label{wmassatlas}
\end{figure}

\begin{figure}[htbp]
  \centering
\vspace{-3cm}
  \begin{minipage}{.75\textwidth}
    \centering
    \includegraphics[width=\textwidth]{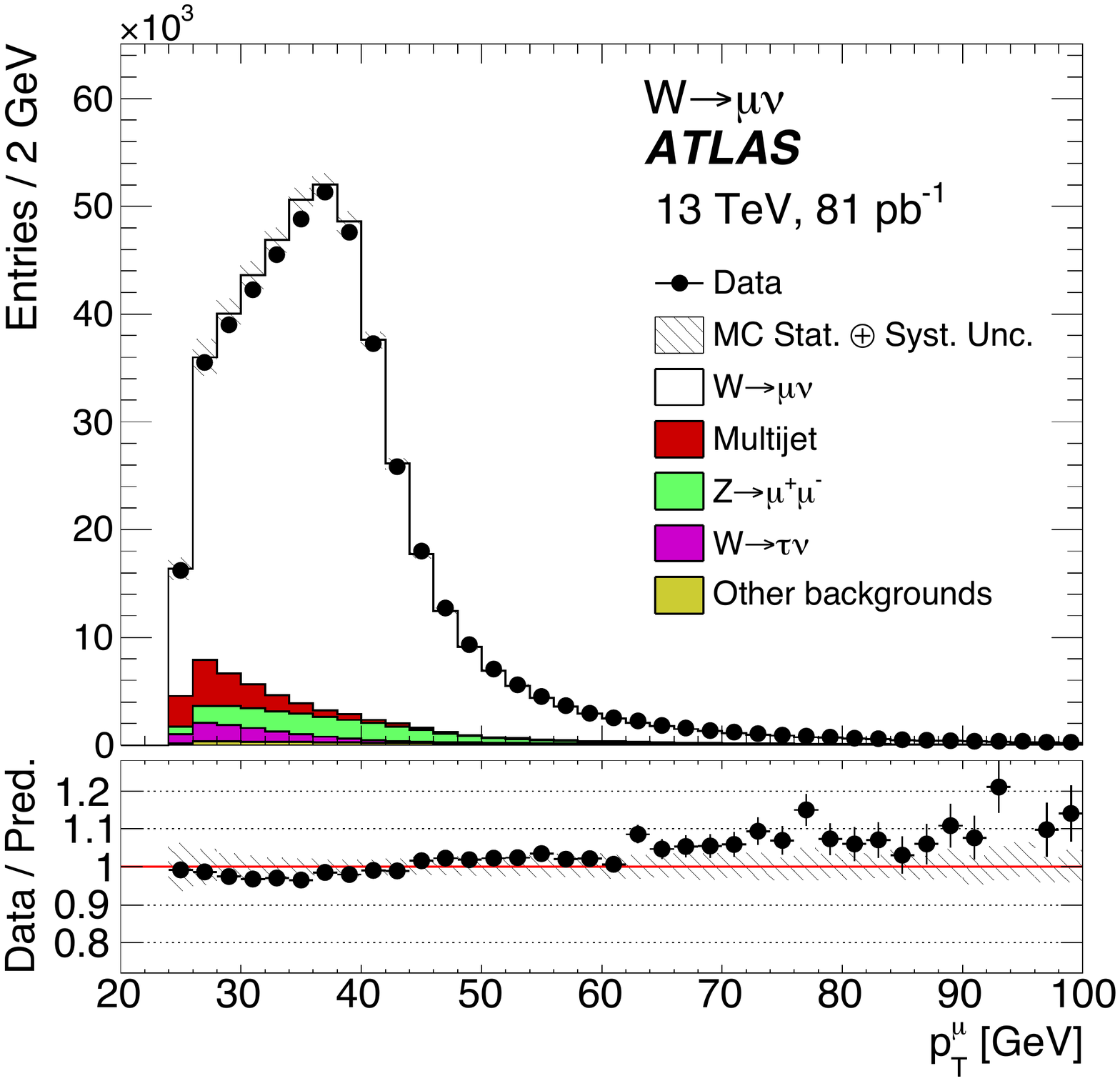}
\vspace{-3cm}
  \end{minipage}
  \begin{minipage}{.75\textwidth}
    \centering
    \includegraphics[width=\textwidth]{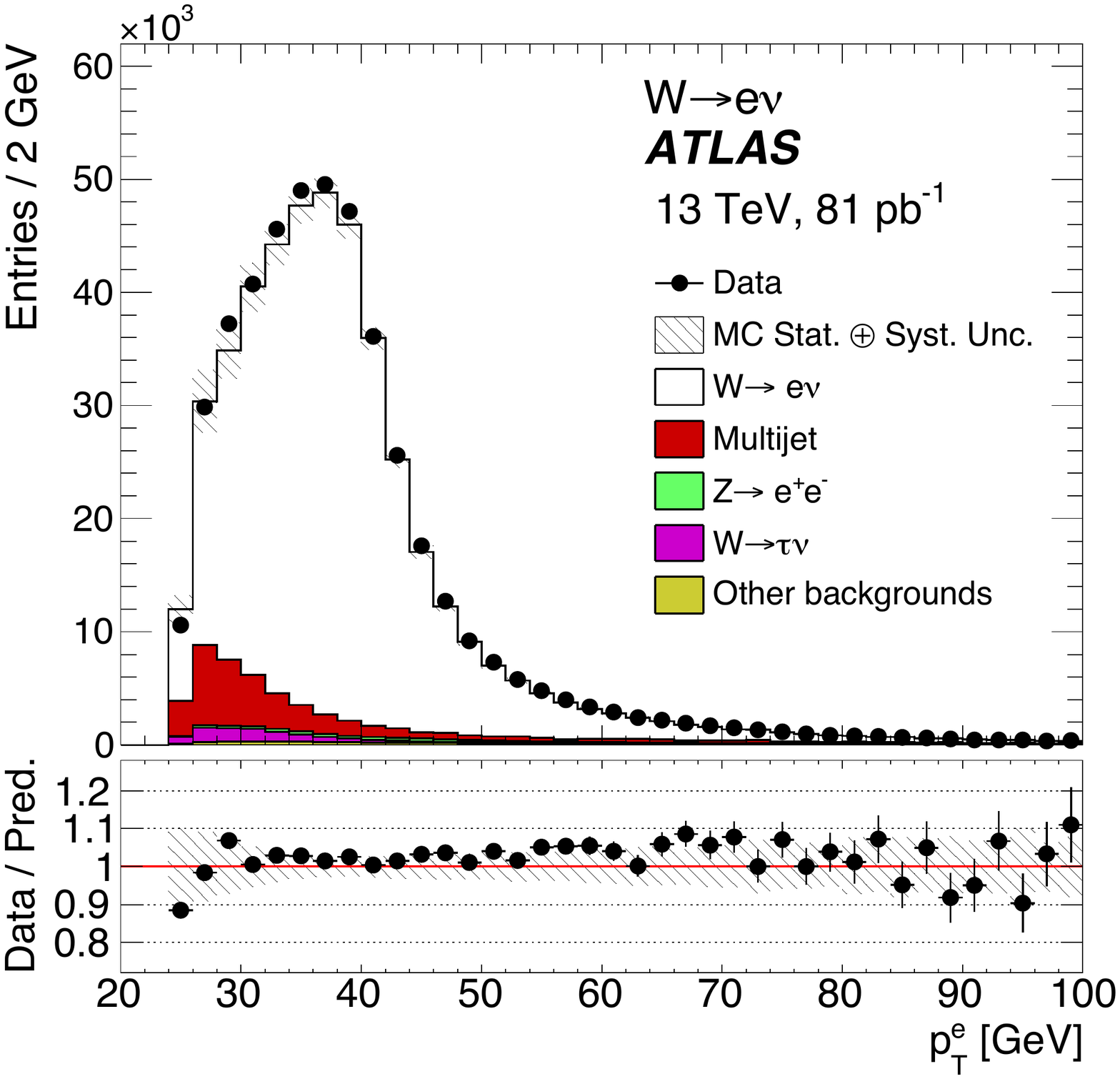}
\vspace{-2cm}
  \end{minipage}
\caption{Transverse momentum distributions in an ATLAS sample of 
$W$+jets events for the $\mu\nu$ and $e\nu$ decay channels~\cite{wzprodatlas}.
The full circles correspond to the data, while the histograms 
represent the MC predictions for all the physics processes with final states 
and kinematics passing the selection criteria. Error bars are statistical
and the shaded band shows the systematic uncertainty in the data 
measurement.}
\label{wmomentumatlas}
\end{figure}

For the ATLAS experiment, Figs.~\ref{zmassatlas},~\ref{zmomentumatlas},~\ref{wmassatlas},~\ref{wmomentumatlas} depict the di-lepton invariant mass 
distributions in the $Z$+jets sample and the 
$M_T=\sqrt{2p_T^lp_T^{\nu}[1-{\rm cos}(\phi_l-\phi_{\nu})]}$ 
distribution for the $W$+jets sample for the 
$Z \rightarrow \mu^{+}\mu^{-}/e^{+}e^{-}$ and 
$W \rightarrow \mu\nu/e\nu$ decay channels~\cite{wzprodatlas}.
The full circles correspond to the data, while the histograms 
represent the MC predictions for all the physics processes with final states 
and kinematics passing the selection criteria. The MC prediction
agrees with the data within the systematic uncertainties in all cases,
an impressive result given that the uncertainties are $<10\%$ for most 
distributions in the domain ranges with good statistics.

\subsubsection{Missing Transverse Energy Distributions} \label{misstransmom}

The missing transverse energy, denoted \met\ or $E^{\rm miss}_T$, 
is defined as the negative 
vector sum of the transverse components of the energy of all particles in 
the event. Although this term is physically incorrect because
energy is a scalar quantity, it is widely used in high-energy particle
physics because calorimeters measure energy (not momentum) and, in most
events of interest in current collider experiments, particle masses are 
negligible with respect to their total
energy. In this limit of negligible mass, energy equals 
momentum and missing transverse momentum may be approximated by 
\met\ as defined above. 
Modeling \met\ is one of the most challenging simulation tasks
because this event level quantity depends on accurate simulation of all 
types of particles, 
including hadronic showers from jets, as well as unclustered energy not 
assigned to any particle in the event.
Challenging as it is,
accurate modeling of \met\ in simulation is of paramount importance to 
the quality of BSM searches for SUSY and Extra Dimensions (ED), 
as well as in collider-based searches for dark matter. Simulation of \met\ 
also played a crucial role in the discovery and 
characterization of the Higgs boson, particularly in the 
$H \rightarrow \tau\tau$ final state, and channels with a 
$W\rightarrow l\nu$ or a $Z \rightarrow \nu\nu$. 

\begin{figure}[htbp]
  \centering
\vspace{-2cm}
  \begin{minipage}{.7\textwidth}
    \centering
    \includegraphics[width=\textwidth]{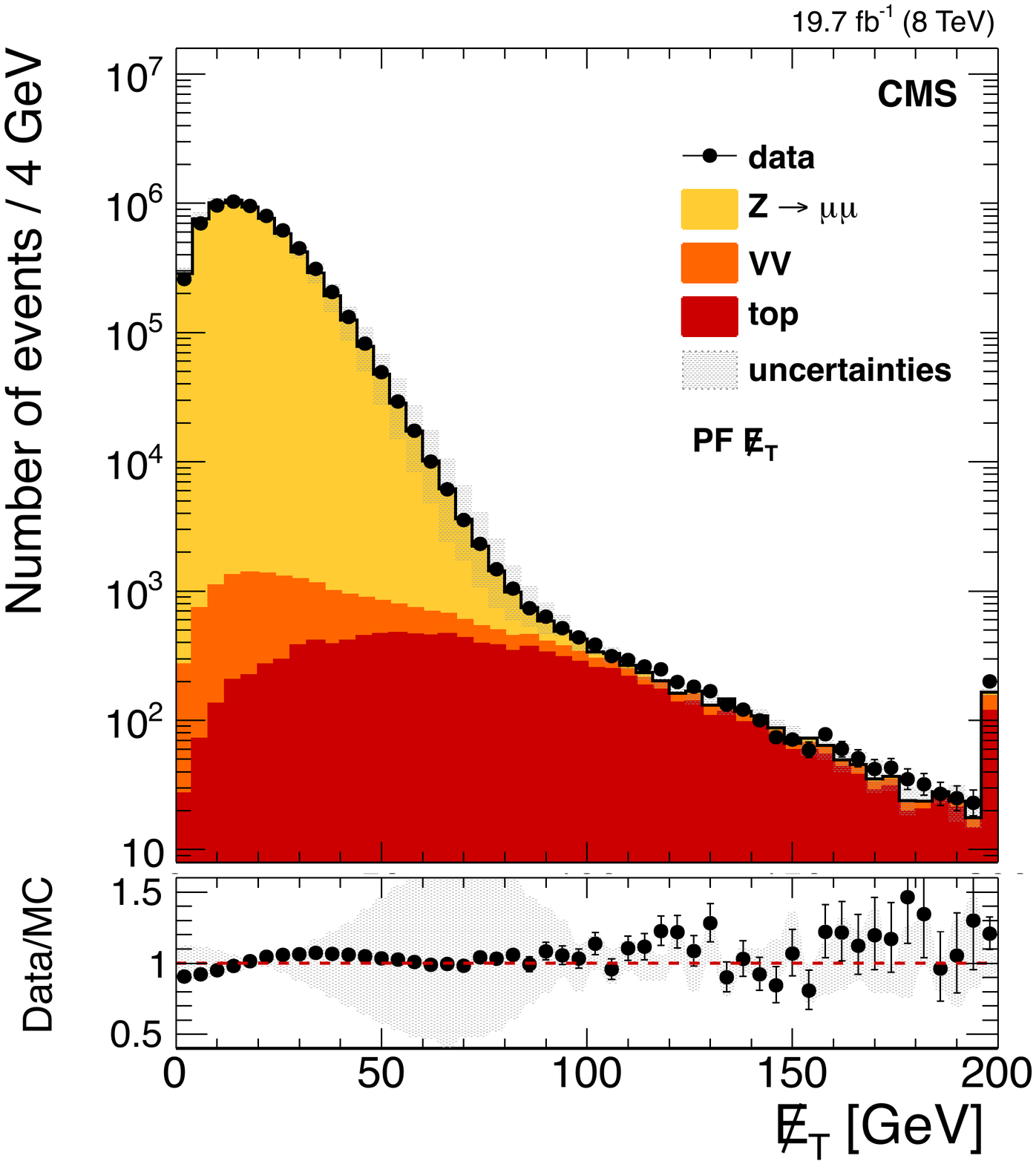}
\vspace{-2.5cm}
  \end{minipage}
  \begin{minipage}{.7\textwidth}
    \centering
    \includegraphics[width=\textwidth]{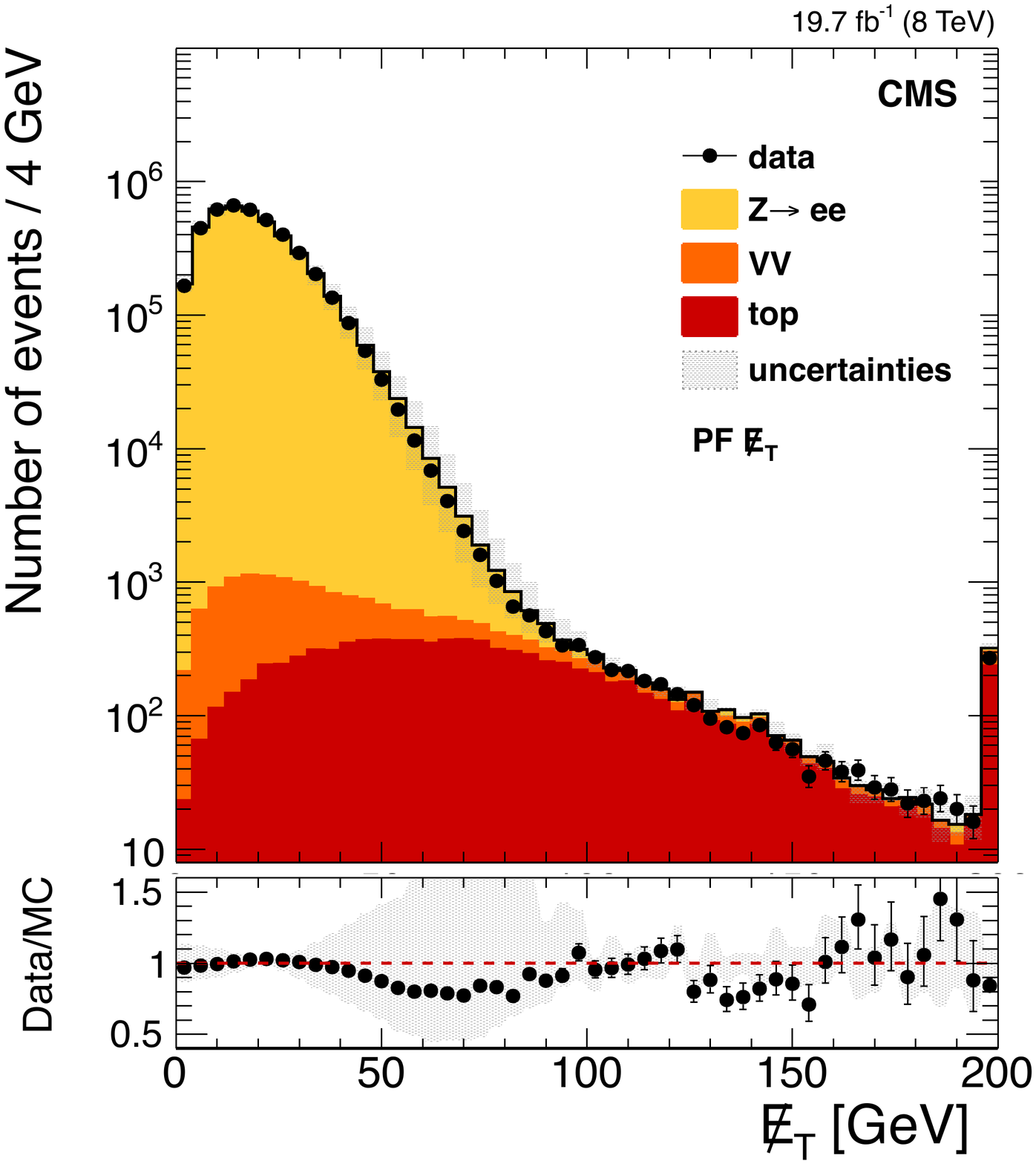}
\vspace{-1.5cm}
  \end{minipage}
\caption{Missing transverse energy distribution in a CMS $Z$+jets 
sample for the  
$\mu\mu$ and $e^{+}e^{-}$ decay channels~\cite{metperf}.
The full circles correspond to the data, while the histograms 
represent the MC predictions for all the physics processes with final states 
and kinematics passing the selection criteria. Error bars are statistical
and the shaded band shows the systematic uncertainty in the data 
measurement.}
\label{zmetcms}
\end{figure}

\begin{figure}[htbp]
  \centering
\vspace{-3cm}
  \begin{minipage}{.75\textwidth}
    \centering
    \includegraphics[width=\textwidth]{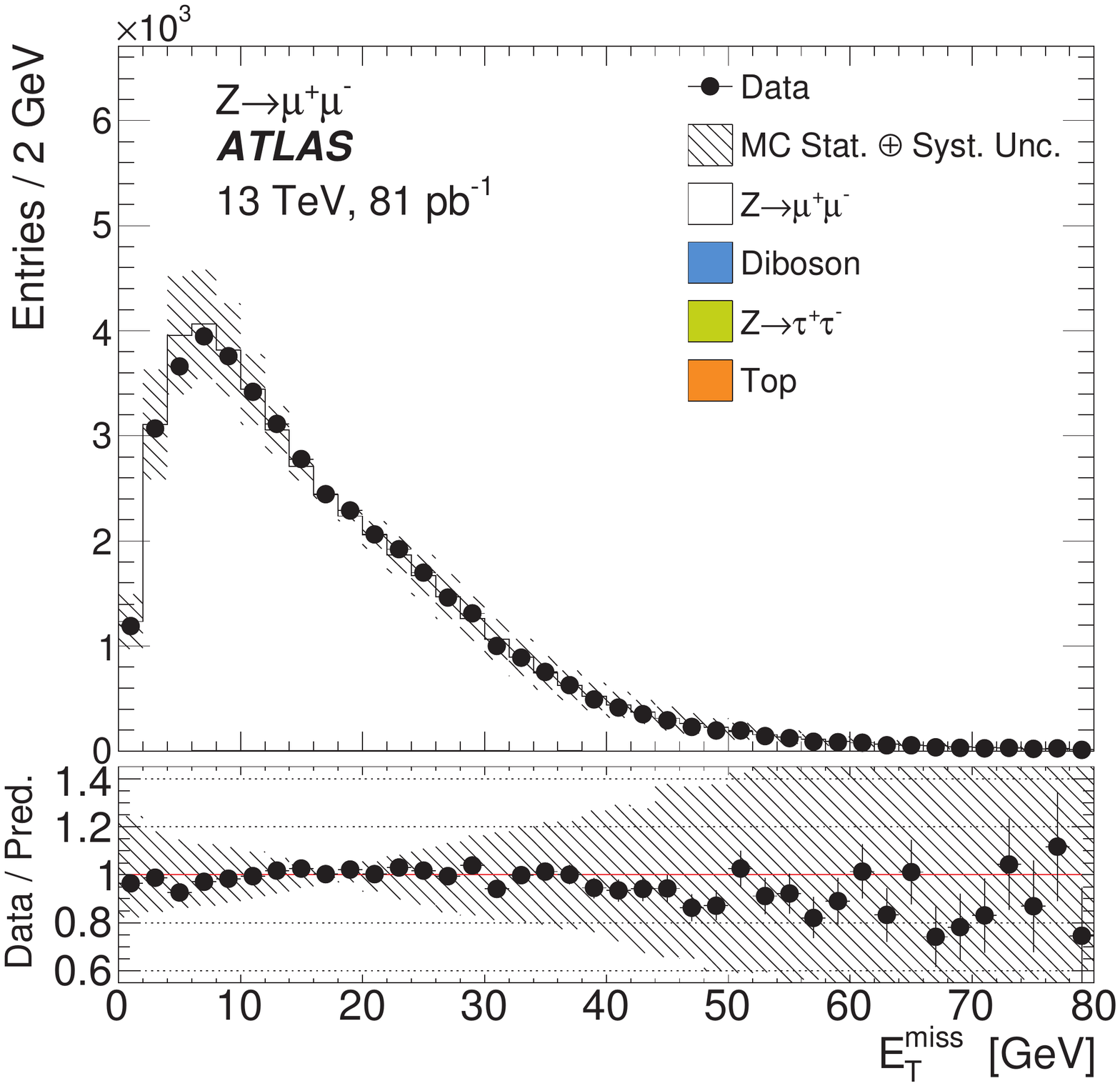}
\vspace{-3cm}
  \end{minipage}
  \begin{minipage}{.75\textwidth}
    \centering
    \includegraphics[width=\textwidth]{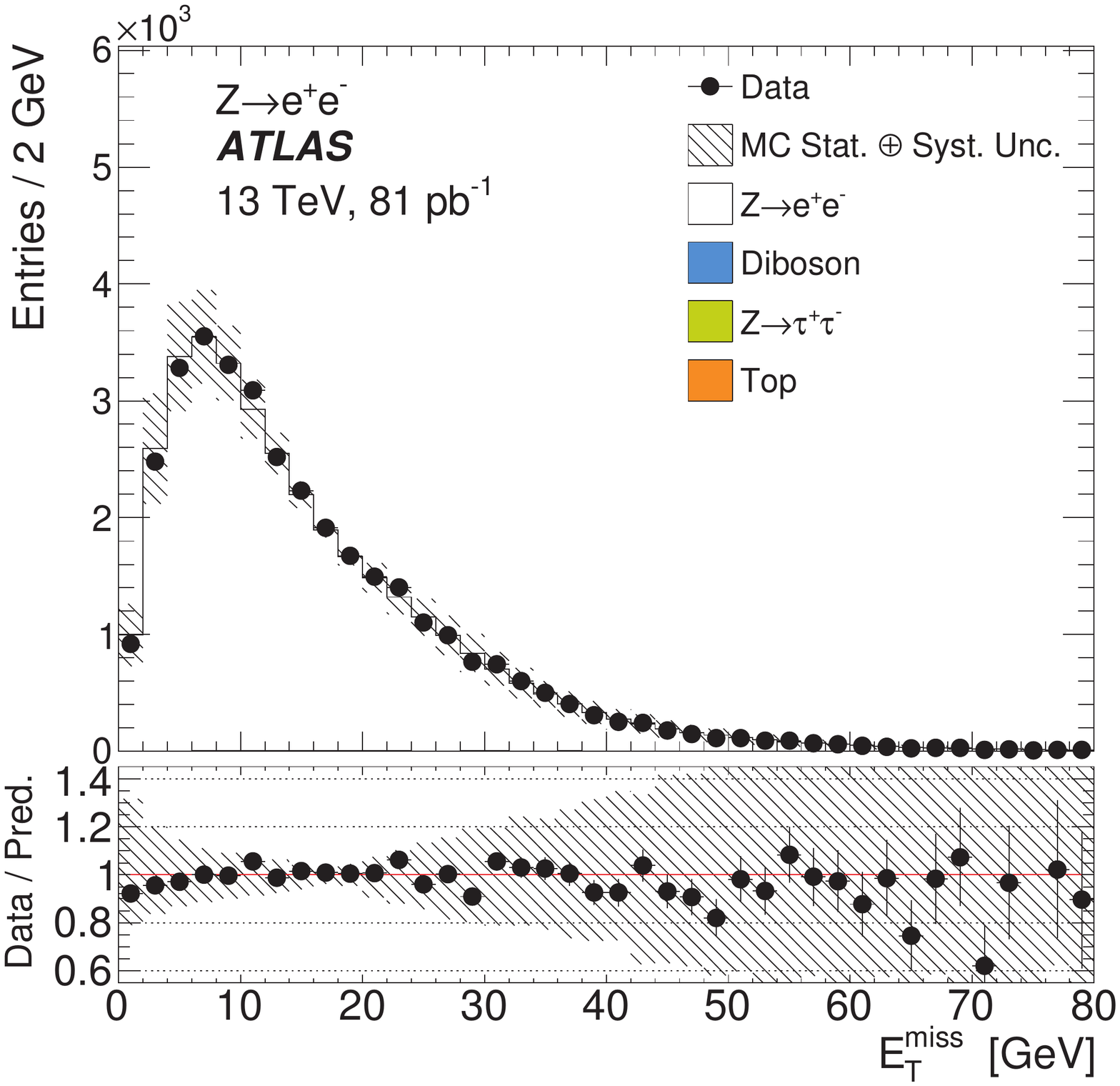}
\vspace{-2cm}
  \end{minipage}
\caption{Missing transverse energy distributions in an ATLAS $Z$+jets 
sample for the  
$\mu\mu$ and $e^{+}e^{-}$ decay channels~\cite{wzprodatlas}.
The full circles correspond to the data, while the histograms 
represent the MC predictions for all the physics processes with final states 
and kinematics passing the selection criteria. Error bars are statistical
and the band shows the systematic uncertainty in the data 
measurement.}
\label{zmetatlas}
\end{figure}

\begin{figure}[htbp]
  \centering
\vspace{-3cm}
  \begin{minipage}{.75\textwidth}
    \centering
    \includegraphics[width=\textwidth]{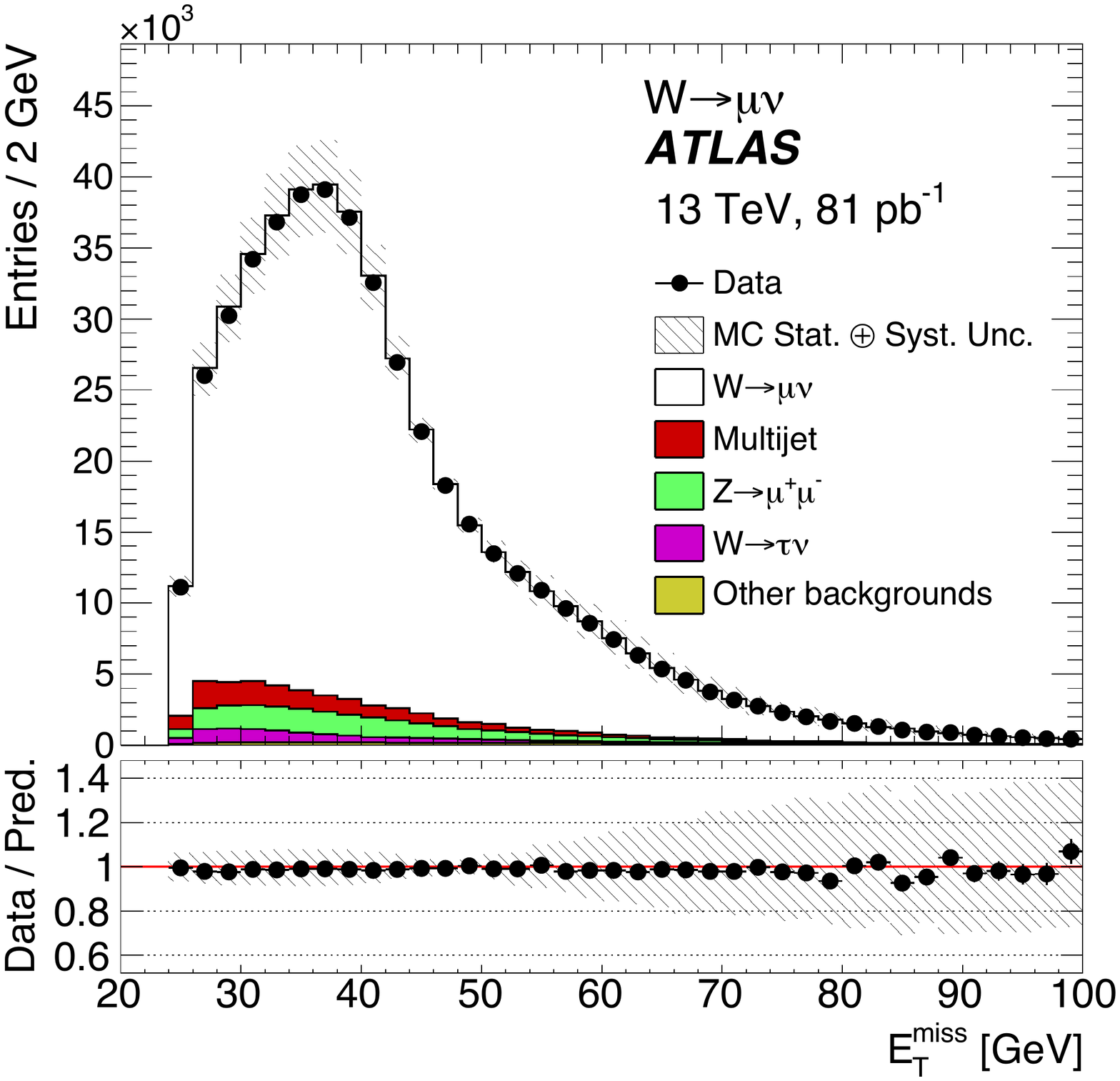}
\vspace{-3cm}
  \end{minipage}
  \begin{minipage}{.75\textwidth}
    \centering
    \includegraphics[width=\textwidth]{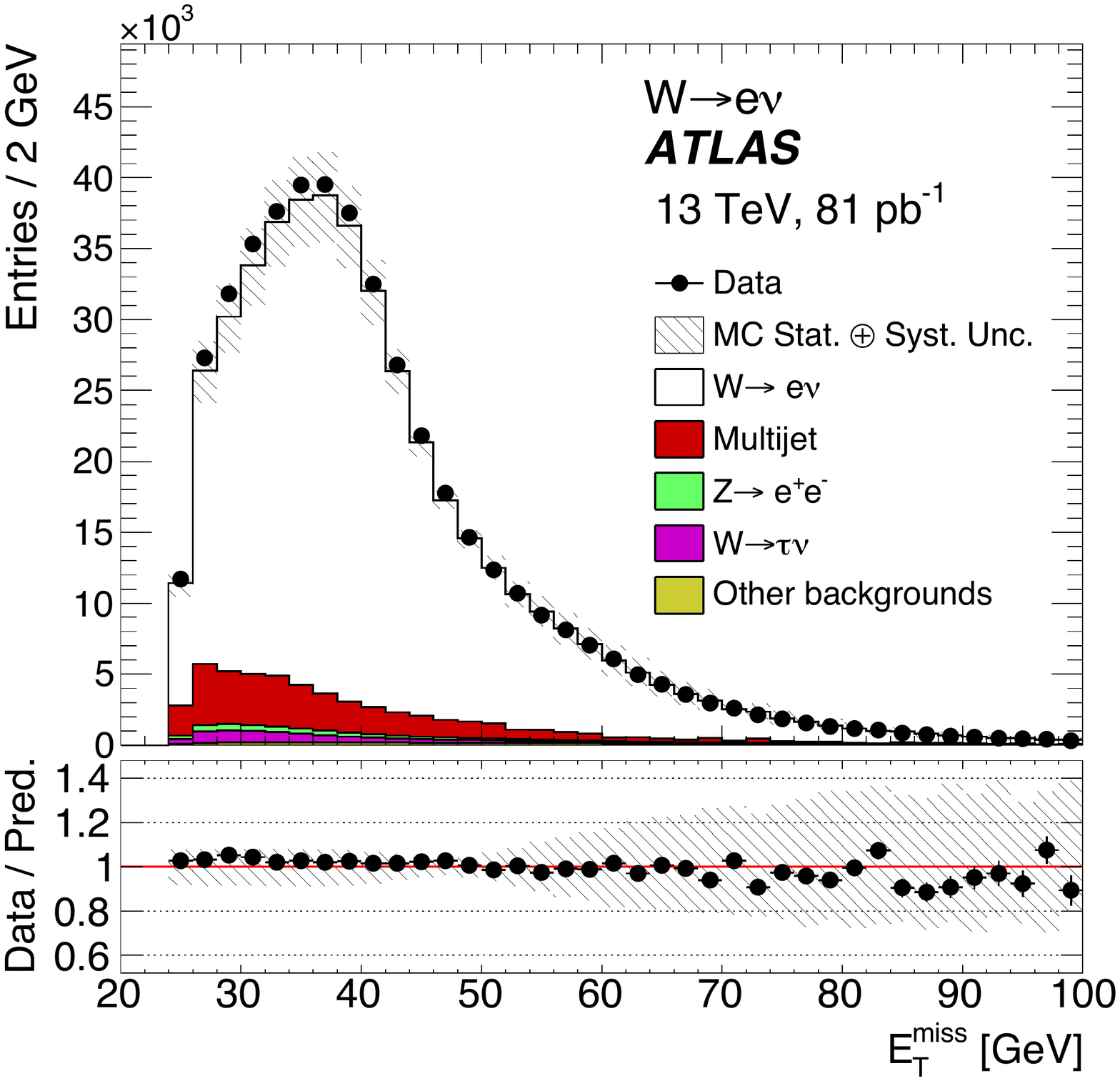}
\vspace{-2cm}
  \end{minipage}
\caption{Missing transverse energy distributions in an ATLAS $W$+jets 
sample for the  
$\mu\nu$ and $e\nu$ decay channels~\cite{wzprodatlas}.
The full circles correspond to the data, while the histograms 
represent the MC predictions for all the physics processes with final states 
and kinematics passing the selection criteria. Error bars are statistical
and the band shows the systematic uncertainty in the data 
measurement.}
\label{wmetatlas}
\end{figure}

The \met\ distibution in $W/Z$+jets events is a key ingredient 
of many SM measurements and BSM searches.
Events with a $W$ or a $Z$ and many jets have intrinsic \met\ when 
the $W$ decays to $e\nu/\mu \nu$ or the $Z$ decays to $\nu\nu$, and 
spurious \met\ coming from jet energy resolution effects. As a result,
these processes contribute significant background to most searches for signals
with weakly interacting particles because the latter have a 
large \met\ signature, and the $W/Z$+jets events may have large fake \met.
The level of understanding of $W/Z$+jets \met\ distributions
and the quality of their simulation modeling in CMS and ATLAS is illustrated in
Figs.~\ref{zmetcms},~\ref{zmetatlas},~\ref{wmetatlas} for 
$Z \rightarrow  \mu^{+}\mu^{-}/e^{+}e^{-}$
and $W \rightarrow \mu\nu/e\nu$ decays\cite{metperf,wzprodatlas}. 
The data-to-MC ratios show that the
nominal differences are less than 20\% for CMS $Z$+jets \met\ distributions, 
and less than 10\% for ATLAS $Z/W$+jets distributions. In both experiments,
systematic uncertainties grow above 50\% in different ranges of the \met\
domain. In CMS, uncertainties are largest in the 50-90~GeV range where the
contribution of hadronic shower mis-measurement dominates.

All-jet events resulting from strong production of highly collimated beams
of particles are among the most difficult to simulate and so is their
associated \met. To model the high-energy tail of the
\met\ distribution of these multi-jet events, the simulation needs to 
include a high degree of detail in the development and fluctuations of 
particle showers in the detector, as well as 
modeling of rare occurrences of detector signal processing malfunctions.
Consequently, MC-truth predictions of \met\ distributions are 
typically not reliable, to the level 
required in modern collider experiments, to make accurate estimates of
the background contributions of these QCD events to measurements of other
SM processes and potential BSM signals. 
The \met\ simulation challenge in multi-jet events originates in the fact
that jets are composed of many particles with energies ranging from a 
few GeV to 
hundreds of GeV, which shower into hundreds of more particles as they traverse 
the detector material. Different models for electromagnetic and nuclear 
interactions and a careful handling of the model-to-model transitions
are required to simulate these showers, depending on the 
particle type, energy, and material involved. Small changes in the
modeling of energy fluctuations translate into large differences in the
transverse momentum imbalance observed in 
multi-jet events, which ranges from mild to severe and 
result in small to very large fake missing transverse momentum. 

More examples of backgrounds to SM and BSM measurements 
with \met\ are presented below.
Backgrounds are irreducible when they come from a physics process
that results in a final state indistinguishable from the one for the signal.
Reducible backgrounds are those that are distinguishable from
the signal due to distinctive physics properties.
Irreducible QCD background to $t \overline{t}$ production occurs for the
final state where both top quarks decay hadronically to two b jets plus light 
jets. In this case, the multi-jet event observed in the detector is
indistinguishable from an event with the same light-jet and b-jet 
multiplicity which originates in the strong production 
and subsequent hadronization and fragmentation of light quarks, b quarks, and 
gluons.
Instead, when a $t \overline{t}$ pair decays semi-leptonically,
QCD background arises from the measurement process, when jets in the 
tails of the response distribution cause fake \met\ that mimics the 
$W \rightarrow l \nu$ process in top decays. 
A SUSY search with all jets in the final state, where
events with large \met\ are selected, is another example of instrumental
QCD background. Events in the tails of the jet 
response distribution, observed in Fig.~\ref{cmsjetresponse},
make large contributions to SUSY SR's due to 
their large production cross sections. The source and rates of these rare
events with extremely large \met\ are very difficult to
identify, evaluate, and eventually simulate. 

\begin{figure}[htbp]
\centering
\includegraphics[width=0.9\linewidth]{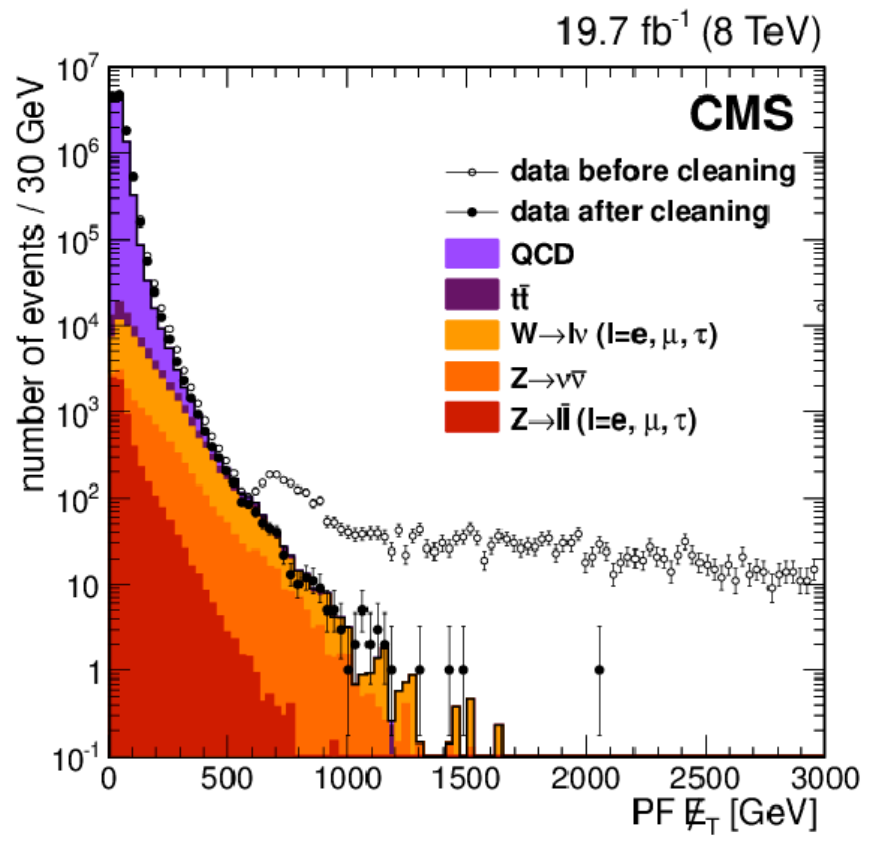}
\caption{Particle Flow \met\ distributions for CMS di-jet events with (full
circles) and without (open circles) the fake events removal algorithms 
applied. The SM contributions to the simulated \met\ distribution are shown 
in histograms~\cite{metperf}.}
\label{metdistrib}
\end{figure}

Data-to-MC comparisons of \met\ distributions are presented next to 
illustrate the level of simulation accuracy achieved in modern 
experiments despite the many challenges described above.
Fig.~\ref{metdistrib} shows the \met\ distribution for CMS di-jet events 
before and after applying the software algorithms to remove events with 
spurious \met~\cite{metperf} . 
Excellent agreement is observed in the $>500$~GeV range, even in the 
tail of the distribution. Agreement deteriorates below $500$~GeV as the
contribution of QCD events in the sample increases and eventually becomes
dominant. In spite of the excellent agreement observed in the high \met\
range, data-driven methods are preferred over 
MC predictions of multi-jet backgrounds with high \met, particularly in searches
with a SR in the tail of the \met\ distribution, and simulation-based
closure tests are utilized to demonstrate accuracy. The reason is that it is
basically impossible to demonstrate that all sources of spurious events in this
region of low statistics have been identified, understood, and modeled in the 
MC with the correct rates of occurrence. One example of these rare
occurrencies is when a high-energy
particle hits directly a photo-diode in the detector readout circuit. 

\begin{figure}[htbp]
  \centering
  \begin{minipage}{.6\textwidth}
    \centering
    \includegraphics[width=\textwidth]{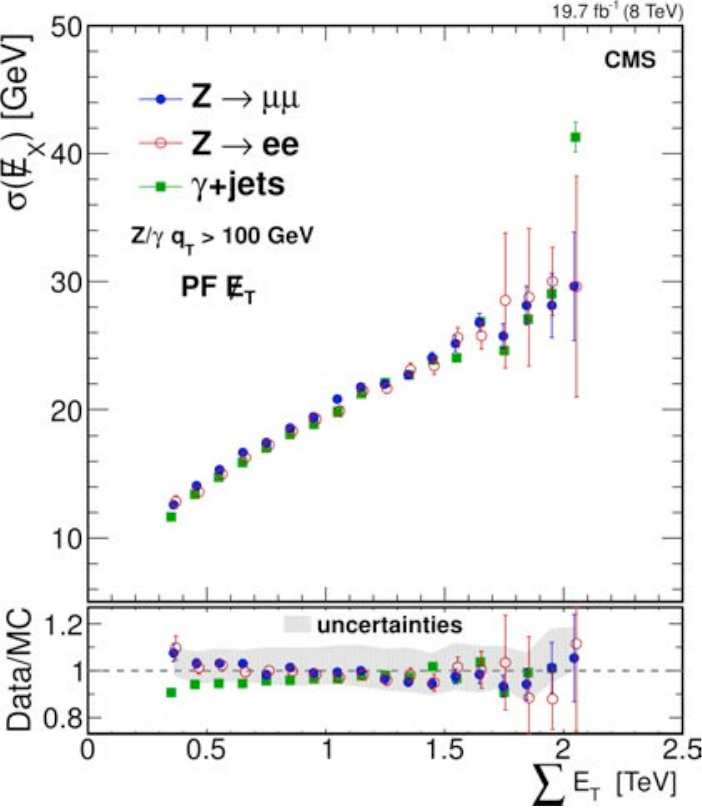}
  \end{minipage}
  \begin{minipage}{.6\textwidth}
    \centering
    \includegraphics[width=\textwidth]{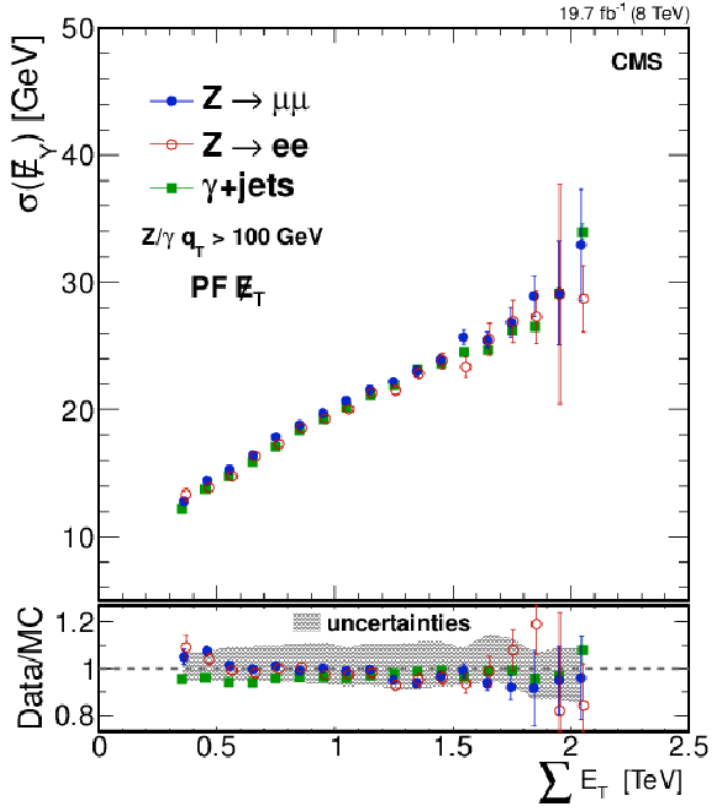}
  \end{minipage}
\caption{Resolutions for the CMS \met\ projections along the $x$ and $y$ 
axes as a 
function of the scalar sum of the transverse energy, $\Sigma E_T$, of all the 
reconstructed particles, except for the $\gamma$ in the $\gamma$+jets sample 
or the leptons from the decay of the $Z$ boson candidate in the $Z$+jets 
sample~\cite{metperf}. The error bars are statistical and the shaded band
represents the systematic uncertainties.}
\label{metresol}
\end{figure}

Fig.~\ref{metresol} shows the resolution of the CMS particle flow \met\ 
projections along 
the $x$ and $y$ axes as a function of the scalar sum of the transverse energy, 
$\Sigma E_T$, of all the 
reconstructed particles in the event, except for the 
$\gamma$ in the $\gamma$+jets sample or the leptons 
from the decay of the $Z$-boson candidate in the $Z$+jets 
sample~\cite{metperf}. 
\met\ resolutions in $Z$+jets events for the $e^{+}e^{-}$ and $\mu^{+}\mu^{-}$
decay channels
are described in the simulation within a 10\% accuracy, well within the
statistical and systematic uncertainties of the measurements. 
A good modeling of the energy resolutions
of measured particles and event quantities such as \met\ is important because
small data-to-MC discrepancies would cause a different amount of
distribution ``smearing'' and therefore bin-to-bin migration of events. A
poor modeling of the resolution smearing effect would render the MC of limited 
use in physics analysis. Migration effects may be large and are challenging to
simulate, particularly in the case of the jet $p_T$ spectrum,
which reflects the rapidly falling $p_T$ dependence of the QCD cross sections
for jet production. In this case, the event \met\ distribution would be
significantly affected in a detector with poor energy resolution because 
low $p_T$ jets produced with large cross sections would populate the high
\met\ tail, reducing the purity in the highest \met\ bins of the distribution.

\begin{figure}[htbp]
\centering
\vspace{-3cm}
\includegraphics[width=\textwidth]{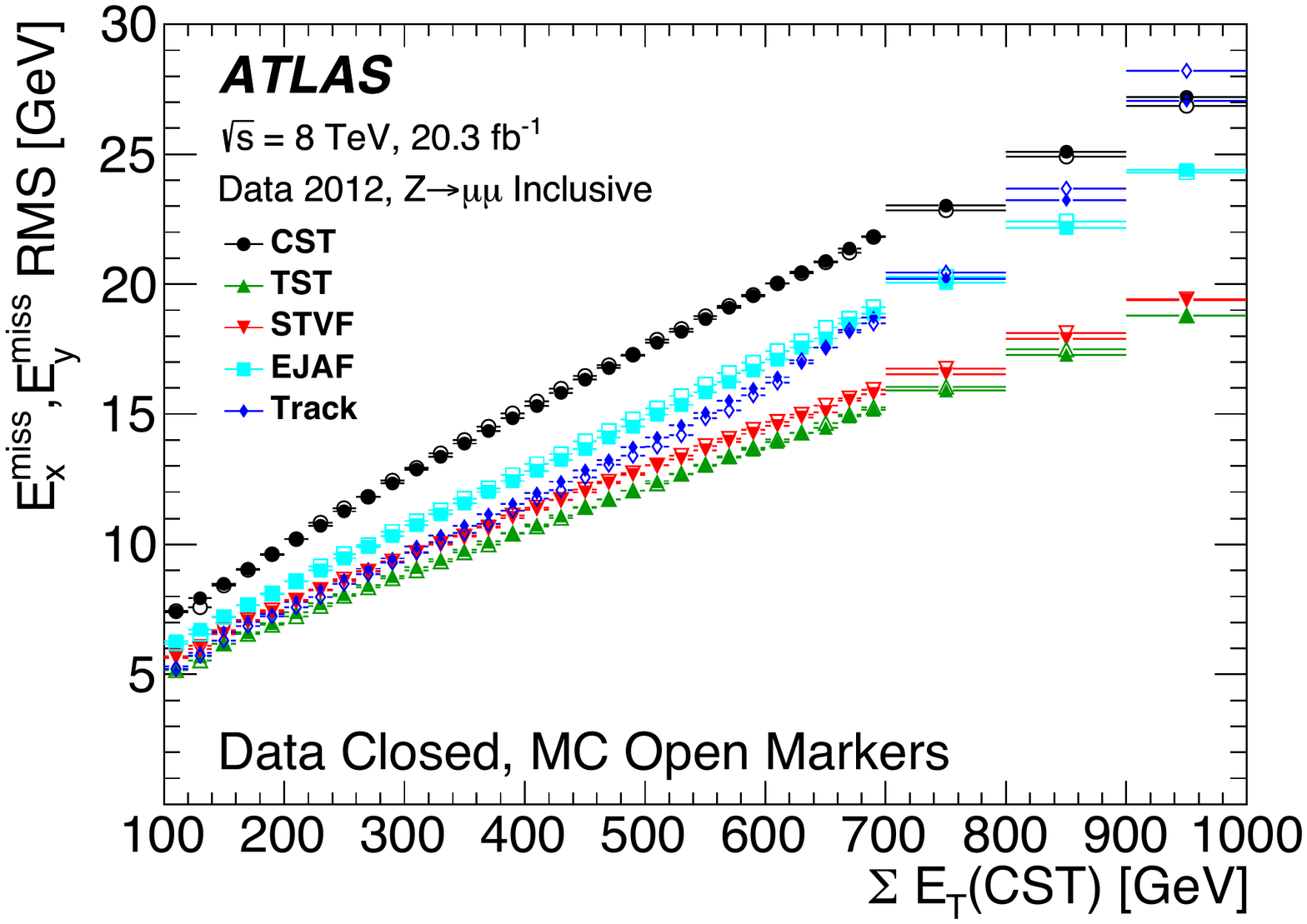}
\vspace{-4cm}
\caption{RMS of the ATLAS distribution obtained from the
combination of the $x$ and $y$ components of \met\ as a 
function of the scalar sum of the transverse energy, $\Sigma E_T$, of all 
reconstructed physics objects in an ATLAS $Z\rightarrow \mu^{+}\mu^{-}$+jets
sample~\cite{metperfatlas}.}
\label{metresolatlas}
\end{figure}

For the ATLAS experiment, Fig.~\ref{metresolatlas} shows a 
plot of the RMS obtained from the combined distribution of the 
$x$ and $y$ components of \met\ versus the scalar sum of the
$E_T$ of the physics objects in a $Z$+jets sample~\cite{metperfatlas}. 
The data is presented in
full markers and the MC in open markers for different alternative
\met\ calculations.
The \met\ is reconstructed as the negative vector sum of calibrated physics 
objects ($e$'s, $\gamma$'s, $\tau$'s, jets, $\mu$'s) and a soft term that 
comprises all the detector signals not matched to physics objects. CST,
TST, STVF, EJAF, and Track refers to different algorithms to reconstruct and 
calibrate the soft term, based on different combinations of tracker and 
calorimeter information, and pile-up subtraction techniques. 
Over the whole $\Sigma E_T$
domain, the agreement of MC with data is always better than 5\%.

\subsection{Simulation and  Jet Cross Sections} \label{jetcrosssections}

The example of jet cross sections and QCD jet measurements in general is 
particularly useful to illustrate the impact of simulation in data 
measurements because of its dependence on a single dominant source of 
systematic uncertainty, the jet energy correction, which in turn relies to
a large extent on how well the hadronic response and energy resolutions are 
modeled in the simulation. In particular, 
the energy response to low-energy hadrons ($E=1-10$~GeV) is the most difficult 
to model and affects even high-energy jets given that the energy of the jet 
constituents grows slowly, approximately as the square root of the jet energy. 

\begin{figure}[htbp]
  \centering
  \begin{minipage}{.9\textwidth}
    \centering
    \includegraphics[width=\textwidth]{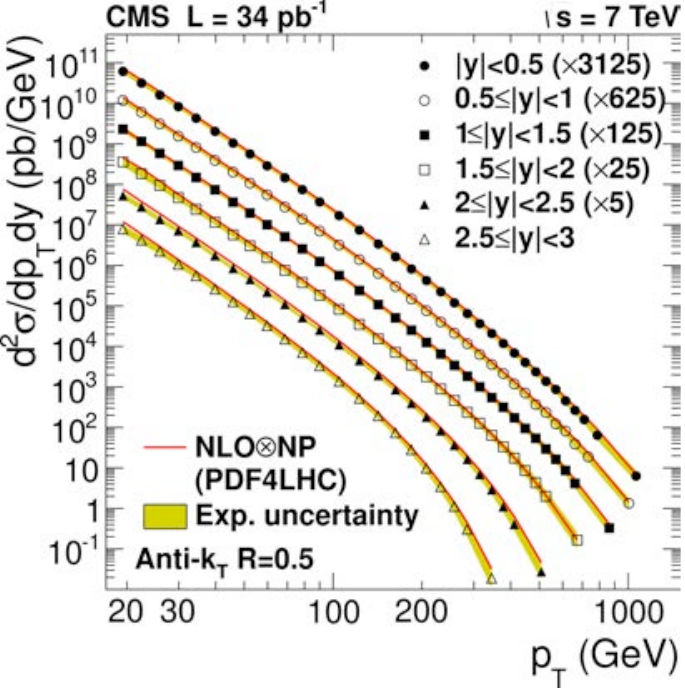}
  \end{minipage}
  \begin{minipage}{.9\textwidth}
    \centering
    \includegraphics[width=\textwidth]{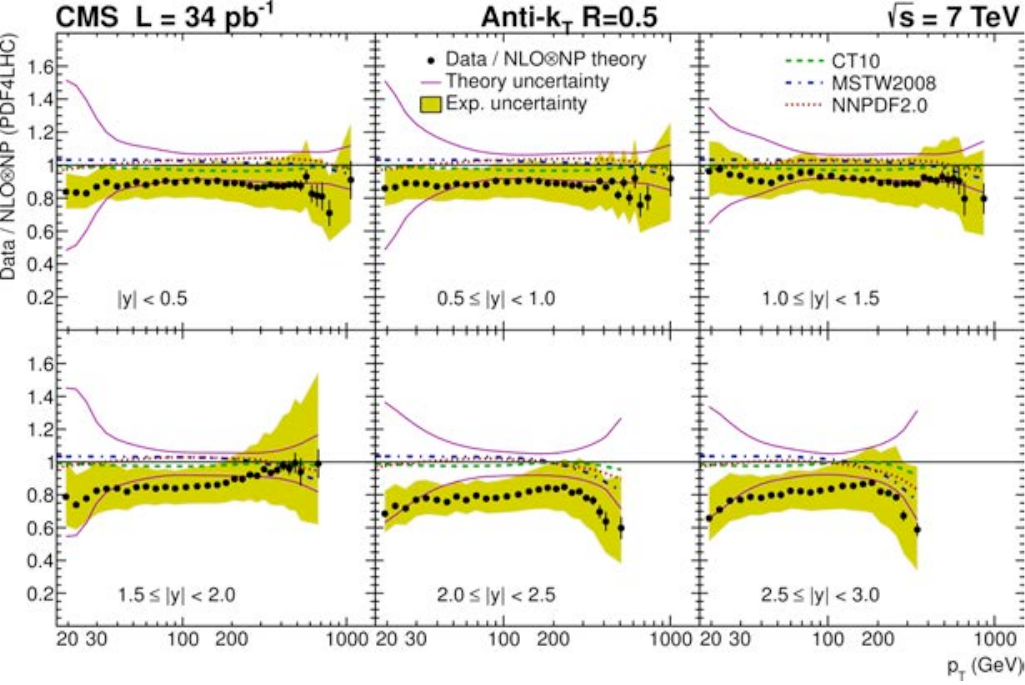}
  \end{minipage}
\caption{First inclusive jet cross sections published by the CMS experiment, 
based on the full 
2010 (Run~1) sample~\cite{cmsjetxs11}.}
\label{cmsjetxs11plot}
\end{figure}

\begin{figure}[htbp]
  \centering
  \begin{minipage}{.9\textwidth}
    \centering
    \includegraphics[width=\textwidth]{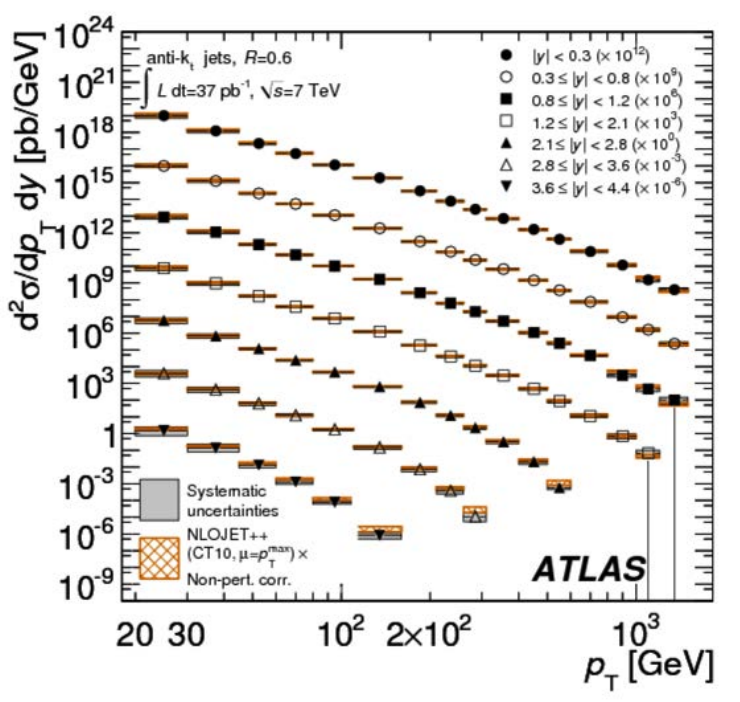}
  \end{minipage}
  \begin{minipage}{.45\textwidth}
    \centering
    \includegraphics[width=\textwidth]{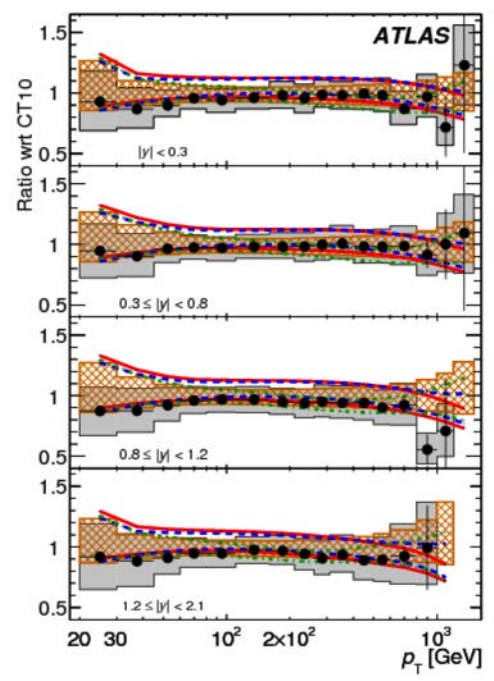}
  \end{minipage}
\begin{minipage}{.45\textwidth}
    \centering
    \includegraphics[width=\textwidth]{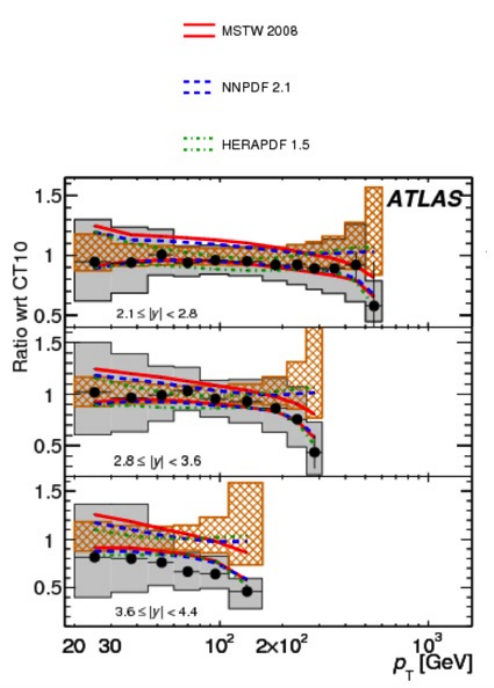}
  \end{minipage}
\caption{First inclusive jet cross sections published by the ATLAS
experiment, based on the full 
2010 (Run~1) sample~\cite{atlasjetxs12}.}
\label{atlasjetxs12plot}
\end{figure}

\begin{figure}[htbp]
  \centering
  \begin{minipage}{.8\textwidth}
    \centering
    \includegraphics[width=\textwidth]{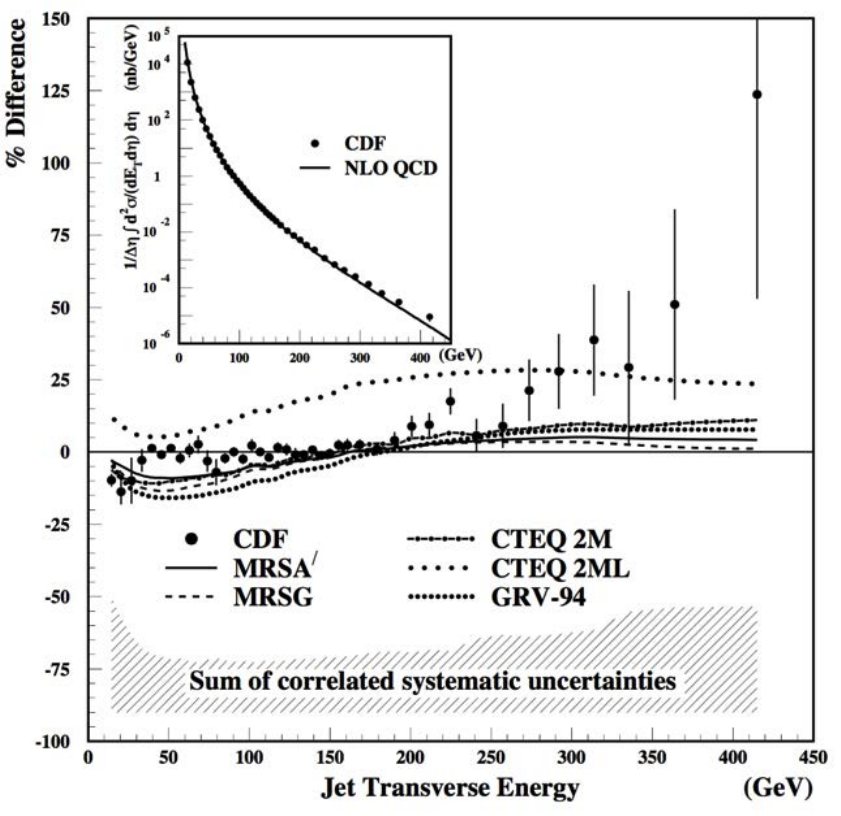}
  \end{minipage}
  \begin{minipage}{.7\textwidth}
    \centering
    \includegraphics[width=\textwidth]{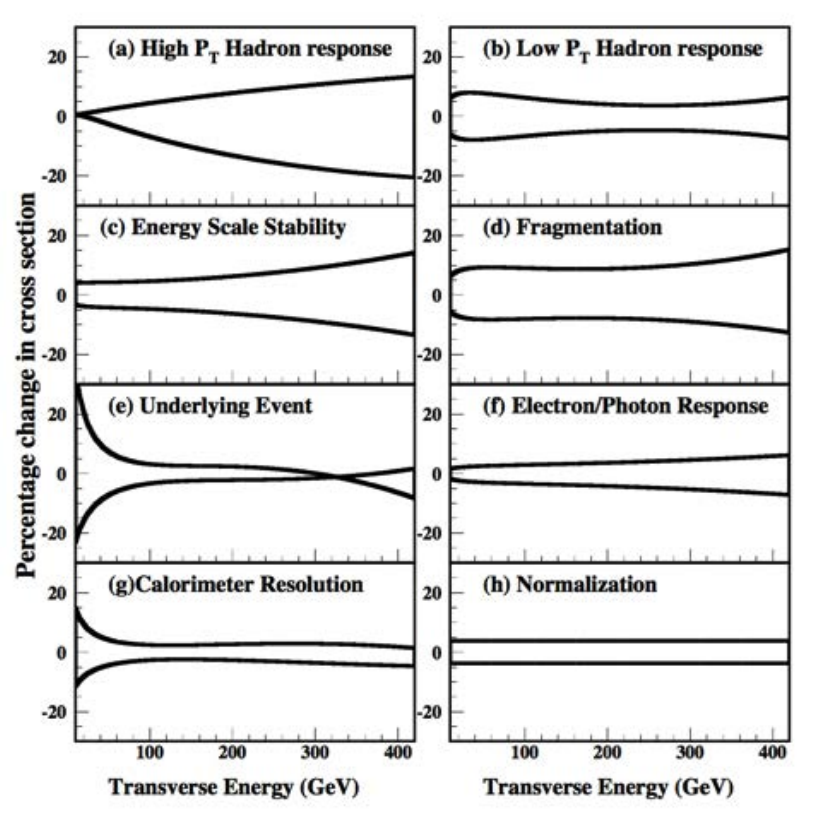}
  \end{minipage}
\caption{First inclusive jet cross sections published by the CDF experiment,
based on the full 1992-1993 (Run 1a) sample~\cite{cdfjetxs96}.}
\label{cdfjetxs96plot}
\end{figure}

\begin{figure}[htbp]
  \centering
  \begin{minipage}{.85\textwidth}
    \centering
    \includegraphics[width=\textwidth]{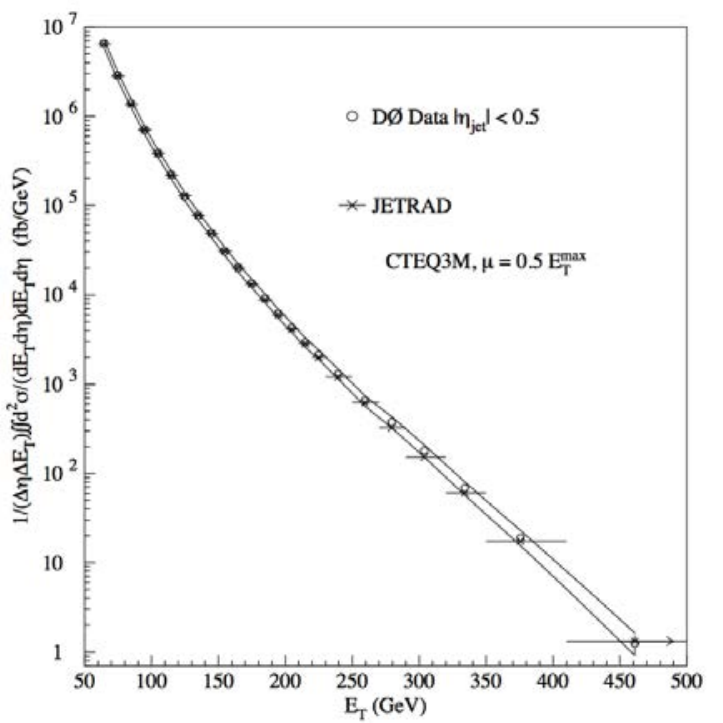}
  \end{minipage}
  \begin{minipage}{.8\textwidth}
    \centering
    \includegraphics[width=\textwidth]{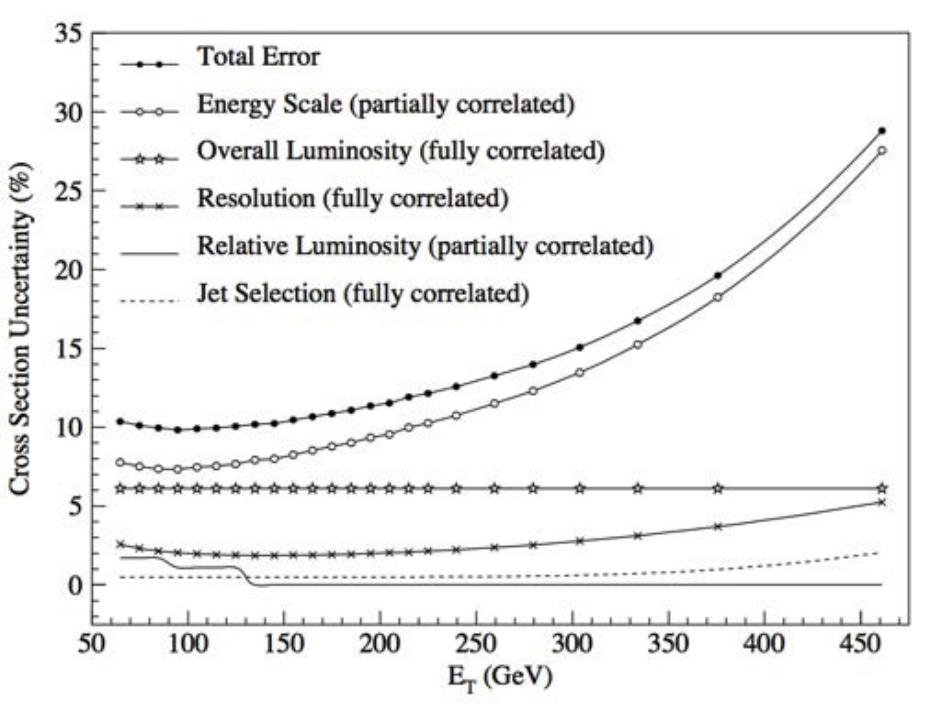}
  \end{minipage}
\caption{First inclusive jet cross sections published by the D0 experiment, 
based on the full 
1993-1995 (Run 1b) sample~\cite{d0jetxs99}.}
\label{d0jetxs99plot}
\end{figure}

Figs.~\ref{atlasjetxs12plot}, \ref{cmsjetxs11plot}, \ref{cdfjetxs96plot}, 
\ref{d0jetxs99plot} 
do not show the latest and most precise measurements 
of the ATLAS~\cite{atlasjetxs12}, CMS~\cite{cmsjetxs11}, 
CDF~\cite{cdfjetxs96}, and D0~\cite{d0jetxs99} inclusive jet cross sections. 
Instead, they 
show the results in the first publications of each experiment based on the 
full 2010 Run 1 ATLAS and CMS samples, the full 1992-1993 Run 1a CDF 
sample and the full 1993-1995 Run 1b D0 sample. 
The goal here is to analyze the impact of simulation in the 
publication process timeline and the precision of the result. 
There is an important 
difference between the comparisons for jet measurements and the rest of the 
data-to-MC comparisons in this article. 
While the latter are comparisons between 
detector-level measured quantities and predicted quantities based on events 
generated and passed through detector simulation software, the former are 
comparisons between measured quantities corrected to the particle level and 
NLO-QCD parton level theoretical predictions which, sometimes, contain
non-perturbative hadronization corrections. In the inclusive jet 
measurements, all detector effects such as jet energy response and resolution 
smearing have been removed, in average, as part of the analysis procedure. 
Therefore, theoretical 
predictions do not need to be passed through detector modeling software in 
order to be on the same footing for comparison with data. While
comparisons in previous sections give information about the quality of the 
event generators and detector simulation software tools, 
the jet cross section comparisons 
evaluate the accuracy of the QCD theoretical predictions, which depend on the 
order of the calculation, the choice of factorization and renormalization 
parameters, and the parton distribution functions (PDFs). The aspect of the 
jet cross section measurements to highlight in this section is the 
relationship between the size of the systematic uncertainty (measurement 
precision), the role of
simulation, and the 
publication timeline (publication turnaround). The capabilities of the 
detectors as well as the quality of the FullSim and ParSim tools, 
utilized to either design the 
data-driven jet correction derivation 
methods or directly extract the corrections after thorough tuning and 
validation, dominate the accuracy of the 
measurements and the publication timeline. 

Fig.~\ref{cmsjetxs11plot} shows the CMS inclusive jet cross section 
measurements based on the full $34$~pb$^{-1}$ data collected in 2010. 
The data taking period started in 
March 2010 and the results were published in June 2011, only 
seven months after the end of the run. The jet cross sections shown in 
Fig.~\ref{atlasjetxs12plot} are based on the full 2010 data set and were
published by the ATLAS experiment in April 2012. ATLAS also published an 
intermediate result~\cite{atlasjetxs10}, based 
on half the data-set, in October 2010, before the end of the run. The CMS 
measurement extends up to rapidities of $|y|=3$ with uncertainties in the range 
of $10-20\%$ in the most central region and $15-30\%$ in the most forward bin. 
ATLAS measured the jet cross sections up to rapidities of $|y|=4.4$ with 
uncertainties similar to those in the CMS measurement for the central bin 
and in the $12-40\%$ range for 
the most forward bin. Rapidity is defined as 
$y=\frac{1}{2}{\rm ln}[(E+p_z)/(E-p_z)]$, where $E$ is the jet energy 
and $p_z$ its momentum component along the beam axis.

The CDF experiment published the $19.5$~pb$^{-1}$ Run 1a data-set in January 
1996, almost five years after the start of the run at the Tevatron. The 
measurement, shown in Fig.~\ref{cdfjetxs96plot}, covers only the central 
pseudorapidity region, $0.1<|\eta|<0.7$, 
with uncertainties in the $20-35\%$ range. D0's first inclusive
jet cross section measurement, shown in Fig.~\ref{d0jetxs99plot}, was
published in 1999, eight years after the start of Run 1. 
It was based on the full $92$~pb$^{-1}$ Run 1b data-set,
restricted to the $|\eta|<0.5$ range, and reported uncertainties in the 
$10-30\%$ range. A few years later, in 2001, D0 extended the Run 1 
measurement to the forward pseudorapidity region, up to 
$|\eta|=3$~\cite{d0jetxs01}. 
Both experiments published the Run 2 inclusive jets cross sections, CDF in
$|\eta|<2.1$ (2008)~\cite{cdfjetxs08}, and D0 in a slightly larger, 
$|\eta|<2.4$, region (2011)~\cite{d0jetxs12}. The reason for the CDF delay
in extending the $\eta$ coverage is that the experiment 
initially tuned the ParSim only for the central calorimeter. The End Plug
Calorimeter was not incorporated until a GFLASH based approach was undertaken 
in 2002-2003. Jet energy calibration in forward regions relies on di-jet 
balance techniques, which are significantly affected by resolution biases and 
can be understood in detail only with large and accurate MC samples. In the 
case of D0, a ParSim approach was not viable due to the absence of a 
solenoidal magnetic field in the tracker and scarce test beam data. 
The in situ 
calibration approach based on data-driven methods applied to collider data had 
to be developed from scratch and without the aid of large and accurate FullSim 
samples for studies and closure tests. Consequently, D0 could not deliver a
result with competitive systematic uncertainties until 1996, while 
CDF published jet cross sections with large uncertainties in 1989 
(Run 0)~\cite{cdfjetxs89} and 1992. The latter was an 
intermediate Run 1 result based on early data~\cite{cdfjetxs92}.

The Tevatron inclusive jet cross section story is one of limited test beam 
programs, complex tuning of parametrized simulations, and a lengthy process 
of developing data-driven techniques with little aid from full simulation. 
The LHC experiments benefited from new generation detectors with excellent 
capabilities, mature data-driven techniques and expertise, simulation of 
unprecedented quality, and a computing infrastructure with the capacity to 
generate not hundreds of thousands but billions of MC events. While it took
months to CMS and ATLAS to publish jet cross section results with 
uncertainties on the order of 10-40\% (10-20\% in the most central region),
it took years to D0 and CDF to achieve a level of precision that was a
factor of two inferior.

\section{Simulation and Publication Turnaround} \label{turnaround}

\begin{figure}[htbp]
  \centering
  \begin{minipage}{.75\textwidth}
    \centering
    \includegraphics[width=\textwidth]{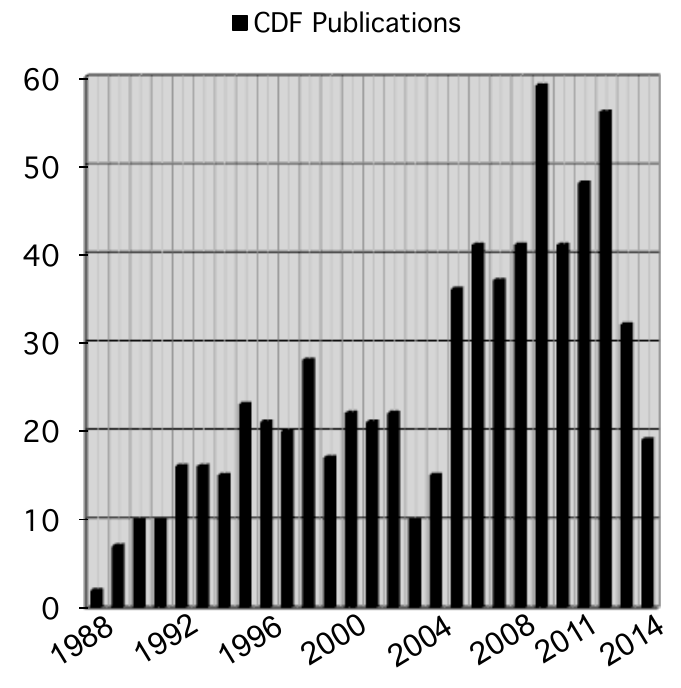}
  \end{minipage}
  \begin{minipage}{.75\textwidth}
    \vspace{1cm}
    \centering
    \includegraphics[width=\textwidth]{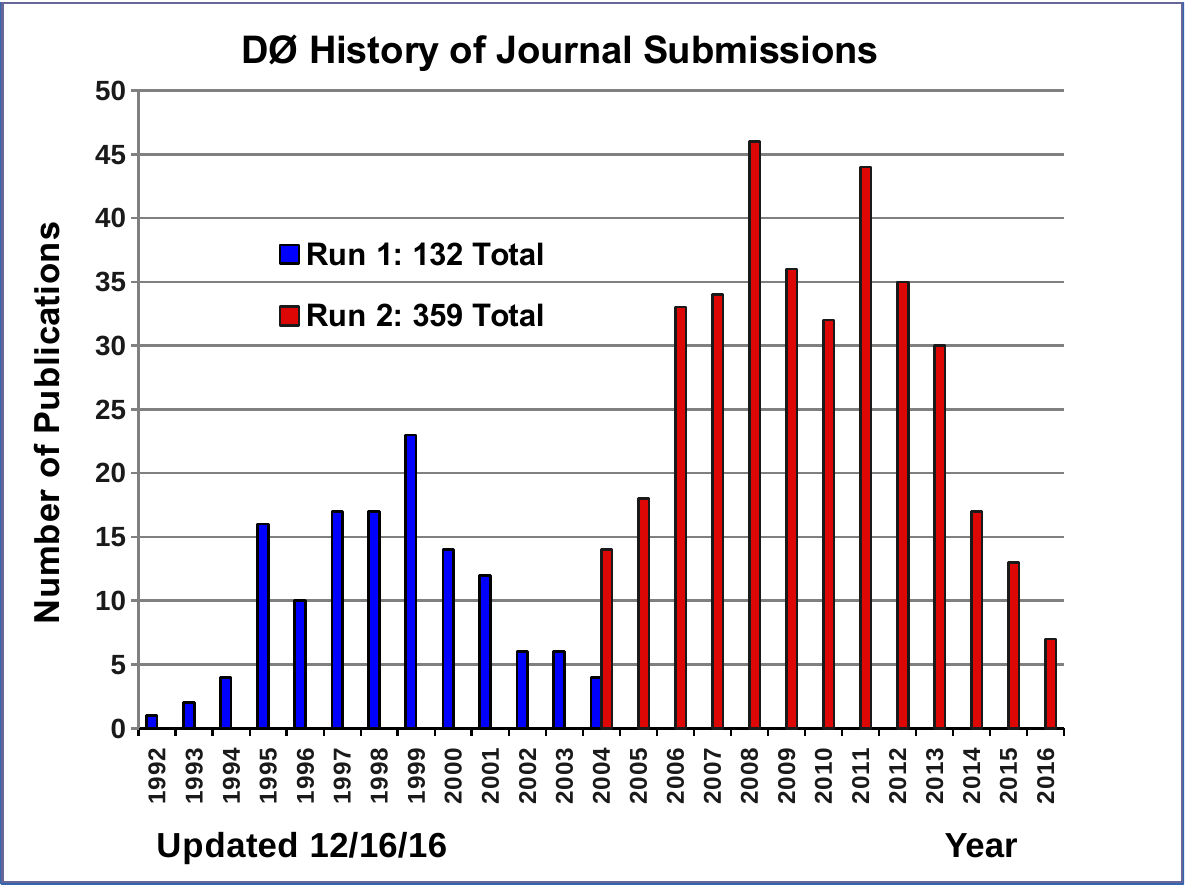}
  \end{minipage}
\vspace{0.5cm}
\caption{Number of publications per year for the 
CDF~\cite{cdfpubtime} and D0~\cite{d0pubtime} experiments at the Tevatron.}
\label{publicationwithtime1}
\end{figure}

\begin{figure}[htbp]
  \centering
  \begin{minipage}{.75\textwidth}
    \centering
    \includegraphics[width=\textwidth]{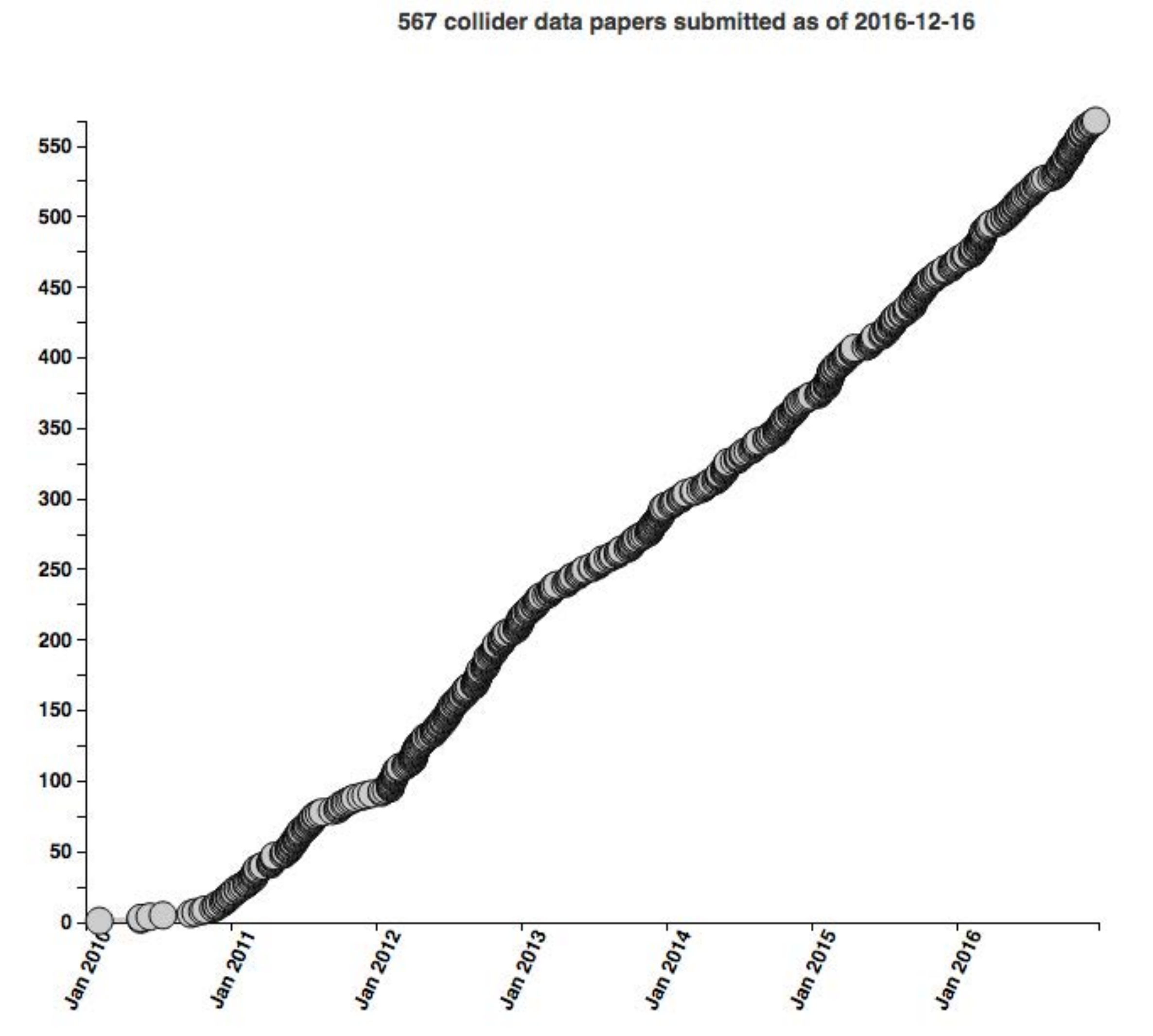}
  \end{minipage}
\caption{Integrated number of publications as a function of time for the 
CMS~\cite{cmspubtime} experiment at the LHC.}
\label{publicationwithtime2}
\end{figure}

The process of publication of physics measurements from start-up to paper 
submission has accelerated significantly in modern particle physics 
experiments. Although many technological and human factors account for this 
trend, including the
fact that the LHC experiments have thousands of members and the Tevatron
experiments hundreds at their peak, simulation has 
played a significant role. 
Figs.~\ref{publicationwithtime1},~\ref{publicationwithtime2} show the 
number of publications per year between 1998 (1992) and 2014 (2016) for 
the CDF~\cite{cdfpubtime} (D0~\cite{d0pubtime}) experiment at the
Tevatron, and the 
integrated number of publications as a function of time for the CMS 
experiment at LHC~\cite{cmspubtime}. 
For the Tevatron experiments, Run 1a started 
in June of 1992 and finished by the end of the spring of 1993. Unlike D0, CDF 
had a Run 0 in 1988-1990. As illustrated in Fig.~\ref{publicationwithtime1}, 
while the first D0 
physics paper was published in early 1994, the publications distribution for 
Run 1a peaked in 1995, three years after the start of the run. Run 1b started 
in 1994, the publications distribution peaked in 1998 and began to slow down 
in 2001. The absence of a Run 0 explains D0's delay with respect to CDF in 
submitting the first Run 1a publications, since the experiment had to 
commission the detector, optimize the software algorithms and develop analysis 
techniques using the early data. 

In the absence of fast enough simulation, 
GEANT3 was available but was computationally costly given the speed of
the machines at the time, 
the process of developing data-driven 
techniques from the scratch was a challenging and lengthy process for both 
Tevatron experiments. As discussed in Sec.~\ref{ToyParaFull}, 
CDF used Run 0 data to measure the 
single-track energy response from minimum bias and track triggers to tune a 
fast MC. This parametrized 
simulation approach was preferred during Run 0 and Run 1 over the full 
GEANT3-based option because the latter was prohibitively slow. In Run 1, 
D0 did not have a solenoid magnet wrapped around the tracker to measure the 
momentum of single-charged-particles and tune the simulation. Consequently, the 
experiment relied purely on in situ measurements of calibration factors and 
efficiencies using 
data-driven methods applied to collider events. The process of developing 
these methods and, eventually, improving and tuning their simulation 
software was very lengthy for the experiments because they had to rely on 
small MC samples, of the order of a few ten to a few hundred thousand events, 
to develop the techniques and demonstrate their correctness via closure tests. 
Even with the aid of GEANT3-based simulation software, accuracy was often 
sacrificed for speed by introducing approximations to the sub-detector shapes, 
material, or particle shower modeling. The resulting MC samples 
were only partially useful to develop data-driven techniques, investigate their 
associated biases, and establish closure. At the LHC, hundreds of millions of 
fully simulated events were generated using Geant4-based applications even 
before the start of the first run. Reconstruction algorithms and data-driven 
methods to derive efficiencies and calibration factors were developed using 
these MC samples, and performed, basically, as in design specifications on 
real collider data at start-up. MC truth predictions of calibration curves
and physics observables were 
in such good agreement with data, that they could be used 
``out-of-the-box'' almost 
immediately after start-up, requiring only small corrections and even smaller 
uncertainties derived from comparisons with results from data-driven
methods applied on real collider data. 

The highly accurate simulation software of the LHC experiments, fast computing 
and precise data-driven techniques, which leveraged the Tevatron 
experience, 
contributed to a large extent to the much faster publication turnaround at the 
LHC.

\section{Economic Impact and Cost of Simulation in HEP Experiments}

Simulation, including physics generation, interaction with matter 
(Geant4 or ParSim),
readout modeling, reconstruction and analysis takes a large fraction of the 
computing resources consumed in HEP experiments. The estimate of this number 
for the CMS experiment presented in this article has a large uncertainty 
and it varies significantly year-to-year. Since 
the software commissioning period in preparation for Run 1, the Geant4 part of 
the CMS simulation software chain has taken the largest fraction of the CPU 
time, while the physics generation contribution has been small, 
except in the case of the generation of BSM signal samples in a large
model parameter space. 
Readout modeling takes a relatively small fraction and reconstruction of the 
same order as the Geant4 module. 

From start-up in 2009 through May 2016, CMS simulation as defined in the 
first sentence took approximately 85\% of the total CPU time utilized by
CMS, while the 
Geant4 module took about 40\%. (This information was obtained from the CMS
Dashboard, which is a computing information monitoring source available
to CMS members.)
ATLAS's Geant4 module takes approximately seven times more CPU time than 
CMS's due to the more complex geometry and other 
factors. In CMS, the rest of the CPU cycles were primarily used to 
reconstruct and analyze real collider data. 
The assumption for the 85\% figure is that the 
analysis of simulated data consumes 75\% of the CPU time spent in analysis,
including both simulated and real data, 
and excludes the generation of signal samples for BSM searches. The reason 
why the analysis of simulated data takes a larger fraction of the total 
analysis CPU time than the analysis of real collider data is that the design 
and optimization of the measurements, as well as the development and 
validation of data-driven methods, are all based on MC samples. 

In more detail, CMS spent on simulation 540 thousand core months during 2012 
(860 thousand core months in the May 2015-May 2016 period), corresponding to 
more than 45,000 (70,000) CPU cores at full capacity that cost on the order 
of 5 (8) million US dollars. (This information was obtained from the
CMS Dashboard and from private communication with Oliver Gutsche.) 
These numbers account only 
for purchasing cost though, and a more realistic estimation may be based on 
a value of 0.9 US dollar cents per core hour, which is what Fermilab spends 
on physical hardware including life-cycle, operation and maintenance. 
(The information was obtained from private communication with Oliver Gutsche.) 
An alternative estimate is based on the cost of renting the CPU time from 
industry, at a rate of 
1.4 US dollar cents per core hour. (The information was
obtained from private communication with Oliver Gutsche.)
The 0.9 (1.4) US dollar cents assumption 
puts the annual cost of simulation for CMS in the range of 3.5-6.2 
(5.5-10) million US dollars, half of it spent on executing the Geant4 module. 
A corollary to this discussion is that improvements of 1\%, 10\%, and 35\% in 
the time performance of the Geant4 toolkit would render 50-80k, 500-800k, 
1.8-2.8M US dollars per year of savings to CMS. Improvements on the order of a 
2-5 speed-up factor, as targeted by current R\&D efforts 
(GeantV~\cite{geantv}), would yield savings 
on the order of 2-4 (3-6) million US dollars per year. An important question, 
rarely addressed even by modern experiments at the time of detector design and 
technology selection, is related to the added costs to the detector 
construction, commissioning, and operations that comes from detector choices 
that maximize physics output in exchange for expensive and time-consuming 
simulation and reconstruction operations.

It is important to mention that the LHC experiments expect their computing 
needs to increase by a factor of 10 to 100 in the High-Luminosity LHC (HL-LHC) 
era, depending on the solutions developed to face simulation, pile-up, and 
reconstruction challenges arising from the high-luminosity environment.
In principle, reconstruction would take a larger fraction of the computing 
resources during the HL-LHC era, since the CPU time consumption is predicted
to increase exponentially with the number of pile-up events. However,
while simulation code is highly optimized and offers few non-revolutionary
time performance improvement opportunities, the reconstruction code 
under development for the upgraded or new HL-LHC sub-detector systems 
still offers low hanging fruit to exploit, at least in the case of CMS.
Consequently, 
a significant improvement in simulation computing 
performance is a need in present times of flat budgets, and so are the 
research efforts with that goal in mind.

\begin{figure}[htbp]
  \centering
  \begin{minipage}{.85\textwidth}
    \centering
    \includegraphics[width=\textwidth]{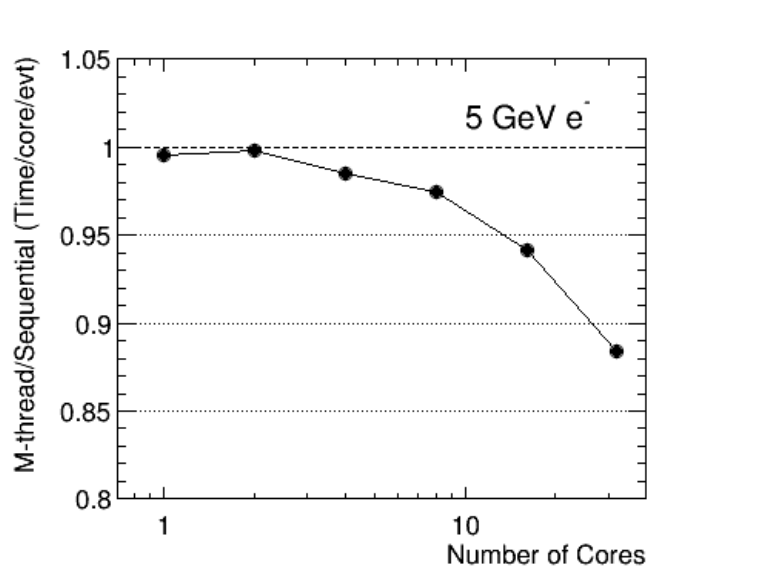}
  \end{minipage}
  \begin{minipage}{.85\textwidth}
    \centering
    \includegraphics[width=\textwidth]{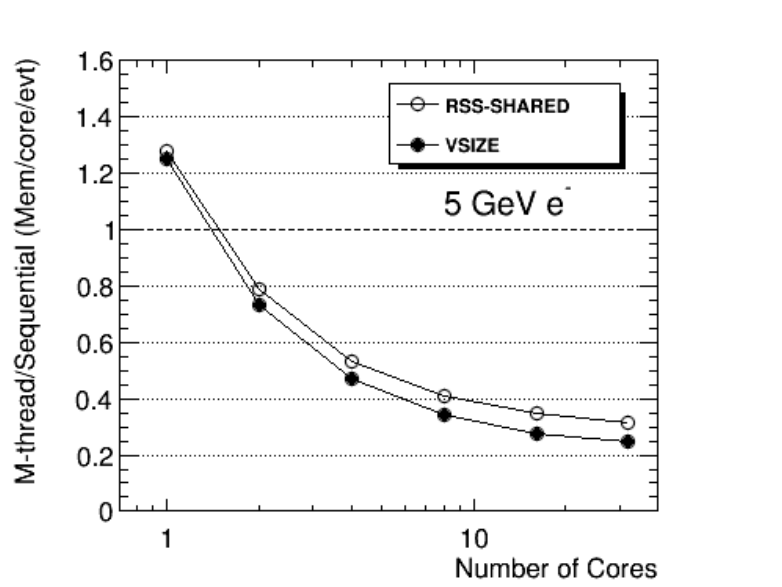}
  \end{minipage}
\caption{CPU time (top) and memory (bottom) ratios for a CMS standalone 
simulation application (outside of the CMS software framework) 
based on Geant4 version 10.1.p02 executed for 5~GeV electrons in 
multithreaded and sequential modes~\cite{g4qa}.}
\label{g4cpumemory}
\end{figure}

The Geant4 Collaboration has gone to great lengths to improve the 
toolkit computing performance during the last few years, as code was reviewed 
and optimized. In 2013, the introduction of event-level multithreading 
capabilities in Geant4 brought significant memory savings, as 
illustrated in Fig.~\ref{g4cpumemory}, which shows the CPU time (top) and 
memory (bottom) consumption ratios of a CMS standalone
simulation application (outside of the CMS software framework), 
based on Geant4 version 10.1.p02, executed for 
5~GeV electrons 
in multithreaded and sequential modes. While time performance does not 
improve, deviating from perfect scaling by approximately 10\% when executed 
on 30 cores, memory consumption improves significantly with 170~MB used in 
the first event, and only 30~MB per event used by each additional 
thread~\cite{g4qa}. 

\begin{figure}[htbp]
  \centering
  \begin{minipage}{1.0\textwidth}
    \centering
    \includegraphics[width=\textwidth]{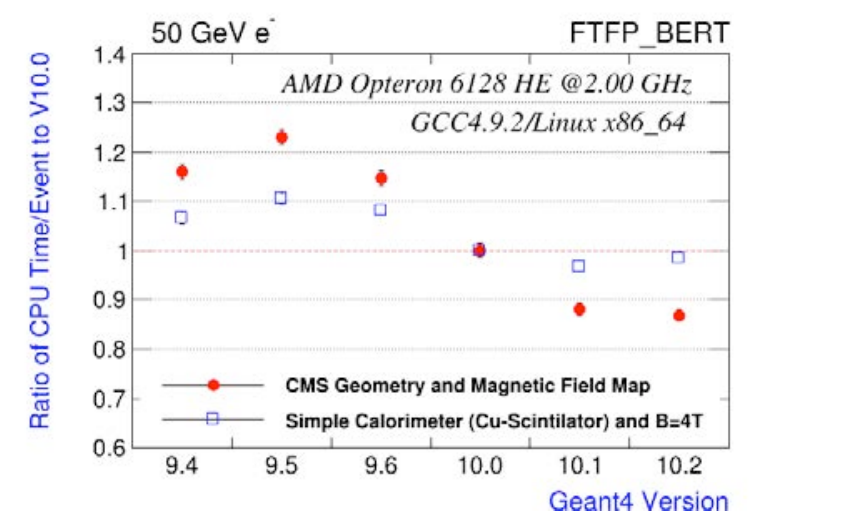}
  \end{minipage}
  \begin{minipage}{1.0\textwidth}
    \centering
    \includegraphics[width=\textwidth]{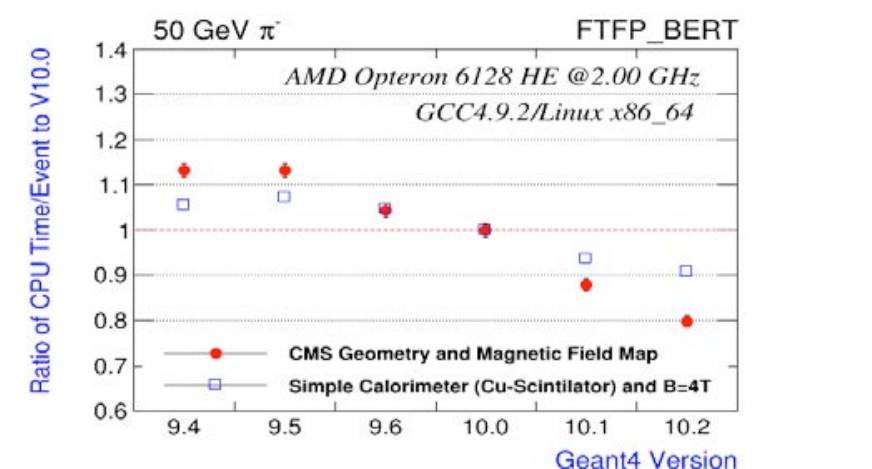}
  \end{minipage}
\vspace{0.5cm}
\caption{Percentage change in CPU time performance with respect to 
Geant4 version 10.0, starting with Geant4 version 9.4.p02 (2010) and ending 
with Geant4 version 10.2 (2015), 
for a standalone CMS application (outside of the CMS software framework) 
and a simple calorimeter configuration made 
of Cu-Scintillator in a 4~Tesla magnetic field. The study is performed for  
50~GeV $e^{-}$ and $\pi^{-}$~\cite{g4qa}.}
\label{g4cpuimprovement1}
\end{figure}

\begin{figure}[htbp]
  \centering
\begin{minipage}{1.0\textwidth}
    \centering
    \includegraphics[width=\textwidth]{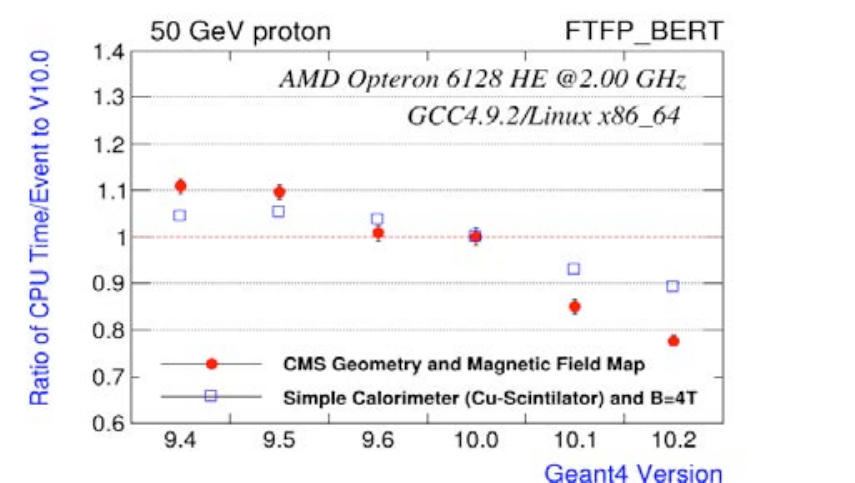}
  \end{minipage}
\begin{minipage}{1.0\textwidth}
    \centering
    \includegraphics[width=\textwidth]{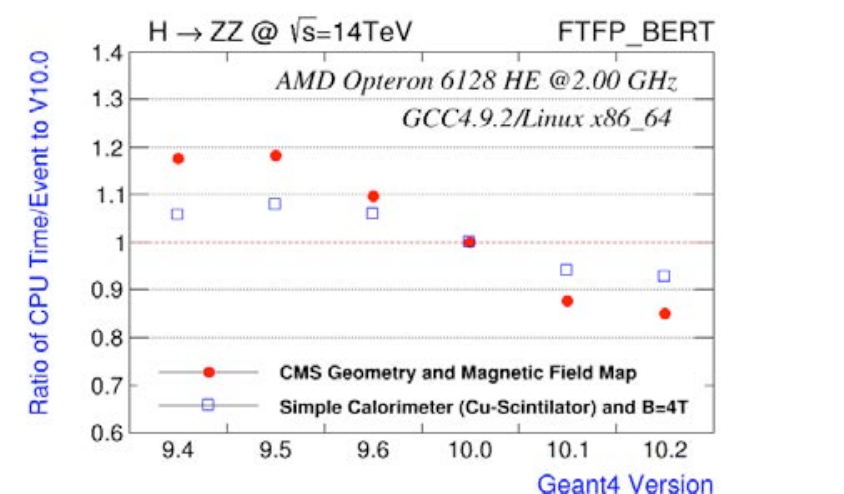}
  \end{minipage}
\vspace{0.5cm}
\caption{Percentage change in CPU time performance with respect to 
Geant4 version 10.0, starting with Geant4 version 9.4.p02 (2010) and ending 
with Geant4 version 10.2 (2015), 
for a standalone CMS application (outside of the CMS software framework) 
and a simple calorimeter configuration made 
of Cu-Scintillator in a 4~Tesla magnetic field. The study is performed for  
protons and $H \rightarrow ZZ$ events~\cite{g4qa}.}
\label{g4cpuimprovement2}
\end{figure}

Figs.~\ref{g4cpuimprovement1},~\ref{g4cpuimprovement2} show the percentage 
change in CPU time performance taking Geant4 version 10.0 as a reference, 
starting with Geant4 version 9.4.p02 (2010) and ending with 
Geant4 version 10.2 (2015) for the standalone CMS application and a simple 
calorimeter configuration made of Cu-Scintillator in a 4~Tesla magnetic 
field~\cite{g4qa}. The study 
is performed for 50~GeV $e^{-}$, $\pi^{-}$, and protons, as well as for
$H \rightarrow ZZ $ events. In average, the time performance 
improvement through the life of the LHC 
experiments (2010-2015) is of the order of 35\%. All tests were performed on 
the same hardware (AMC Opteron 6128 HE @ 2 GHz) using the same operating 
system. Remarkably, the percentage time performance improvement during the 
period of time shown in the plots is in the double digits, even as the physics 
models were improved significantly for accuracy, something that typically comes 
associated with a time performance penalty.

The cost of simulation presented before refers only to the purchase, 
life-cycle, operations, and maintenance of the computing resources allocated
to the task. The values do not include the design, development, validation, 
operation, and support of the simulation tool-kits, such as Geant4, or of the 
simulation software in the experiments. As a reference, during its 22 years 
of existence, the person-power investment in Geant4 has totaled more than 
500 person-years, equivalent to about 100~M US dollars, including fringe 
benefits and overhead. Additional investments of the same 
order have been made by the major $21^{st}$ century HEP experiments on detector 
specific simulation and reconstruction software. An interesting corollary is 
that the cost of the physics software amounts to a significant fraction of the 
cost of the detectors.

It would be an interesting exercise to estimate the cost of running a 
modern HEP experiment with and without efficient simulation tool-kits and full 
simulation software. The truth is that the experiments as we know them 
today would simply not exist without these tools. 
 How much physics would be lost to a deficient detector design or 
poor optimization? How would the design and operation of systems 
such as data acquisition, distribution, storage, and analysis workflows be 
affected? How accurate, efficient, and fast would reconstruction 
algorithms, calibration and analysis methods be? How much person-power and how 
many years of delay in delivering scientific publications would it cost to 
reproduce, without good quality simulation, the level of accuracy in 
physics measurements achieved by modern experiments? Is it possible at all to 
deliver physics of the quality and accuracy we produce today without 
simulation? 

\section{The Future} \label{future}

The accuracy of the simulation software developed for
the current generation of high-energy physics detectors,
coupled with the speed of contemporary computers, has enabled the 
experiments to perform tasks that scientists could only 
have dreamed of before. Simulation helps physicists to design and optimize 
detectors for best physics performance, stress-test the computing 
infrastructure, program data reconstruction algorithms that perform almost
as in design specifications at the begining of the experiment run, 
develop data-driven techniques for calibration and
physics analysis, and produce data samples with the properties predicted
by many candidate theories to describe currently unexplained physical
phenomena.  

Modern HEP experiments generate and handle an enormous amount of real and 
simulated data. For its size and complexity, these data has earned a place in 
the world of what is known as Big Data. The experiments at the CERN
Large Hadron Collider (LHC) have produced, reconstructed, stored, 
transferred, and analyzed tens of billion of simulated events during the
first two runs.
According to Ref.~\cite{wiredbigdata}, the amount of data collected and stored 
by the LHC experiments through the end of 2013 was of the order of 
15~PB/year, not so far from the 
180~PB/year uploaded to Facebook, the 98~PB of data in the Google search 
index, or the 15~PB/year in videos uploaded to YouTube. Integrated on time
over the last two decades, the cost of simulation and reconstruction in large 
modern HEP experiments exceeded the one hundred million dollars mark.

The high instantaneous luminosity required at the LHC experiments, needed to 
reach the 3000~fb$^{-1}$ integrated luminosity milestone associated with
the high-luminosity LHC physics program, will tax heavily the performance of 
the reconstruction algorithms. Through the end of the 2030's, 
the experiments expect to collect 150 times more data than in Run 1.
The 50~PB of raw data produced in 2016 will grow to approximately 
600~PB in 2026 while
the CPU needs will increase by a factor of about 60. The exact
numbers will depend on the approximations and the loss of information that
the experiments are willing to tolerate to keep computing performance
within the limits established by the available resources. 
Thus, the effort to improve the computing performance of the
simulation and reconstruction software requires immediate attention, in
order to restrain the 
increasing demand of computing power within the limits of flat budgets. 
Although transistor density growth is more or less 
keeping up with Moore's law, doubling every couple of years, clock speed has 
been flat since approximately 2003. Consequently, solutions must be found 
elsewhere, leveraging the core count growth in multicore machines, using new 
generation coprocessors, and re-engineering code under new 
programming paradigms based on concurrency and parallel programming. 
Coprocessors, or accelerators, specialize in operations such as floating point 
math or graphics. A hybrid computing model would allow to share work across 
a mixture of computers with different architectures. Each processor type could 
be used to perform different tasks depending on its nature.  

In parallel, experiments are transforming their software frameworks to support 
event 
multithreading and task-level parallelization. In the specific case of Geant4, 
the release of the first version with multithreading capability in 2013 
allowed significant savings in memory, although not in time performance. For 
the latter, expert teams are invested in R\&D programs to explore the potential 
of multithread track-level (particle-level) parallelization, improved 
instruction pipelining, data locality, and vectorization for single 
instruction multiple data. One example is the 
GeantV~\cite{geantv} project to develop the next generation detector simulation
toolkit, with a goal set to achieve a speedup factor of 2 to 5 with respect
to Geant4,
while enhancing the physics accuracy of the code and offering fast simulation
options that include machine learning techniques for fast and precise
tuning.

Breakthroughs in the design of simulation and reconstruction code, 
exploiting the benefits of fine granularity parallelism in applications 
running in modern computer architectures, will be essential to address the 
software and computing challenges faced by the HEP experiments of the 
21st century. 

\section*{Acknowledgements}

Funding:  Fermilab is operated by Fermi Research Alliance, LLC
under Contract No. DE-AC02-07CH11359 with the U.S. Department of Energy. 

I would like to express my gratitude to the many people I have consulted on 
the material to include in this review. First and foremost, I am eternally
grateful to John Harvey, former
leader of the software group at CERN, who initially asked me a few 
questions about the impact 
of simulation in experimental particle physics, eventually encouraged me to 
write this article, and was the first person to read and comment on a draft. 
I am also grateful to my Geant4 co-collaborators who built on previous 
Geant experience and developed the most widespread and successful detector 
simulation toolkit the HEP field has ever known. Their work has yet to 
receive the recognition and support it deserves. I want to thank 
Federico Carminati,
Guenther Dissertori, and Paris
Sphicas for reading and commenting on a draft. Thanks to my
ATLAS colleagues John Chapman, Andrea Dotti,  Zach Marshall, and 
Ariel Schwartzman for
pointing me to outstanding ATLAS material on simulation, test beams, jets,
and missing transverse energy.  
In order of topic appearance in 
the paper, I would like to thank Julia Yarba and Alberto Ribon for pointing 
me to specific Geant4 physics validation plots and associated information, 
Sunanda Banerjee for useful discussions and the figures with the CMS test beam 
and single particle response comparisons, Mike Tartaglia and Harrison Prosper 
for trying hard to recall details of the D0 test beam experiments, my many 
D0 colleagues with whom we navigated the difficult waters of producing high 
quality physics measurements with neither a solenoidal field nor 
fast-enough high-quality simulation software during Run 1, 
Liz Sexton-Kennedy and Robert Harris for 
discussions on the CDF simulation software and first inclusive jet cross 
section results, Soon Young Jun for providing precise information and 
documentation on the CDF Monte Carlo tuning effort and for the Geant4 
computing performance plots, Marjorie Shapiro for private communication
on the CDF QFL fast simulation software, my CMS colleagues with whom we 
experienced the thrill of developing outstanding simulation code and 
producing the most precise physics measurements ever in a hadron collider, 
Kevin Burkett for 
pointing me to CDF publications data statistics, 
Oliver Gutsche for providing most of the 
information for the cost evaluation of the simulation operation in CMS. Last 
but not least, I want to thank Krzysztof Genser and the rest of the 
members of my Fermilab Physics and Detector Simulation group (PDS)
for their hard work in the area of simulation software research, 
development 
and support, as well as the Fermilab Scientific Computing Division and the US 
Department of Energy for their continued support of the Fermilab Geant 
operations and research programs. 





\bibliography{ImpDSPPCE-DE-Accepted}

\begin{thebibliography}{10}
\expandafter\ifx\csname url\endcsname\relax
  \def\url#1{\texttt{#1}}\fi
\expandafter\ifx\csname urlprefix\endcsname\relax\def\urlprefix{URL }\fi
\expandafter\ifx\csname href\endcsname\relax
  \def\href#1#2{#2} \def\path#1{#1}\fi

\bibitem{geant4}
{\rm S. Agostineli et al. (Geant4 Collaboration)}, {\rm Gean4-a simulation
  toolkit}, Nucl. Instrum. Meth. A 506 (2003) 250--303.
\newblock \href {http://dx.doi.org/10.1016/S0168-9002(03)01368-8}
  {\path{doi:10.1016/S0168-9002(03)01368-8}}.

\bibitem{geant4R}
{\rm J. Allison et al. (Geant4 Collaboration)}, {\rm Recent developments in
  Geant4}, Nucl. Instrum. Meth. A 835 (2016) 186--225.
\newblock \href {http://dx.doi.org/10.1016/j.nima.2016.06.125}
  {\path{doi:10.1016/j.nima.2016.06.125}}.

\bibitem{egs}
R.~Ford, W.~Nelson, {\rm The EGS Code System - Version 3}, Stanford Linear
  Accelerator Center Report SLAC-210.

\bibitem{geant3}
R.~Brun, F.~Bruyant, M.~Maire, A.~McPherson, P.~Zanarini, {\rm Geant3}
  CERN-DD-EE-84-1.

\bibitem{atlashiggs}
{\rm The ATLAS Collaboration}, {\rm Observation of a new particle in the search
  for the Standard Model Higgs boson with the ATLAS detector at the LHC}, Phys.
  Lett. B 716 (2012) 1--29.
\newblock \href {http://dx.doi.org/10.1016/j.physletb.2012.08.020}
  {\path{doi:10.1016/j.physletb.2012.08.020}}.

\bibitem{cmshiggs}
{\rm The CMS Collaboration}, {\rm Observation of a new boson at a mass of
  125~GeV with the CMS experiment at the LHC}, Phys. Lett. B 716 (2012) 30--61.
\newblock \href {http://dx.doi.org/10.1016/j.physletb.2012.08.021}
  {\path{doi:10.1016/j.physletb.2012.08.021}}.

\bibitem{cmssimulationL}
{\rm D.J. Lange et al. for the CMS Collaboration}, {\rm Upgrades for the CMS
  full simulation}, J. Phys. Conf. Ser. 608 (2015) 012056.

\bibitem{cmssimulationB}
{\rm S. Banerjee et al. for the CMS Collaboration}, {\rm Validation and tuning
  of the CMS simulation}, J. Phys. Conf. Ser. 331 (2011) 032015.

\bibitem{cmssimulationE}
{\rm V. Daniel Elvira for the CMS Collaboration}, {\rm Readiness of the CMS
  detector simulation}, NSS/MIC 2007 C07-10-28 (2007) 2081--2085.
\newblock \href {http://dx.doi.org/10.1109/NSSMIC.2007.4436563}
  {\path{doi:10.1109/NSSMIC.2007.4436563}}.

\bibitem{atlassimulationC}
{\rm P. J. Clark for the ATLAS Collaboration}, {\rm The ATLAS Detector
  Simulation}, Nucl. Phys. B - Proceedings Supplements 215 (2011) 85--88.
\newblock \href {http://dx.doi.org/10.1016/j.nuclphysbps.2011.03.142}
  {\path{doi:10.1016/j.nuclphysbps.2011.03.142}}.

\bibitem{atlassimulationM}
{\rm Z. Marshall for the ATLAS Collaboration}, {\rm Validation and performance
  studies for the ATLAS simulation}, J. Phys. Conf. Ser. 219 (2010) 032016.

\bibitem{atlassimulationR}
{\rm A. Rimoldi for the ATLAS Collaboration}, {\rm The ATLAS detector
  simulation application}, Nucl. Phys. B - Proceedings Supplements 172 (2007)
  49--52.
\newblock \href {http://dx.doi.org/10.1016/j.nuclphysbps.2007.07.005}
  {\path{doi:10.1016/j.nuclphysbps.2007.07.005}}.

\bibitem{gflash}
{\rm G. Grindhammer et al.}, {\rm The Fast Simulation of Electromagnetic and
  Hadronic Showers}, Nucl. Instrum. Meth. A 290 (1990) 469.

\bibitem{fluka}
A.~Ferrari, P.~Sala, A.~Fasso, , J.~Ranft, {\rm FLUKA: a multi-particle
  transport code}, CERN-2005-10 (2005), INFN/TC\_05/11, SLAC-R-773.

\bibitem{mars}
{\rm The MARS Code System} http://mars.fnal.gov.

\bibitem{calice}
{\rm C. Adloff et al.}, {\rm Study of the interactions of pions in the CALICE
  silicon-tungsten calorimeter prototype}, JINST 5 (2010) 05007.
\newblock \href {http://dx.doi.org/10.1088/1748-0221/5/05/P05007}
  {\path{doi:10.1088/1748-0221/5/05/P05007}}.

\bibitem{harp1}
{\rm M. Apollonio et al. (HARP Collaboration)}, {\rm Forward production of
  charged pions with incident $\pi^+/-$ on nuclear targets measured at the CERN
  PS}, Nucl. Phys. A 821 (2009) 118--192.
\newblock \href {http://dx.doi.org/10.1016/j.nuclphysa.2009.01.080}
  {\path{doi:10.1016/j.nuclphysa.2009.01.080}}.

\bibitem{harp2}
{\rm M. Apollonio et al. (HARP Collaboration)}, {\rm Large-angle production of
  charged pions with incident pion beams on nuclear targets}, Phys. Rev. C 80
  (2009) 065207.
\newblock \href {http://dx.doi.org/10.1103/PhysRevC.80.065207}
  {\path{doi:10.1103/PhysRevC.80.065207}}.

\bibitem{harp3}
{\rm M. G. Catanesi et al. (HARP Collaboration)}, {\rm Large-angle production
  of charged pions with 3-12~GeV/c incident protons on nuclear targets}, Phys.
  Rev. C 77 (2008) 055207.
\newblock \href {http://dx.doi.org/10.1103/PhysRevC.77.055207}
  {\path{doi:10.1103/PhysRevC.77.055207}}.

\bibitem{na49a}
{\rm C. Alt et al. (NA49 Collaboration)}, {\rm Inclusive production of charged
  pions in p+C collisions at 158~GeV/c beam momentum}, Eur. Phys. J. C 49
  (2007) 897--917.
\newblock \href {http://dx.doi.org/10.1140/epjc/s10052-006-0165-7}
  {\path{doi:10.1140/epjc/s10052-006-0165-7}}.

\bibitem{na49b}
{\rm C. Alt et al. (NA49 Collaboration)}, {\rm Inclusive production of charged
  pions in p+p collisions at 158~GeV/c beam momentum}, Eur. Phys. J. C 45
  (2006) 343--381.
\newblock \href {http://dx.doi.org/10.1140/epjc/s2005-02391-9}
  {\path{doi:10.1140/epjc/s2005-02391-9}}.

\bibitem{na49c}
{\rm T. Anticic et al. (NA49 Collaboration)}, {\rm Inclusive production of
  protons, anti-protons and neutrons in p+p collisions at 158~GeV/c beam
  momentum}, Eur. Phys. J. C 65 (2010) 9--63.
\newblock \href {http://dx.doi.org/10.1140/epjc/s10052-009-1172-2}
  {\path{doi:10.1140/epjc/s10052-009-1172-2}}.

\bibitem{na61a}
{\rm N. Abgrall et al. (NA61/SHINE Collaboration)}, {\rm Measurement of
  production properties of positively charged kaons in proton-carbon
  interactions at 31~GeV/c}, Phys. Rev. C 85 (2012) 035210.
\newblock \href {http://dx.doi.org/10.1103/PhysRevC.85.035210}
  {\path{doi:10.1103/PhysRevC.85.035210}}.

\bibitem{na61b}
{\rm N. Abgrall et al. (NA61/SHINE Collaboration)}, {\rm Measurements of cross
  sections and charged pion spectra in proton-carbon interactions at 31~GeV/c},
  Phys. Rev. C 84 (2011) 034604.
\newblock \href {http://dx.doi.org/10.1103/PhysRevC.84.034604}
  {\path{doi:10.1103/PhysRevC.84.034604}}.

\bibitem{fritiof}
{\rm V.V. Uzhinsky for the Geant4 Hadronic Working Group}, {\rm The Fritiof
  (FTF) Model in Geant4}, International Conference on Calorimetry for the High
  Energy Frontier (CHEF 2013) C13-04-22.4 (2013) 260--264.

\bibitem{bertini}
D.~Wright, M.~Kelsey, {\rm The Geant4 Bertini Cascade}, Nucl. Instrum. Meth. A
  804 (2015) 175--188.
\newblock \href {http://dx.doi.org/10.1016/j.nima.2015.09.058}
  {\path{doi:10.1016/j.nima.2015.09.058}}.

\bibitem{itep}
{\rm Yu. D. Bayukov et al.}, {\rm Angular Dependences Of Inclusive Nucleon
  Production In Nuclear Reactions At High-energies And Separation Of
  Contributions From Quasifree And Deep Inelastic Nuclear Processes}, Sov. J.
  Nucl. Phys. 42 (1985) 116.

\bibitem{cmsg4valid}
{\rm S. Banerjee for the CMS Collaboration}, {\rm Validation of Physics Models
  of Geant4 using data from CMS experiment}, To be published in Journal of
  Physics Conference Series (JPCS). Proceedings of the CHEP 2016 Conference,
  San Francisco
  https://twiki.cern.ch/twiki/bin/viewauth/CMS/Geant4Validation2016.

\bibitem{atlasg4valid}
{\rm The ATLAS Collaboration}, {\rm A measurement of the calorimeter response
  to single hadrons and determination of the jet energy scale uncertainty using
  LHC Run-1 pp-collision data with the ATLAS detector}, Eur. Phys. J. C (2017)
  77:26.
\newblock \href {http://dx.doi.org/10.1140/epjc/s10052-016-4580-0}
  {\path{doi:10.1140/epjc/s10052-016-4580-0}}.

\bibitem{atlastb}
ATLAS Public Results https://twiki.cern.ch/twiki/bin/view/AtlasPublic/
  ApprovedPlotsTileTestBeamResults.

\bibitem{d0testbeam}
{\rm The D0 Collaboration}, {\rm Beam tests of the D0 uranium liquid argon end
  calorimeters}, Nucl. Instrum. Meth. A 324 (1993) 53--76.
\newblock \href {http://dx.doi.org/10.1016/0168-9002(93)90965-K}
  {\path{doi:10.1016/0168-9002(93)90965-K}}.

\bibitem{d0jes1}
{\rm The D0 Collaboration}, {\rm Jet energy scale determination in the D0
  experiment}, Nucl. Instrum. Meth. A 763 (2014) 442--475.
\newblock \href {http://dx.doi.org/10.1016/j.nima.2014.05.044}
  {\path{doi:10.1016/j.nima.2014.05.044}}.

\bibitem{d0jes2}
{\rm The D0 Collaboration}, {\rm Determination of the Absolute Jet Energy Scale
  in the DZERO Calorimeters}, Nucl. Instrum. Meth. A 424 (1999) 352--394.
\newblock \href {http://dx.doi.org/10.1016/S0168-9002(98)01368-0}
  {\path{doi:10.1016/S0168-9002(98)01368-0}}.

\bibitem{cdfjes}
{\rm A. Bhatti et al.}, {\rm Determination of the jet energy scale at the
  collider detector at Fermilab}, Nucl. Instum. Meth. A 566 (2006) 375--412.
\newblock \href {http://dx.doi.org/10.1016/j.nima.2006.05.269}
  {\path{doi:10.1016/j.nima.2006.05.269}}.

\bibitem{cmsjes}
{\rm The CMS Collaboration}, {\rm Determination of Jet Energy Calibration and
  Transverse Momentum Resolution in CMS}, JINST 6 (2011) 11002.
\newblock \href {http://dx.doi.org/10.1088/1748-0221/6/11/P11002}
  {\path{doi:10.1088/1748-0221/6/11/P11002}}.

\bibitem{atlasjes}
{\rm The ATLAS Collaboration}, {\rm Jet energy measurement with the ATLAS
  detector in proton--proton collisions at $\sqrt{s} = 7$~TeV}, Eur. Phys. J. C
  73 (2013) 2304.
\newblock \href {http://dx.doi.org/10.1140/epjc/s10052-013-2304-2}
  {\path{doi:10.1140/epjc/s10052-013-2304-2}}.

\bibitem{pflow}
{\rm Florian Beaudette for the CMS Collaboration}, {\rm The CMS Particle Flow
  Algorithm}, International Conference on Calorimetry for the High Energy
  Frontier (CHEF 2013) C13-04-22.4 (2013) 295--304.

\bibitem{atlasjesplots}
{\rm The ATLAS Collaboration}, {\rm Jet energy measurement and its systematic
  uncertainty in proton--proton collisions at $\sqrt{s}=7$~TeV with the ATLAS
  detector}, Eur. Phys. J. C 75 (2015) 17.
\newblock \href {http://dx.doi.org/10.1140/epjc/s10052-014-3190-y}
  {\path{doi:10.1140/epjc/s10052-014-3190-y}}.

\bibitem{atlasjerplots}
{\rm The ATLAS Collaboration}, {\rm Jet energy resolution in proton-proton
  collisions at $\sqrt{s}=7$~TeV recorded in 2010 with the ATLAS detector},
  Eur. Phys. J. C 73 (2013) 2306.
\newblock \href {http://dx.doi.org/10.1140/epjc/s10052-013-2306-0}
  {\path{doi:10.1140/epjc/s10052-013-2306-0}}.

\bibitem{susy}
S.~P. Martin, {\rm A Supersymmetry Primer} arXiv:hep-ph/9709356.

\bibitem{ra2}
{\rm The CMS Collaboration}, {\rm Search for New Physics with Jets and Missing
  Transverse Momentum in pp collisions at $\sqrt{s} = 7$~TeV}, JHEP 8 (2011)
  155.
\newblock \href {http://dx.doi.org/10.1007/JHEP08(2011)155}
  {\path{doi:10.1007/JHEP08(2011)155}}.

\bibitem{tschum}
T.~Schum, {\rm QCD Processes and Search for Supersymmetry at the LHC},
  DESY-THESIS-2012-029.

\bibitem{muonpog}
{\rm The CMS Collaboration}, {\rm Performance of CMS muon reconstruction in pp
  collision events at $\sqrt{s} = 7$~TeV}, JINST 7 (2012) 10002.
\newblock \href {http://dx.doi.org/10.1088/1748-0221/7/10/P10002}
  {\path{doi:10.1088/1748-0221/7/10/P10002}}.

\bibitem{electronpog}
{\rm The CMS Collaboration}, {\rm Performance of electron reconstruction and
  selection with the CMS detector in proton-proton collisions at $\sqrt{s} =
  8$~TeV}, JINST 10 (2015) 06005.
\newblock \href {http://dx.doi.org/10.1088/1748-0221/10/06/P06005}
  {\path{doi:10.1088/1748-0221/10/06/P06005}}.

\bibitem{atlasmuon}
{\rm The ATLAS Collaboration}, {\rm Measurement of the muon reconstruction
  performance of the ATLAS detector using 2011 and 2012 LHC proton-proton
  collision data}, Eur. Phys. J. C 74 (2014) 3130.
\newblock \href {http://dx.doi.org/10.1140/epjc/s10052-014-3130-x}
  {\path{doi:10.1140/epjc/s10052-014-3130-x}}.

\bibitem{atlaselectron}
{\rm The ATLAS Collaboration}, {\rm Electron efficiency measurements with the
  ATLAS detector using 2012 LHC proton-proton collision data}, Submitted to
  Eur. Phys. J. C arXiv:1612.01456 [hep-ex].

\bibitem{susymultijets}
{\rm The CMS Collaboration}, {\rm Search for new physics in the multijet and
  missing transverse momentum final state in proton-proton collisions at
  $\sqrt{s} = 8$~TeV}, JHEP 6 (2014) 055.
\newblock \href {http://dx.doi.org/10.1007/JHEP06(2014)055}
  {\path{doi:10.1007/JHEP06(2014)055}}.

\bibitem{cmsupgrade}
{\rm The CMS Collaboration}, {\rm CMS Phase II Upgrade Scope Document},
  CERN-LHCC-2015-019 https://cds.cern.ch/record/2055167.

\bibitem{delphes}
{\rm J.de Favereau et al.}, {\rm DELPHES 3: a modular framework for fast
  simulation of a generic collider experiment}, JHEP 02 (2014) 057.
\newblock \href {http://dx.doi.org/10.1007/JHEP02(2014)057}
  {\path{doi:10.1007/JHEP02(2014)057}}.

\bibitem{atlasupgrade}
{\rm The ATLAS Collaboration}, {\rm ATLAS Phase II Upgrade Scope Document},
  CERN-LHCC-2015-020 https://cds.cern.ch/record/2055248.

\bibitem{cerngrid}
{\rm WLCG: Worldwide LHC Computing Grid},
  http://wlcg-public.web.cern.ch/tier-centres.

\bibitem{csa2008}
D.~Futyan, R.~Mankel, {\rm Christoph Paus for the CMS Collaboration}, {\rm The
  CMS Computing, Software and Analysis Challenge}, Journal of Physics:
  Conference Series 219 (2010) 032008.
\newblock \href {http://dx.doi.org/10.1088/1742-6596/219/3/032008}
  {\path{doi:10.1088/1742-6596/219/3/032008}}.

\bibitem{trackperf}
{\rm The CMS Collaboration}, {\rm Description and performance of track and
  primary-vertex reconstruction with the CMS tracker}, JINST 9 (2014) 10009.
\newblock \href {http://dx.doi.org/10.1088/1748-0221/9/10/P10009}
  {\path{doi:10.1088/1748-0221/9/10/P10009}}.

\bibitem{trackmat}
{\rm The CMS Collaboration}, {\rm Studies of Tracker Material}, CMS Public
  Note, CMS-PAS-TRK-10-003 https://cds.cern.ch/record/1279138.

\bibitem{atlasphotconv}
{\rm The ATLAS Collaboration}, {\rm Photon Conversions at $\sqrt{s}=900$~GeV
  measured with the ATLAS Detector}, ATLAS Conference Note, ATLAS-CONF-2010-007
  https://cds.cern.ch/record/1274001.

\bibitem{atlaskshort}
{\rm The ATLAS Collaboration}, {\rm Study of the Material Budget in the ATLAS
  Inner Detector with $K_{S}^{0}$ decays in collision data at
  $\sqrt{s}=900$~GeV}, ATLAS Conference Note, ATLAS-CONF-2010-019
  https://cds.cern.ch/record/1277651.

\bibitem{btaggingperf}
{\rm The CMS Collaboration}, {\rm Identification of b-quark jets with the CMS
  experiment}, JINST 8 (2013) 04013.
\newblock \href {http://dx.doi.org/10.1088/1748-0221/8/04/P04013}
  {\path{doi:10.1088/1748-0221/8/04/P04013}}.

\bibitem{btaggingperfatlas}
{\rm The ATLAS Collaboration}, {\rm Performance of b-jet identification in the
  ATLAS experiment}, JINST 11 (2016) 04008.
\newblock \href {http://dx.doi.org/10.1088/1748-0221/11/04/P04008}
  {\path{doi:10.1088/1748-0221/11/04/P04008}}.

\bibitem{pythia1}
{\rm Torbjorn Sjostrand et al.}, {\rm An Introduction to PYTHIA 8.2},
  arXiv:1410.3012 [hep-ph]\href {http://dx.doi.org/10.1016/j.cpc.2015.01.024}
  {\path{doi:10.1016/j.cpc.2015.01.024}}.

\bibitem{pythia2}
T.~Sjostrand, S.~Mrenna, P.~Skands, {\rm PYTHIA 6.4 Physics and Manual}, JHEP05
  (2006) 026\href {http://dx.doi.org/10.1088/1126-6708/2006/05/026}
  {\path{doi:10.1088/1126-6708/2006/05/026}}.

\bibitem{madgraph}
J.~Alwall, R.~Frederix, S.~Frixione, V.~Hirschi, F.~Maltoni, O.~Mattelaer,
  H.-S. Shao, T.~Stelzer, P.~Torrielli, M.~Zaro, {\rm The automated computation
  of tree-level and next-to-leading order differential cross sections, and
  their matching to parton shower simulations}, JHEP07 (2014) 079\href
  {http://dx.doi.org/10.1007/JHEP07(2014)079}
  {\path{doi:10.1007/JHEP07(2014)079}}.

\bibitem{metperf}
{\rm The CMS Collaboration}, {\rm Performance of the CMS missing transverse
  energy reconstruction in $pp$ data at $\sqrt{s} = 8$~TeV}, JINST 10 (2015)
  2006.
\newblock \href {http://dx.doi.org/10.1088/1748-0221/10/02/P02006}
  {\path{doi:10.1088/1748-0221/10/02/P02006}}.

\bibitem{wzprodatlas}
{\rm The ATLAS Collaboration}, {\rm Measurement of $W^{\pm}$ and $Z$-boson
  production cross sections in pp collisions at $\sqrt{s}=13$~TeV with the
  ATLAS detector}, Phys. Lett. B 759 (2016) 601.
\newblock \href {http://dx.doi.org/10.1016/j.physletb.2016.06.023}
  {\path{doi:10.1016/j.physletb.2016.06.023}}.

\bibitem{metperfatlas}
{\rm The ATLAS Collaboration}, {\rm Performance of algorithms that reconstruct
  missing transverse momentum in $\sqrt{s}=8$~TeV proton–proton collisions in
  the ATLAS detector}, Submitted to Eur. Phys. J. C arXiv:1609.09324 [hep-ex].

\bibitem{atlasjetxs12}
{\rm The ATLAS Collaboration}, {\rm Measurement of inclusive jet and dijet
  production in $pp$ collisions at $\sqrt{s} = 7$~TeV using the ATLAS
  detector}, Phys. Rev. D 86 (2012) 014022.
\newblock \href {http://dx.doi.org/10.1103/PhysRevD.86.014022}
  {\path{doi:10.1103/PhysRevD.86.014022}}.

\bibitem{cmsjetxs11}
{\rm The CMS Collaboration}, {\rm Measurement of the Inclusive Jet Cross
  Section in $pp$ Collisions at $\sqrt{s} = 7$~TeV}, Phys. Rev. Lett. 107
  (2011) 132001.
\newblock \href {http://dx.doi.org/10.1103/PhysRevLett.107.132001}
  {\path{doi:10.1103/PhysRevLett.107.132001}}.

\bibitem{cdfjetxs96}
{\rm The CDF Collaboration}, {\rm Inclusive Jet Cross Section in $\overline{p}
  p$ Collisions at $\sqrt{s}=1.8$~TeV}, Phys. Rev. Lett. 77 (1996) 438.
\newblock \href {http://dx.doi.org/10.1103/PhysRevLett.77.438}
  {\path{doi:10.1103/PhysRevLett.77.438}}.

\bibitem{d0jetxs99}
{\rm The D0 Collaboration}, {\rm Inclusive Jet Cross Section in $\overline{p}
  p$ Collisions at $\sqrt{s}=1.8$~TeV}, Phys. Rev. Lett. 82 (1999) 2451.
\newblock \href {http://dx.doi.org/10.1103/PhysRevLett.82.2451}
  {\path{doi:10.1103/PhysRevLett.82.2451}}.

\bibitem{atlasjetxs10}
{\rm The ATLAS Collaboration}, {\rm Measurement of inclusive jet and dijet
  cross sections in proton-proton collisions at 7~TeV centre-of-mass energy
  with the ATLAS detector}, Eur. Phys. J. C 71 (2011) 1512.
\newblock \href {http://dx.doi.org/10.1140/epjc/s10052-010-1512-2}
  {\path{doi:10.1140/epjc/s10052-010-1512-2}}.

\bibitem{d0jetxs01}
{\rm The D0 Collaboration}, {\rm Inclusive Jet Production in $\overline{p} p$
  Collisions}, Phys. Rev. Lett. 86 (2001) 1707--1712.
\newblock \href {http://dx.doi.org/10.1103/PhysRevLett.86.1707}
  {\path{doi:10.1103/PhysRevLett.86.1707}}.

\bibitem{cdfjetxs08}
{\rm The CDF Collaboration}, {\rm Measurement of the inclusive jet cross
  section at the Fermilab Tevatron $p \overline{p}$ collider using a cone-based
  jet algorithm}, Phys. Rev. D 78 (2008) 052006.
\newblock \href {http://dx.doi.org/10.1103/PhysRevD.78.052006}
  {\path{doi:10.1103/PhysRevD.78.052006}}.

\bibitem{d0jetxs12}
{\rm The D0 Collaboration}, {\rm Measurement of the inclusive jet cross section
  in $p \overline{p}$ collisions at $\sqrt{s}=1.96$~TeV}, Phys. Rev. D 85
  (2012) 052006.
\newblock \href {http://dx.doi.org/10.1103/PhysRevD.85.052006}
  {\path{doi:10.1103/PhysRevD.85.052006}}.

\bibitem{cdfjetxs89}
{\rm The CDF Collaboration}, {\rm Measurement of the Inclusive Jet Cross
  Section in $\overline{p} p$ Collisions at $\sqrt{s}=1.8$~TeV}, Phys. Rev.
  Lett. 62 (1989) 613.
\newblock \href {http://dx.doi.org/10.1103/PhysRevLett.62.613}
  {\path{doi:10.1103/PhysRevLett.62.613}}.

\bibitem{cdfjetxs92}
{\rm The CDF Collaboration}, {\rm Inclusive jet cross section in $\overline{p}
  p$ collisions at $\sqrt{s} = 1.8$~TeV}, Phys. Rev. Lett. 68 (1992) 1104.
\newblock \href {http://dx.doi.org/10.1103/PhysRevLett.68.1104}
  {\path{doi:10.1103/PhysRevLett.68.1104}}.

\bibitem{cdfpubtime}
{\rm Summary Table} https://www-cdf.fnal.gov/physics/physics.html.

\bibitem{d0pubtime}
{\rm Year-to-year history of paper submissions}
  https://www-d0.fnal.gov/Run2Physics/WWW/results.htm.

\bibitem{cmspubtime}
{\rm CMS Publications versus Time}
  http://cms-results.web.cern.ch/cms-results/public-results/publications-vs-ti%
me/.

\bibitem{geantv}
{\rm J. Apostolakis et al. (GeantV Development Group)}, {\rm Adaptive track
  scheduling to optimize concurrency and vectorization in GeantV}, J. Phys.:
  Conf. Ser. 608 (2015) 012003.
\newblock \href {http://dx.doi.org/10.1088/1742-6596/608/1/012003}
  {\path{doi:10.1088/1742-6596/608/1/012003}}.

\bibitem{g4qa}
{\rm Source: Geant4 Testing and Quality Assurance Working Group}
  https://g4cpt.fnal.gov/g4p/prplots/cpu\_by\_version.html.

\bibitem{wiredbigdata}
J.~Pearlstein, {\rm Information Revolution: Big Data Has Arrived at an Almost
  Unimaginable Scale}, Wired Magazine https://www.wired.com/2013/04/bigdata.

\end{thebibliography}

\end{document}